\documentclass[10pt,twocolumn,letterpaper]{article}

\usepackage[pagenumbers]{cvpr} 

\usepackage{graphicx}
\usepackage{amsmath}
\usepackage{amssymb}
\usepackage{booktabs}

\usepackage{mathtools}
\usepackage{amsthm}

\usepackage{multirow}

\usepackage[dvipsnames]{xcolor}

\theoremstyle{plain}

\theoremstyle{definition}

\theoremstyle{remark}

\definecolor{cvprblue}{rgb}{0.21,0.49,0.74}
\usepackage[pagebackref,breaklinks,colorlinks,citecolor=cvprblue]{hyperref}

\usepackage[accsupp]{axessibility}  

\usepackage{xcolor, soul}
\sethlcolor{pink}

\usepackage{pgfplots}
    \usepgfplotslibrary{colorbrewer}
    \pgfplotsset{
        cycle list/Dark2,
        cycle multiindex* list={
            mark list*\nextlist
            Dark2\nextlist
        },
    }
\pgfplotsset{compat=1.14}

\definecolor{Cerulean_dark}{rgb}{0.0, 0.30, 0.47}
\definecolor{goldenpoppy2}{rgb}{0.95, 0.72, 0.0}
\definecolor{flamingopink_dark}{rgb}{0.75, 0.32, 0.43}
\definecolor{smokyblack2}{rgb}{0.25, 0.25, 0.25}

\definecolor{ForestGreen}{RGB}{34,139,34}
\definecolor{Cerulean}{rgb}{0.0, 0.48, 0.65}
\definecolor{OrangeRed}{RGB}{255,69,0}
\definecolor{OrangeRed2}{RGB}{240,9,30}
\definecolor{goldenpoppy}{rgb}{0.99, 0.76, 0.0}
\definecolor{goldenpoppy3}{rgb}{0.99, 0.46, 0.1}
\definecolor{skyblue}{rgb}{0.53, 0.81, 0.92}
\definecolor{skyblue2}{rgb}{0.55, 0.83, 0.94}
\definecolor{red-violet}{rgb}{0.78, 0.08, 0.52}
\definecolor{darkcerulean}{rgb}{0.03, 0.33, 0.55}
\definecolor{darkcerulean2}{rgb}{0.00, 0.20, 0.35}
\definecolor{flamingopink}{rgb}{0.99, 0.56, 0.67}
\definecolor{caribbeangreen}{rgb}{0.0, 0.8, 0.6}
\definecolor{caribbeangreen2}{rgb}{0.0, 0.825, 0.625}
\definecolor{darkpastelpurple}{rgb}{0.69, 0.34, 0.84}
\definecolor{darkpastelpurple2}{rgb}{0.68, 0.33, 0.83}
\definecolor{smokyblack}{rgb}{0.1, 0.1, 0.1}
\usepackage{standalone}

\usepackage{rotating}

\usepackage{booktabs}
\usepackage{array}
\newcolumntype{P}[1]{>{\raggedright\arraybackslash}p{#1}}
\usepackage{makecell}      

\title{Evaluating Computational Pathology Foundation Models for\\ Prostate Cancer Grading under Distribution Shifts}

\author{Fredrik K. Gustafsson$^{1,2}$
\and
Mattias Rantalainen$^{*,1}$\vspace{2.0mm}
\and
$^{1}$Department of Medical Epidemiology and Biostatistics, Karolinska Institutet, Stockholm, Sweden\\
$^{2}$Department of Engineering Science, University of Oxford, Oxford, UK\\
\small$^{*}$Corresponding Author, {\tt mattias.rantalainen@ki.se}
}

\begin{document}

\maketitle

\begin{abstract}
    Pathology foundation models (PFMs) have emerged as powerful pretrained encoders for computational pathology, but their robustness under clinically relevant distribution shifts remains insufficiently understood. We benchmark the robustness of recent PFMs in the setting of prostate cancer grading from whole-slide images (WSIs). Using the PANDA dataset, we evaluate PFMs as frozen patch-level feature extractors within weakly supervised slide-level grading models, and assess robustness to two important forms of distribution shift: shifts in WSI image appearance across collection sites, and shifts in the label distribution over cancer grade groups. Across in-distribution settings, PFMs consistently achieve strong performance and clearly outperform a natural-image baseline. Under cross-site transfer from Radboud to Karolinska, however, performance drops substantially for all models, showing that large-scale pretraining alone does not guarantee robust downstream generalization. In contrast, PFMs are less sensitive to label-distribution shift, indicating that visually grounded domain shift is the dominant challenge. Representation analysis further supports these findings by revealing persistent domain separation between sites across all PFMs. While grade-related structure is present, it is comparatively weak, indicating that domain-related variation dominates in the learned feature space. Together, these results provide a comprehensive benchmark of PFMs under distribution shift and highlight an important practical message: although PFMs provide strong representations, generalisability remains constrained by the quality and diversity of the data used to train downstream prediction models.

\end{abstract}
\vspace{-10.0mm}

\section*{}

Computational pathology uses machine learning and computer vision methods to automatically extract useful information from histopathology whole-slide images (WSIs)~\cite{acs2020artificial, echle2021deep, cifci2022artificial, shmatko2022artificial, campanella2019clinical}. Important applications include histological grading~\cite{strom2020artificial, wang2022improved, bulten2022artificial}, risk stratification~\cite{hohn2023colorectal, jiang2024end, volinsky2024prediction}, and biomarker prediction~\cite{coudray2018classification, wang2021predicting, weitz2022transcriptome, niehues2023generalizable, arslan2024systematic}. Foundation models, i.e.\ large deep learning models trained on large amounts of data using self-supervised or multimodal learning~\cite{bommasani2021opportunities, azad2023foundational, moor2023foundation}, have recently become a major research direction within the field~\cite{chen2024uni, conch2024, gigapath2024, virchow2024, virchow22024, filiot2023scaling, nechaev2024hibou, aben2024towards, bilal2025foundation, li2025survey}. These pathology foundation models (PFMs) are intended to serve as general-purpose feature extractors, with the promise of transferring well across diverse downstream tasks. 

A central challenge, however, is that real-world pathology image data exhibits substantial variability, arising from differences in tissue processing, staining protocols, scanner hardware, and local clinical practice~\cite{Shah2025TissueThickness, lin2025institutionbias, Jahanifar2025DomainGen, thiringer2026scanner}. A practically useful PFM should therefore not only perform well in-distribution, but also remain robust under the distribution shifts that arise naturally during deployment~\cite{quionero2009dataset, finlayson2021clinician}. While the broader machine learning literature has repeatedly shown that deep learning systems can be highly sensitive to distribution shifts~\cite{hendrycks2018benchmarking, ovadia2019can, koh2021wilds, gustafsson2023how}, this question remains comparatively underexplored for PFMs.

\begin{figure*}[t]
    \centering
    \begin{tikzpicture}
  \centering
  \begin{axis}[
        ybar=2.5pt, axis on top,
        title={},
        height=4.925cm, width=23.0cm, 
        bar width=0.26cm, 
        ymajorgrids, tick align=inside,
        enlarge y limits={value=.1,upper},
        ymin=0, ymax=0.9,
        set layers,
        ytick={0.2, 0.4, 0.6, 0.8},
        axis x line*=bottom,
        axis y line*=left,
        y axis line style={opacity=0},
        tickwidth=0pt,
        enlarge x limits=0.20, 
        extra y tick style={
            grid style={
                solid,
                line width = 2.5pt,
                /pgfplots/on layer=axis background,
            },
        },
        tick style={
            grid style={
                dashed,
                /pgfplots/on layer=axis background,
            },
        },
        legend style={
            at={(0.5,-0.2)},
            anchor=north,
            legend columns=-1,
            /tikz/every even column/.append style={column sep=0.325cm},
            font=\footnotesize
        },
        ylabel={Kappa ($\uparrow$)},
        ylabel style={font=\footnotesize},
        symbolic x coords={
           PANDA,
           Karolinska,
           Radboud, 
           Radboud $\rightarrow$ Karolinska},
       xtick=data,
       xticklabel=\empty,
       x tick label style={text width = 2.25cm, align=center, font=\footnotesize},
    ]
    \addplot [fill=OrangeRed2, draw=OrangeRed2, error bars/.cd, y dir=both, y explicit, error bar style={thick}, error bar style=black] coordinates {
      (PANDA, 0.773048) +- (0, 0.013836)
      (Karolinska, 0.750227) +- (0, 0.0271252)
      (Radboud, 0.726182) +- (0, 0.0236357)
      (Radboud $\rightarrow$ Karolinska, 0.185957) +- (0, 0.0414844)};

    \addplot [fill=darkgray, draw=darkgray, error bars/.cd, y dir=both, y explicit, error bar style={thick}, error bar style=black] coordinates {
      (PANDA, 0.874011) +- (0, 0.0113699)
      (Karolinska, 0.874871) +- (0, 0.0112448)
      (Radboud, 0.837963) +- (0, 0.00868408)
      (Radboud $\rightarrow$ Karolinska, 0.168457) +- (0, 0.0897553)};
    
    \addplot [fill=flamingopink, draw=flamingopink, line width=0mm, error bars/.cd, y dir=both, y explicit, error bar style={thick}, error bar style=black] coordinates { 
      (PANDA, 0.8884) +- (0, 0.013359)
      (Karolinska, 0.886496 ) +- (0, 0.00971835)
      (Radboud, 0.853788) +- (0, 0.0262305)
      (Radboud $\rightarrow$ Karolinska, 0.24719) +- (0, 0.138006)};

    \addplot [fill=red-violet, draw=red-violet, line width=0mm, error bars/.cd, y dir=both, y explicit, error bar style={thick}, error bar style=black] coordinates { 
      (PANDA, 0.885085) +- (0, 0.0123753)
      (Karolinska, 0.885949) +- (0, 0.0159321)
      (Radboud, 0.85256) +- (0, 0.0152281)
      (Radboud $\rightarrow$ Karolinska, 0.270728) +- (0, 0.143959)};

    \addplot [fill=caribbeangreen2, draw=caribbeangreen2, error bars/.cd, y dir=both, y explicit, error bar style={thick}, error bar style=black] coordinates {
      (PANDA, 0.866334) +- (0, 0.00907365)
      (Karolinska, 0.86257) +- (0, 0.015705)
      (Radboud, 0.818953) +- (0, 0.0213635)
      (Radboud $\rightarrow$ Karolinska, 0.0245775) +- (0, 0.0183994)};

    \addplot [fill=ForestGreen, draw=ForestGreen, error bars/.cd, y dir=both, y explicit, error bar style={thick}, error bar style=black] coordinates {
      (PANDA, 0.858159) +- (0, 0.00987344)
      (Karolinska, 0.86872) +- (0, 0.0181902)
      (Radboud, 0.815367) +- (0, 0.02357)
      (Radboud $\rightarrow$ Karolinska, 0.723546) +- (0, 0.037291)};

    \addplot [fill=skyblue2, draw=skyblue2, error bars/.cd, y dir=both, y explicit, error bar style={thick}, error bar style=black] coordinates {
      (PANDA, 0.889037) +- (0, 0.01254)
      (Karolinska, 0.89135) +- (0, 0.0195419)
      (Radboud, 0.860933) +- (0, 0.0227627)
      (Radboud $\rightarrow$ Karolinska, 0.472049) +- (0, 0.105997)};

    \addplot [fill=Cerulean, draw=Cerulean, line width=0mm, error bars/.cd, y dir=both, y explicit, error bar style={thick}, error bar style=black] coordinates { 
      (PANDA, 0.886692) +- (0, 0.0169589)
      (Karolinska, 0.882394) +- (0, 0.014148)
      (Radboud, 0.863841) +- (0, 0.0261018)
      (Radboud $\rightarrow$ Karolinska, 0.428641) +- (0, 0.0988533)};

    \addplot [fill=darkcerulean2, draw=darkcerulean2, line width=0mm, error bars/.cd, y dir=both, y explicit, error bar style={thick}, error bar style=black] coordinates { 
      (PANDA, 0.885995) +- (0, 0.0134458)
      (Karolinska, 0.884609) +- (0, 0.0129327)
      (Radboud, 0.859427) +- (0, 0.0152547)
      (Radboud $\rightarrow$ Karolinska, 0.405437) +- (0, 0.0840904)};

    \addplot [fill=darkpastelpurple2, draw=darkpastelpurple2, error bars/.cd, y dir=both, y explicit, error bar style={thick}, error bar style=black] coordinates {
      (PANDA, 0.8862) +- (0, 0.0116586)
      (Karolinska, 0.874592) +- (0, 0.017091)
      (Radboud, 0.857914) +- (0, 0.0195372)
      (Radboud $\rightarrow$ Karolinska, 0.459026) +- (0, 0.0749562)};

    \addplot [fill=goldenpoppy, draw=goldenpoppy, error bars/.cd, y dir=both, y explicit, error bar style={thick}, error bar style=black] coordinates {
      (PANDA, 0.867677) +- (0, 0.0135075)
      (Karolinska, 0.863586) +- (0, 0.0270691)
      (Radboud, 0.838156) +- (0, 0.0233552)
      (Radboud $\rightarrow$ Karolinska, 0.61698) +- (0, 0.0590592)};

    \addplot [fill=goldenpoppy3, draw=goldenpoppy3, error bars/.cd, y dir=both, y explicit, error bar style={thick}, error bar style=black] coordinates {
      (PANDA, 0.876571) +- (0, 0.0117124)
      (Karolinska, 0.878736) +- (0, 0.013313)
      (Radboud, 0.847051) +- (0, 0.0178447)
      (Radboud $\rightarrow$ Karolinska, 0.504072) +- (0, 0.146641)};
      
  \end{axis}
  \end{tikzpicture}\vspace{-3.0mm}
    \begin{tikzpicture}
  \centering
  \begin{axis}[
        ybar=2.5pt, axis on top,
        title={},
        height=4.925cm, width=23.0cm, 
        bar width=0.26cm, 
        ymajorgrids, tick align=inside,
        enlarge y limits={value=.1,upper},
        ymin=0, ymax=0.9,
        set layers,
        ytick={0.2, 0.4, 0.6, 0.8},
        axis x line*=bottom,
        axis y line*=left,
        y axis line style={opacity=0},
        tickwidth=0pt,
        enlarge x limits=0.20, 
        extra y tick style={
            grid style={
                solid,
                line width = 2.5pt,
                /pgfplots/on layer=axis background,
            },
        },
        tick style={
            grid style={
                dashed,
                /pgfplots/on layer=axis background,
            },
        },
        legend style={
            at={(0.5,-0.2)},
            anchor=north,
            legend columns=-1,
            /tikz/every even column/.append style={column sep=0.325cm},
            font=\footnotesize
        },
        ylabel={Kappa ($\uparrow$)},
        ylabel style={font=\footnotesize},
        symbolic x coords={
           PANDA,
           Karolinska,
           Radboud, 
           Radboud $\rightarrow$ Karolinska},
       xtick=data,
       xticklabel=\empty,
       x tick label style={text width = 2.25cm, align=center, font=\footnotesize},
    ]
    \addplot [fill=OrangeRed2, draw=OrangeRed2, error bars/.cd, y dir=both, y explicit, error bar style={thick}, error bar style=black] coordinates {
      (PANDA, 0.610517) +- (0, 0.0309128)
      (Karolinska, 0.484289) +- (0, 0.0414748)
      (Radboud, 0.565746) +- (0, 0.0239493)
      (Radboud $\rightarrow$ Karolinska, 0.216677) +- (0, 0.0230846)};

    \addplot [fill=darkgray, draw=darkgray, error bars/.cd, y dir=both, y explicit, error bar style={thick}, error bar style=black] coordinates {
      (PANDA, 0.789349) +- (0, 0.0215116)
      (Karolinska, 0.744444) +- (0, 0.0361009)
      (Radboud, 0.7775) +- (0, 0.0219794)
      (Radboud $\rightarrow$ Karolinska, 0.212961) +- (0, 0.0533835)};
    
    \addplot [fill=flamingopink, draw=flamingopink, line width=0mm, error bars/.cd, y dir=both, y explicit, error bar style={thick}, error bar style=black] coordinates { 
      (PANDA, 0.84559) +- (0, 0.0208975)
      (Karolinska, 0.779771) +- (0, 0.0216045)
      (Radboud, 0.846202) +- (0, 0.015932)
      (Radboud $\rightarrow$ Karolinska, 0.489885) +- (0, 0.0596735)};

    \addplot [fill=red-violet, draw=red-violet, line width=0mm, error bars/.cd, y dir=both, y explicit, error bar style={thick}, error bar style=black] coordinates { 
      (PANDA, 0.835556) +- (0, 0.0115376)
      (Karolinska, 0.764095) +- (0, 0.0362231)
      (Radboud, 0.82717) +- (0, 0.025873)
      (Radboud $\rightarrow$ Karolinska, 0.335084) +- (0, 0.0431102)};

    \addplot [fill=caribbeangreen2, draw=caribbeangreen2, error bars/.cd, y dir=both, y explicit, error bar style={thick}, error bar style=black] coordinates {
      (PANDA, 0.777674) +- (0, 0.0198599)
      (Karolinska, 0.714824) +- (0, 0.0230472)
      (Radboud, 0.743565) +- (0, 0.0255462)
      (Radboud $\rightarrow$ Karolinska, 0.0261816 ) +- (0, 0.0150573)};

    \addplot [fill=ForestGreen, draw=ForestGreen, error bars/.cd, y dir=both, y explicit, error bar style={thick}, error bar style=black] coordinates {
      (PANDA, 0.768022) +- (0, 0.0174022)
      (Karolinska, 0.704212) +- (0, 0.0273398)
      (Radboud, 0.727299) +- (0, 0.0206647)
      (Radboud $\rightarrow$ Karolinska, 0.556405) +- (0, 0.0145983)};

    \addplot [fill=skyblue2, draw=skyblue2, error bars/.cd, y dir=both, y explicit, error bar style={thick}, error bar style=black] coordinates {
      (PANDA, 0.841891) +- (0, 0.0163428)
      (Karolinska, 0.782086) +- (0, 0.0264358)
      (Radboud, 0.832701) +- (0, 0.0163559)
      (Radboud $\rightarrow$ Karolinska, 0.359176) +- (0, 0.0426587)};

    \addplot [fill=Cerulean, draw=Cerulean, line width=0mm, error bars/.cd, y dir=both, y explicit, error bar style={thick}, error bar style=black] coordinates { 
      (PANDA, 0.839958) +- (0, 0.0160809)
      (Karolinska, 0.780891) +- (0, 0.0263926)
      (Radboud, 0.829961) +- (0, 0.0204068)
      (Radboud $\rightarrow$ Karolinska, 0.515981) +- (0, 0.0245259)};

    \addplot [fill=darkcerulean2, draw=darkcerulean2, line width=0mm, error bars/.cd, y dir=both, y explicit, error bar style={thick}, error bar style=black] coordinates { 
      (PANDA, 0.822077) +- (0, 0.018959)
      (Karolinska, 0.761503) +- (0, 0.0308394)
      (Radboud, 0.806233) +- (0, 0.0249663)
      (Radboud $\rightarrow$ Karolinska, 0.489018) +- (0, 0.0232659)};

    \addplot [fill=darkpastelpurple2, draw=darkpastelpurple2, error bars/.cd, y dir=both, y explicit, error bar style={thick}, error bar style=black] coordinates {
      (PANDA, 0.832176) +- (0, 0.0189604)
      (Karolinska, 0.750182) +- (0, 0.0345796)
      (Radboud, 0.83313) +- (0, 0.0140878)
      (Radboud $\rightarrow$ Karolinska, 0.416334) +- (0, 0.066083)};

    \addplot [fill=goldenpoppy, draw=goldenpoppy, error bars/.cd, y dir=both, y explicit, error bar style={thick}, error bar style=black] coordinates {
      (PANDA, 0.808804) +- (0, 0.0201696)
      (Karolinska, 0.747719) +- (0, 0.031465)
      (Radboud, 0.791368) +- (0, 0.0292407)
      (Radboud $\rightarrow$ Karolinska, 0.340052) +- (0, 0.0686435)};

    \addplot [fill=goldenpoppy3, draw=goldenpoppy3, error bars/.cd, y dir=both, y explicit, error bar style={thick}, error bar style=black] coordinates {
      (PANDA, 0.827421) +- (0, 0.0133668)
      (Karolinska, 0.774161) +- (0, 0.0232635)
      (Radboud, 0.809273) +- (0, 0.0220188)
      (Radboud $\rightarrow$ Karolinska, 0.222778) +- (0, 0.0738892)};
  \end{axis}
  \end{tikzpicture}\vspace{-3.5mm}
    \begin{tikzpicture}
  \centering
  \begin{axis}[
        ybar=2.5pt, axis on top,
        title={},
        height=4.925cm, width=23.0cm, 
        bar width=0.26cm, 
        ymajorgrids, tick align=inside,
        enlarge y limits={value=.1,upper},
        ymin=0, ymax=0.9,
        set layers,
        ytick={0.2, 0.4, 0.6, 0.8},
        axis x line*=bottom,
        axis y line*=left,
        y axis line style={opacity=0},
        tickwidth=0pt,
        enlarge x limits=0.20, 
        extra y tick style={
            grid style={
                solid,
                line width = 2.5pt,
                /pgfplots/on layer=axis background,
            },
        },
        tick style={
            grid style={
                dashed,
                /pgfplots/on layer=axis background,
            },
        },
        legend style={
            at={(0.5,-0.225)},
            anchor=north,
            legend columns=-1,
            /tikz/every even column/.append style={column sep=0.175cm},
            font=\footnotesize
        },
        ylabel={Kappa ($\uparrow$)},
        ylabel style={font=\footnotesize},
        symbolic x coords={
           PANDA,
           Karolinska,
           Radboud, 
           Radboud $\rightarrow$ Karolinska},
       xtick=data,
       x tick label style={text width = 5.0cm, align=center, font={\bfseries}},
    ]
    \addplot [fill=OrangeRed2, draw=OrangeRed2, error bars/.cd, y dir=both, y explicit, error bar style={thick}, error bar style=black] coordinates {
      (PANDA, 0.613157) +- (0, 0.016445)
      (Karolinska, 0.461135) +- (0, 0.0302627)
      (Radboud, 0.630168) +- (0, 0.0422488)
      (Radboud $\rightarrow$ Karolinska, 0.0705993) +- (0, 0.0134034)};

    \addplot [fill=darkgray, draw=darkgray, error bars/.cd, y dir=both, y explicit, error bar style={thick}, error bar style=black] coordinates {
      (PANDA, 0.687072) +- (0, 0.0202816)
      (Karolinska, 0.556356) +- (0, 0.0316019)
      (Radboud, 0.727437) +- (0, 0.0197546)
      (Radboud $\rightarrow$ Karolinska, 0.126199) +- (0, 0.0605454)};
    
    \addplot [fill=flamingopink, draw=flamingopink, line width=0mm, error bars/.cd, y dir=both, y explicit, error bar style={thick}, error bar style=black] coordinates { 
      (PANDA, 0.727169) +- (0, 0.0182163)
      (Karolinska, 0.582271) +- (0, 0.0332399)
      (Radboud, 0.747012) +- (0, 0.0270309)
      (Radboud $\rightarrow$ Karolinska, 0.214366) +- (0, 0.0376718)};

    \addplot [fill=red-violet, draw=red-violet, line width=0mm, error bars/.cd, y dir=both, y explicit, error bar style={thick}, error bar style=black] coordinates { 
      (PANDA, 0.719051) +- (0, 0.0217944)
      (Karolinska, 0.560601) +- (0, 0.0475562)
      (Radboud, 0.757702) +- (0, 0.0320785)
      (Radboud $\rightarrow$ Karolinska, 0.21015) +- (0, 0.0267073)};

    \addplot [fill=caribbeangreen2, draw=caribbeangreen2, error bars/.cd, y dir=both, y explicit, error bar style={thick}, error bar style=black] coordinates {
      (PANDA, 0.690939) +- (0, 0.0214606)
      (Karolinska, 0.590366) +- (0, 0.0260291)
      (Radboud, 0.687728) +- (0, 0.0414731)
      (Radboud $\rightarrow$ Karolinska, 0.208055) +- (0, 0.0128485)};

    \addplot [fill=ForestGreen, draw=ForestGreen, error bars/.cd, y dir=both, y explicit, error bar style={thick}, error bar style=black] coordinates {
      (PANDA, 0.703755) +- (0, 0.0167515)
      (Karolinska, 0.612679) +- (0, 0.0267452)
      (Radboud, 0.695315) +- (0, 0.0419093)
      (Radboud $\rightarrow$ Karolinska, 0.344483) +- (0, 0.0133134)};

    \addplot [fill=skyblue2, draw=skyblue2, error bars/.cd, y dir=both, y explicit, error bar style={thick}, error bar style=black] coordinates {
      (PANDA, 0.724572) +- (0, 0.0149668)
      (Karolinska, 0.587677) +- (0, 0.037649)
      (Radboud, 0.751543) +- (0, 0.0341701)
      (Radboud $\rightarrow$ Karolinska, 0.294091) +- (0, 0.012755)};

    \addplot [fill=Cerulean, draw=Cerulean, line width=0mm, error bars/.cd, y dir=both, y explicit, error bar style={thick}, error bar style=black] coordinates { 
      (PANDA, 0.734141) +- (0, 0.0190904)
      (Karolinska, 0.60592) +- (0, 0.0270284)
      (Radboud, 0.754972) +- (0, 0.0333241)
      (Radboud $\rightarrow$ Karolinska, 0.374623) +- (0, 0.0199929)};

    \addplot [fill=darkcerulean2, draw=darkcerulean2, line width=0mm, error bars/.cd, y dir=both, y explicit, error bar style={thick}, error bar style=black] coordinates { 
      (PANDA, 0.687414) +- (0, 0.0244893)
      (Karolinska, 0.540177) +- (0, 0.0366539)
      (Radboud, 0.711327) +- (0, 0.0364132)
      (Radboud $\rightarrow$ Karolinska, 0.346756) +- (0, 0.00713873)};

    \addplot [fill=darkpastelpurple2, draw=darkpastelpurple2, error bars/.cd, y dir=both, y explicit, error bar style={thick}, error bar style=black] coordinates {
      (PANDA, 0.721069) +- (0, 0.0251075)
      (Karolinska, 0.567143) +- (0, 0.0311424)
      (Radboud, 0.744075) +- (0, 0.0198549)
      (Radboud $\rightarrow$ Karolinska, 0.228048) +- (0, 0.0367065)};

    \addplot [fill=goldenpoppy, draw=goldenpoppy, error bars/.cd, y dir=both, y explicit, error bar style={thick}, error bar style=black] coordinates {
      (PANDA, 0.674131) +- (0, 0.0188681)
      (Karolinska, 0.555084) +- (0, 0.0419075)
      (Radboud, 0.672634) +- (0, 0.0195404)
      (Radboud $\rightarrow$ Karolinska, 0.025) +- (0, 0.0644499)};

    \addplot [fill=goldenpoppy3, draw=goldenpoppy3, error bars/.cd, y dir=both, y explicit, error bar style={thick}, error bar style=black] coordinates {
      (PANDA, 0.694544) +- (0, 0.0116528)
      (Karolinska, 0.594539) +- (0, 0.0296609)
      (Radboud, 0.696469) +- (0, 0.0268229)
      (Radboud $\rightarrow$ Karolinska, 0.378458) +- (0, 0.0143491)};
    \legend{
            Resnet-IN,
            Phikon-v2,
            UNI,
            UNI2-h,
            CONCH,
            CONCHv1.5,
            H-optimus-0,
            H-optimus-1,
            H0-mini,
            Prov-GigaPath,
            Virchow,
            Virchow2
            }
  \end{axis}
  \end{tikzpicture}\vspace{0.0mm}
    \caption{\textbf{Main model comparison across PANDA and its primary subsets.} Performance of all twelve evaluated models across the full PANDA dataset and its two site-specific subsets (Karolinska, Radboud), when used as frozen patch-level feature extractors in the \textit{ABMIL} \textbf{(top)}, \textit{Mean Feature} \textbf{(middle)}, and \textit{kNN} \textbf{(bottom)} ISUP grade classification models. Across in-distribution settings (PANDA, Karolinska, Radboud), all PFMs achieve strong performance and consistently outperform the natural-image baseline Resnet-IN. In contrast, under cross-site transfer (\textit{Radboud $\rightarrow$ Karolinska}), performance drops substantially for all models. All results are reported as mean $\pm$ standard deviation over $10$ cross-validation folds. Corresponding results in terms of MAE are provided in Figure~\ref{fig:main_results3a_12fms_mae} in the supplementary material.}
    \label{fig:main_results3a_12fms_kappa}
\end{figure*}
\begin{figure*}[t]
    \centering
    \begin{tikzpicture}
  \centering
  \begin{axis}[
        ybar=2.5pt, axis on top,
        title={},
        height=4.925cm, width=23.0cm, 
        bar width=0.26cm, 
        ymajorgrids, tick align=inside,
        enlarge y limits={value=.1,upper},
        ymin=0, ymax=0.9,
        set layers,
        ytick={0.2, 0.4, 0.6, 0.8},
        axis x line*=bottom,
        axis y line*=left,
        y axis line style={opacity=0},
        tickwidth=0pt,
        enlarge x limits=0.20, 
        extra y tick style={
            grid style={
                solid,
                line width = 2.5pt,
                /pgfplots/on layer=axis background,
            },
        },
        tick style={
            grid style={
                dashed,
                /pgfplots/on layer=axis background,
            },
        },
        legend style={
            at={(0.5,-0.2)},
            anchor=north,
            legend columns=-1,
            /tikz/every even column/.append style={column sep=0.325cm},
            font=\footnotesize
        },
        ylabel={Kappa ($\uparrow$)},
        ylabel style={font=\footnotesize},
        symbolic x coords={
           Radboud-U, 
           Radboud-U $\rightarrow$ Karolinska-U,
           Radboud-L, 
           Radboud-L $\rightarrow$ Radboud-R},
       xtick=data,
       xticklabel=\empty,
       x tick label style={text width = 2.25cm, align=center, font=\footnotesize},
    ]
    \addplot [fill=OrangeRed2, draw=OrangeRed2, error bars/.cd, y dir=both, y explicit, error bar style={thick}, error bar style=black] coordinates {
      (Radboud-U, 0.725059) +- (0, 0.0187998)
      (Radboud-U $\rightarrow$ Karolinska-U, 0.380423) +- (0, 0.0353908)
      (Radboud-L, 0.70861) +- (0, 0.0466765)
      (Radboud-L $\rightarrow$ Radboud-R, 0.545218) +- (0, 0.0142131)};

    \addplot [fill=darkgray, draw=darkgray, error bars/.cd, y dir=both, y explicit, error bar style={thick}, error bar style=black] coordinates {
      (Radboud-U, 0.839083) +- (0, 0.0319653)
      (Radboud-U $\rightarrow$ Karolinska-U, 0.507484) +- (0, 0.0579279)
      (Radboud-L, 0.817192) +- (0, 0.0346085)
      (Radboud-L $\rightarrow$ Radboud-R, 0.714851) +- (0, 0.0166124)};
    
    \addplot [fill=flamingopink, draw=flamingopink, line width=0mm, error bars/.cd, y dir=both, y explicit, error bar style={thick}, error bar style=black] coordinates { 
      (Radboud-U, 0.843966) +- (0, 0.0237577)
      (Radboud-U $\rightarrow$ Karolinska-U, 0.459637) +- (0, 0.112533)
      (Radboud-L, 0.826279) +- (0, 0.0361249)
      (Radboud-L $\rightarrow$ Radboud-R, 0.739375) +- (0, 0.0142154)};

    \addplot [fill=red-violet, draw=red-violet, line width=0mm, error bars/.cd, y dir=both, y explicit, error bar style={thick}, error bar style=black] coordinates { 
      (Radboud-U, 0.855195) +- (0, 0.0186979)
      (Radboud-U $\rightarrow$ Karolinska-U, 0.536572) +- (0, 0.0657804)
      (Radboud-L, 0.840353) +- (0, 0.0261901)
      (Radboud-L $\rightarrow$ Radboud-R, 0.730126) +- (0, 0.0199094)};

    \addplot [fill=caribbeangreen2, draw=caribbeangreen2, error bars/.cd, y dir=both, y explicit, error bar style={thick}, error bar style=black] coordinates {
      (Radboud-U, 0.81895) +- (0, 0.0287215)
      (Radboud-U $\rightarrow$ Karolinska-U, 0.206042) +- (0, 0.0755954)
      (Radboud-L, 0.805051) +- (0, 0.0483934)
      (Radboud-L $\rightarrow$ Radboud-R, 0.712784) +- (0, 0.0147616)};

    \addplot [fill=ForestGreen, draw=ForestGreen, error bars/.cd, y dir=both, y explicit, error bar style={thick}, error bar style=black] coordinates {
      (Radboud-U, 0.809397) +- (0, 0.032308)
      (Radboud-U $\rightarrow$ Karolinska-U, 0.69802) +- (0, 0.0174124)
      (Radboud-L, 0.809747) +- (0, 0.0349779)
      (Radboud-L $\rightarrow$ Radboud-R, 0.689074) +- (0, 0.0186569)};

    \addplot [fill=skyblue2, draw=skyblue2, error bars/.cd, y dir=both, y explicit, error bar style={thick}, error bar style=black] coordinates {
      (Radboud-U, 0.849584) +- (0, 0.0248498)
      (Radboud-U $\rightarrow$ Karolinska-U, 0.656305) +- (0, 0.0318687)
      (Radboud-L, 0.833236) +- (0, 0.0307769)
      (Radboud-L $\rightarrow$ Radboud-R, 0.729401) +- (0, 0.0238521)};

    \addplot [fill=Cerulean, draw=Cerulean, line width=0mm, error bars/.cd, y dir=both, y explicit, error bar style={thick}, error bar style=black] coordinates { 
      (Radboud-U, 0.850576) +- (0, 0.0132259)
      (Radboud-U $\rightarrow$ Karolinska-U, 0.643878) +- (0, 0.0598084)
      (Radboud-L, 0.838932) +- (0, 0.0284458)
      (Radboud-L $\rightarrow$ Radboud-R, 0.734561) +- (0, 0.0187071)};

    \addplot [fill=darkcerulean2, draw=darkcerulean2, line width=0mm, error bars/.cd, y dir=both, y explicit, error bar style={thick}, error bar style=black] coordinates { 
      (Radboud-U, 0.852509) +- (0, 0.0202243)
      (Radboud-U $\rightarrow$ Karolinska-U, 0.64361) +- (0, 0.0204429)
      (Radboud-L, 0.827465) +- (0, 0.0269886)
      (Radboud-L $\rightarrow$ Radboud-R, 0.729103) +- (0, 0.0170106)};

    \addplot [fill=darkpastelpurple2, draw=darkpastelpurple2, error bars/.cd, y dir=both, y explicit, error bar style={thick}, error bar style=black] coordinates {
      (Radboud-U, 0.842319) +- (0, 0.023979)
      (Radboud-U $\rightarrow$ Karolinska-U, 0.675557) +- (0, 0.0598927)
      (Radboud-L, 0.835485) +- (0, 0.0248449)
      (Radboud-L $\rightarrow$ Radboud-R, 0.72981) +- (0, 0.012295)};

    \addplot [fill=goldenpoppy, draw=goldenpoppy, error bars/.cd, y dir=both, y explicit, error bar style={thick}, error bar style=black] coordinates {
      (Radboud-U, 0.829841) +- (0, 0.024966)
      (Radboud-U $\rightarrow$ Karolinska-U, 0.66363) +- (0, 0.0559497)
      (Radboud-L, 0.832021) +- (0, 0.0337764)
      (Radboud-L $\rightarrow$ Radboud-R, 0.720456) +- (0, 0.0152852)};

    \addplot [fill=goldenpoppy3, draw=goldenpoppy3, error bars/.cd, y dir=both, y explicit, error bar style={thick}, error bar style=black] coordinates {
      (Radboud-U, 0.833371) +- (0, 0.0284642)
      (Radboud-U $\rightarrow$ Karolinska-U, 0.660097) +- (0, 0.0460499)
      (Radboud-L, 0.818717) +- (0, 0.0199882)
      (Radboud-L $\rightarrow$ Radboud-R, 0.719605) +- (0, 0.0213754)};
  \end{axis}
  \end{tikzpicture}\vspace{-3.0mm}
    \begin{tikzpicture}
  \centering
  \begin{axis}[
        ybar=2.5pt, axis on top,
        title={},
        height=4.925cm, width=23.0cm, 
        bar width=0.26cm, 
        ymajorgrids, tick align=inside,
        enlarge y limits={value=.1,upper},
        ymin=0, ymax=0.9,
        set layers,
        ytick={0.2, 0.4, 0.6, 0.8},
        axis x line*=bottom,
        axis y line*=left,
        y axis line style={opacity=0},
        tickwidth=0pt,
        enlarge x limits=0.20, 
        extra y tick style={
            grid style={
                solid,
                line width = 2.5pt,
                /pgfplots/on layer=axis background,
            },
        },
        tick style={
            grid style={
                dashed,
                /pgfplots/on layer=axis background,
            },
        },
        legend style={
            at={(0.5,-0.2)},
            anchor=north,
            legend columns=-1,
            /tikz/every even column/.append style={column sep=0.325cm},
            font=\footnotesize
        },
        ylabel={Kappa ($\uparrow$)},
        ylabel style={font=\footnotesize},
        symbolic x coords={
           Radboud-U, 
           Radboud-U $\rightarrow$ Karolinska-U,
           Radboud-L, 
           Radboud-L $\rightarrow$ Radboud-R},
       xtick=data,
       xticklabel=\empty,
       x tick label style={text width = 2.25cm, align=center, font=\footnotesize},
    ]
    \addplot [fill=OrangeRed2, draw=OrangeRed2, error bars/.cd, y dir=both, y explicit, error bar style={thick}, error bar style=black] coordinates {
      (Radboud-U, 0.550741) +- (0, 0.0401625)
      (Radboud-U $\rightarrow$ Karolinska-U, 0.369438) +- (0, 0.00683355)
      (Radboud-L, 0.514023) +- (0, 0.0456437)
      (Radboud-L $\rightarrow$ Radboud-R, 0.425298) +- (0, 0.00646147)};

    \addplot [fill=darkgray, draw=darkgray, error bars/.cd, y dir=both, y explicit, error bar style={thick}, error bar style=black] coordinates {
      (Radboud-U, 0.760223) +- (0, 0.0327461)
      (Radboud-U $\rightarrow$ Karolinska-U, 0.350377) +- (0, 0.0322908)
      (Radboud-L, 0.760892) +- (0, 0.0253868)
      (Radboud-L $\rightarrow$ Radboud-R, 0.629822) +- (0, 0.00907769)};
    
    \addplot [fill=flamingopink, draw=flamingopink, line width=0mm, error bars/.cd, y dir=both, y explicit, error bar style={thick}, error bar style=black] coordinates { 
      (Radboud-U, 0.828197) +- (0, 0.0287752)
      (Radboud-U $\rightarrow$ Karolinska-U, 0.516602) +- (0, 0.018329)
      (Radboud-L, 0.806884) +- (0, 0.0328776)
      (Radboud-L $\rightarrow$ Radboud-R, 0.686432) +- (0, 0.0206927)};

    \addplot [fill=red-violet, draw=red-violet, line width=0mm, error bars/.cd, y dir=both, y explicit, error bar style={thick}, error bar style=black] coordinates { 
      (Radboud-U, 0.815075) +- (0, 0.0347275)
      (Radboud-U $\rightarrow$ Karolinska-U, 0.370097) +- (0, 0.0361245)
      (Radboud-L, 0.802894) +- (0, 0.0221661)
      (Radboud-L $\rightarrow$ Radboud-R, 0.677703) +- (0, 0.0047424)};

    \addplot [fill=caribbeangreen2, draw=caribbeangreen2, error bars/.cd, y dir=both, y explicit, error bar style={thick}, error bar style=black] coordinates {
      (Radboud-U, 0.743879) +- (0, 0.0386241)
      (Radboud-U $\rightarrow$ Karolinska-U, 0.165819) +- (0, 0.0488119)
      (Radboud-L, 0.73137) +- (0, 0.055067)
      (Radboud-L $\rightarrow$ Radboud-R, 0.59083) +- (0, 0.0195734)};

    \addplot [fill=ForestGreen, draw=ForestGreen, error bars/.cd, y dir=both, y explicit, error bar style={thick}, error bar style=black] coordinates {
      (Radboud-U, 0.72103) +- (0, 0.0362594)
      (Radboud-U $\rightarrow$ Karolinska-U, 0.522532) +- (0, 0.0149313)
      (Radboud-L, 0.729058) +- (0, 0.033372)
      (Radboud-L $\rightarrow$ Radboud-R, 0.584524) +- (0, 0.0211163)};

    \addplot [fill=skyblue2, draw=skyblue2, error bars/.cd, y dir=both, y explicit, error bar style={thick}, error bar style=black] coordinates {
      (Radboud-U, 0.823268) +- (0, 0.0363036)
      (Radboud-U $\rightarrow$ Karolinska-U, 0.380124) +- (0, 0.0558697)
      (Radboud-L, 0.80787) +- (0, 0.0274059)
      (Radboud-L $\rightarrow$ Radboud-R, 0.681326) +- (0, 0.00623155)};

    \addplot [fill=Cerulean, draw=Cerulean, line width=0mm, error bars/.cd, y dir=both, y explicit, error bar style={thick}, error bar style=black] coordinates { 
      (Radboud-U, 0.822937) +- (0, 0.0290758)
      (Radboud-U $\rightarrow$ Karolinska-U, 0.493689) +- (0, 0.0187603)
      (Radboud-L, 0.812939) +- (0, 0.0248034)
      (Radboud-L $\rightarrow$ Radboud-R, 0.685654) +- (0, 0.00678213)};

    \addplot [fill=darkcerulean2, draw=darkcerulean2, line width=0mm, error bars/.cd, y dir=both, y explicit, error bar style={thick}, error bar style=black] coordinates { 
      (Radboud-U, 0.788607) +- (0, 0.0386399)
      (Radboud-U $\rightarrow$ Karolinska-U, 0.420712) +- (0, 0.0236664)
      (Radboud-L, 0.799799) +- (0, 0.0347504)
      (Radboud-L $\rightarrow$ Radboud-R, 0.655461) +- (0, 0.0135482)};

    \addplot [fill=darkpastelpurple2, draw=darkpastelpurple2, error bars/.cd, y dir=both, y explicit, error bar style={thick}, error bar style=black] coordinates {
      (Radboud-U, 0.815392) +- (0, 0.0404605)
      (Radboud-U $\rightarrow$ Karolinska-U, 0.472304) +- (0, 0.0337215)
      (Radboud-L, 0.802815) +- (0, 0.023774)
      (Radboud-L $\rightarrow$ Radboud-R, 0.682357) +- (0, 0.00911505)};

    \addplot [fill=goldenpoppy, draw=goldenpoppy, error bars/.cd, y dir=both, y explicit, error bar style={thick}, error bar style=black] coordinates {
      (Radboud-U, 0.782769) +- (0, 0.0274152)
      (Radboud-U $\rightarrow$ Karolinska-U, 0.349537) +- (0, 0.0430824)
      (Radboud-L, 0.784613) +- (0, 0.0266161)
      (Radboud-L $\rightarrow$ Radboud-R, 0.640242) +- (0, 0.00984075)};

    \addplot [fill=goldenpoppy3, draw=goldenpoppy3, error bars/.cd, y dir=both, y explicit, error bar style={thick}, error bar style=black] coordinates {
      (Radboud-U, 0.805385) +- (0, 0.0375729)
      (Radboud-U $\rightarrow$ Karolinska-U, 0.358065) +- (0, 0.0535879)
      (Radboud-L, 0.791257) +- (0, 0.0210861)
      (Radboud-L $\rightarrow$ Radboud-R, 0.65632) +- (0, 0.0127742)};
  \end{axis}
  \end{tikzpicture}\vspace{-3.5mm}
    \begin{tikzpicture}
  \centering
  \begin{axis}[
        ybar=2.5pt, axis on top,
        title={},
        height=4.925cm, width=23.0cm, 
        bar width=0.26cm, 
        ymajorgrids, tick align=inside,
        enlarge y limits={value=.1,upper},
        ymin=0, ymax=0.9,
        set layers,
        ytick={0.2, 0.4, 0.6, 0.8},
        axis x line*=bottom,
        axis y line*=left,
        y axis line style={opacity=0},
        tickwidth=0pt,
        enlarge x limits=0.20, 
        extra y tick style={
            grid style={
                solid,
                line width = 2.5pt,
                /pgfplots/on layer=axis background,
            },
        },
        tick style={
            grid style={
                dashed,
                /pgfplots/on layer=axis background,
            },
        },
        legend style={
            at={(0.5,-0.225)},
            anchor=north,
            legend columns=-1,
            /tikz/every even column/.append style={column sep=0.175cm},
            font=\footnotesize
        },
        ylabel={Kappa ($\uparrow$)},
        ylabel style={font=\footnotesize},
        symbolic x coords={
           Radboud-U, 
           Radboud-U $\rightarrow$ Karolinska-U,
           Radboud-L, 
           Radboud-L $\rightarrow$ Radboud-R},
       xtick=data,
       x tick label style={text width = 5.0cm, align=center, font={\bfseries}},
    ]
    \addplot [fill=OrangeRed2, draw=OrangeRed2, error bars/.cd, y dir=both, y explicit, error bar style={thick}, error bar style=black] coordinates {
      (Radboud-U, 0.594313) +- (0, 0.0487418)
      (Radboud-U $\rightarrow$ Karolinska-U, 0.0980739) +- (0, 0.0298326)
      (Radboud-L, 0.502797) +- (0, 0.0414738)
      (Radboud-L $\rightarrow$ Radboud-R, 0.282891) +- (0, 0.00887223)};

    \addplot [fill=darkgray, draw=darkgray, error bars/.cd, y dir=both, y explicit, error bar style={thick}, error bar style=black] coordinates {
      (Radboud-U, 0.670146) +- (0, 0.0246172)
      (Radboud-U $\rightarrow$ Karolinska-U, 0.10539) +- (0, 0.0530505)
      (Radboud-L, 0.600378) +- (0, 0.0572836)
      (Radboud-L $\rightarrow$ Radboud-R, 0.350412) +- (0, 0.00958643)};
    
    \addplot [fill=flamingopink, draw=flamingopink, line width=0mm, error bars/.cd, y dir=both, y explicit, error bar style={thick}, error bar style=black] coordinates { 
      (Radboud-U, 0.711353) +- (0, 0.0360002)
      (Radboud-U $\rightarrow$ Karolinska-U, 0.283195) +- (0, 0.0345969)
      (Radboud-L, 0.66672) +- (0, 0.0305624)
      (Radboud-L $\rightarrow$ Radboud-R, 0.397871) +- (0, 0.00518936)};

    \addplot [fill=red-violet, draw=red-violet, line width=0mm, error bars/.cd, y dir=both, y explicit, error bar style={thick}, error bar style=black] coordinates { 
      (Radboud-U, 0.705477) +- (0, 0.0383834)
      (Radboud-U $\rightarrow$ Karolinska-U, 0.358436) +- (0, 0.0180455)
      (Radboud-L, 0.673361) +- (0, 0.0298184)
      (Radboud-L $\rightarrow$ Radboud-R, 0.4149) +- (0, 0.00668088)};

    \addplot [fill=caribbeangreen2, draw=caribbeangreen2, error bars/.cd, y dir=both, y explicit, error bar style={thick}, error bar style=black] coordinates {
      (Radboud-U, 0.665943) +- (0, 0.0300574)
      (Radboud-U $\rightarrow$ Karolinska-U, 0.320718) +- (0, 0.0234927)
      (Radboud-L, 0.630585) +- (0, 0.0435856)
      (Radboud-L $\rightarrow$ Radboud-R, 0.38761) +- (0, 0.00848339)};

    \addplot [fill=ForestGreen, draw=ForestGreen, error bars/.cd, y dir=both, y explicit, error bar style={thick}, error bar style=black] coordinates {
      (Radboud-U, 0.665767) +- (0, 0.0482797)
      (Radboud-U $\rightarrow$ Karolinska-U, 0.400034) +- (0, 0.0166285)
      (Radboud-L, 0.635305) +- (0, 0.0365897)
      (Radboud-L $\rightarrow$ Radboud-R, 0.397994) +- (0, 0.00457531)};

    \addplot [fill=skyblue2, draw=skyblue2, error bars/.cd, y dir=both, y explicit, error bar style={thick}, error bar style=black] coordinates {
      (Radboud-U, 0.712543) +- (0, 0.040858)
      (Radboud-U $\rightarrow$ Karolinska-U, 0.30383) +- (0, 0.0111311)
      (Radboud-L, 0.682925) +- (0, 0.0170415)
      (Radboud-L $\rightarrow$ Radboud-R, 0.407121) +- (0, 0.00866511)};

    \addplot [fill=Cerulean, draw=Cerulean, line width=0mm, error bars/.cd, y dir=both, y explicit, error bar style={thick}, error bar style=black] coordinates { 
      (Radboud-U, 0.717829) +- (0, 0.0459769)
      (Radboud-U $\rightarrow$ Karolinska-U, 0.354394) +- (0, 0.00950279)
      (Radboud-L, 0.720773) +- (0, 0.0254158)
      (Radboud-L $\rightarrow$ Radboud-R, 0.44841) +- (0, 0.0081652)};

    \addplot [fill=darkcerulean2, draw=darkcerulean2, line width=0mm, error bars/.cd, y dir=both, y explicit, error bar style={thick}, error bar style=black] coordinates { 
      (Radboud-U, 0.674127) +- (0, 0.0351979)
      (Radboud-U $\rightarrow$ Karolinska-U, 0.314316) +- (0, 0.0137609)
      (Radboud-L, 0.633288) +- (0, 0.0313997)
      (Radboud-L $\rightarrow$ Radboud-R, 0.375735) +- (0, 0.00683095)};

    \addplot [fill=darkpastelpurple2, draw=darkpastelpurple2, error bars/.cd, y dir=both, y explicit, error bar style={thick}, error bar style=black] coordinates {
      (Radboud-U, 0.711507) +- (0, 0.0339886)
      (Radboud-U $\rightarrow$ Karolinska-U, 0.328888) +- (0, 0.0249578)
      (Radboud-L, 0.652109) +- (0, 0.0390751)
      (Radboud-L $\rightarrow$ Radboud-R, 0.379005) +- (0, 0.00637403)};

    \addplot [fill=goldenpoppy, draw=goldenpoppy, error bars/.cd, y dir=both, y explicit, error bar style={thick}, error bar style=black] coordinates {
      (Radboud-U, 0.644478) +- (0, 0.0315855)
      (Radboud-U $\rightarrow$ Karolinska-U, 0.215219) +- (0, 0.0373707)
      (Radboud-L, 0.593542) +- (0, 0.0519831) 
      (Radboud-L $\rightarrow$ Radboud-R, 0.35055) +- (0, 0.00817425)};

    \addplot [fill=goldenpoppy3, draw=goldenpoppy3, error bars/.cd, y dir=both, y explicit, error bar style={thick}, error bar style=black] coordinates {
      (Radboud-U, 0.672094) +- (0, 0.0407603)
      (Radboud-U $\rightarrow$ Karolinska-U, 0.368012) +- (0, 0.0126055)
      (Radboud-L, 0.653078) +- (0, 0.0454299)
      (Radboud-L $\rightarrow$ Radboud-R, 0.386323) +- (0, 0.00664311)};
    \legend{
            Resnet-IN,
            Phikon-v2,
            UNI,
            UNI2-h,
            CONCH,
            CONCHv1.5,
            H-optimus-0,
            H-optimus-1,
            H0-mini,
            Prov-GigaPath,
            Virchow,
            Virchow2
            }
  \end{axis}
  \end{tikzpicture}\vspace{0.0mm}
    \caption{\textbf{Main model comparison under controlled label-distribution and cross-site shifts.} The same performance comparison as in Figure~\ref{fig:main_results3a_12fms_kappa}, evaluated across four additional PANDA subsets designed to isolate the effects of label-distribution shift and domain shift. All results are reported as mean $\pm$ standard deviation over $10$ cross-validation folds. Corresponding results in terms of MAE are in Figure~\ref{fig:main_results3b_12fms_mae}.}
    \label{fig:main_results3b_12fms_kappa}
\end{figure*}

Recent benchmarking studies have shown that PFMs outperform earlier pathology encoders and natural-image baselines across a range of downstream tasks~\citep{marza2025thunder, kasireddy2026comprehensive, breen2025comprehensive, bareja2025evaluating, campanella2025clinical, neidlinger2025benchmarking, ma2025pathbench, gustafsson2026benchmarking}. However, these comparisons have primarily emphasized average performance under standard evaluation settings, rather than robustness under clinically plausible domain shifts. Recent studies have begun to address this gap more directly. In particular, \citet{thiringer2026scanner} showed that PFMs are sensitive to scanner-induced variability under controlled experimental conditions: using identical tissue sections scanned on multiple devices, they observed pronounced scanner-specific structure in the learned feature spaces. Similarly, \citet{carloni2025pathologyfoundationmodelsscanner} reported scanner sensitivity across several recent PFMs, while \citet{chai2026impact} studied the effects of both staining and scanner variability on downstream performance.

While these studies indicate that scanner- and stain-related variability remain important challenges for PFMs, their impact in realistic deployment scenarios remains less well understood. In particular, real-world applications often involve compound distribution shifts, where differences in scanners, staining protocols, tissue processing, and patient populations occur simultaneously. Moreover, downstream prediction models may be trained on limited data from a single institution and then applied to data from other sites with different acquisition characteristics. This setting constitutes a stringent test of deployment robustness, where the training data may not adequately capture the variability encountered at test time. It also raises a broader question: \textit{to what extent do the scale and diversity of modern PFM pretraining actually translate into robust downstream generalization?} Although PFMs are trained on increasingly large and heterogeneous datasets, this does not necessarily guarantee that models built on top of them will generalize across sites when trained on limited or single-site data.

In this work, we study these questions for prostate cancer grading from biopsy WSIs. We benchmark a diverse set of widely used PFMs (Table~\ref{table:fms}), including state-of-the-art models trained on more than one million WSIs. Using the PANDA dataset~\citep{bulten2022artificial}, we evaluate these PFMs as frozen patch-level feature extractors within standardized weakly supervised WSI-level grading models, and assess robustness under two practically relevant forms of shift: cross-site shifts in WSI image appearance, and shifts in the label distribution over cancer grade groups. This setup enables us to separate strong average performance from robustness under deployment-relevant distribution shifts.

\textit{Our study makes three main contributions:}
(1) We present a systematic benchmark of PFMs under clinically relevant distribution shifts, explicitly evaluating robustness to both cross-site image shift and label-distribution shift in a standardized setting.
(2) We provide a large-scale evaluation of widely used models, including recent vision-only PFMs, vision-language PFMs, a compact distilled PFM, and a natural-image baseline.
(3) We show that, despite strong in-distribution performance, PFMs exhibit substantial degradation under cross-site transfer, and that increasing model scale or pretraining data alone does not reliably improve robustness. These results highlight that the quality and diversity of downstream training data remain crucial even in the foundation-model era.

\begin{figure*}[t]
    \centering
    \begin{tikzpicture}
  \centering
  \begin{axis}[
        ybar, axis on top,
        title={},
        height=4.525cm, width=18.0cm, 
        bar width=0.35cm,
        ymajorgrids, tick align=inside,
        enlarge y limits={value=.1,upper},
        ymin=0, ymax=0.9,
        set layers,
        ytick={0, 0.2, 0.4, 0.6, 0.8},
        axis x line*=bottom,
        axis y line*=left,
        y axis line style={opacity=0},
        tickwidth=0pt,
        enlarge x limits=true,
        extra y tick style={
            grid style={
                solid,
                line width = 2.5pt,
                /pgfplots/on layer=axis background,
            },
        },
        tick style={
            grid style={
                dashed,
                /pgfplots/on layer=axis background,
            },
        },
        legend style={
            at={(0.5,-0.30)},
            anchor=north,
            legend columns=-1,
            /tikz/every even column/.append style={column sep=0.325cm},
            font=\footnotesize
        },
        ylabel={Kappa ($\uparrow$)},
        ylabel style={font=\footnotesize},
        symbolic x coords={
           PANDA,
           Karolinska,
           Radboud, 
           Radboud $\rightarrow$\\ Karolinska,
           Radboud-U, 
           Radboud-U $\rightarrow$\\ Karolinska-U,
           Radboud-L, 
           Radboud-L $\rightarrow$\\ Radboud-R},
       xtick=data,
       xticklabel=\empty,
       x tick label style={text width = 2.25cm, align=center, font=\footnotesize},
    ]
    \addplot [fill=skyblue2, draw=skyblue2, line width=0mm, error bars/.cd, y dir=both, y explicit, error bar style={thick}, error bar style=black] coordinates { 
      (PANDA, 0.773048) +- (0, 0.013836)
      (Karolinska, 0.750227) +- (0, 0.0271252)
      (Radboud, 0.726182) +- (0, 0.0236357)
      (Radboud $\rightarrow$\\ Karolinska, 0.185957) +- (0, 0.0414844)
      (Radboud-U, 0.725059) +- (0, 0.0187998)
      (Radboud-U $\rightarrow$\\ Karolinska-U, 0.380423) +- (0, 0.0353908)
      (Radboud-L, 0.70861) +- (0, 0.0466765)
      (Radboud-L $\rightarrow$\\ Radboud-R, 0.545218) +- (0, 0.0142131)};
    \addplot [fill=caribbeangreen2, draw=caribbeangreen2, error bars/.cd, y dir=both, y explicit, error bar style={thick}, error bar style=black] coordinates {
      (PANDA, 0.610517) +- (0, 0.0309128)
      (Karolinska, 0.484289) +- (0, 0.0414748)
      (Radboud, 0.565746) +- (0, 0.0239493)
      (Radboud $\rightarrow$\\ Karolinska, 0.216677) +- (0, 0.0230846)
      (Radboud-U, 0.550741) +- (0, 0.0401625)
      (Radboud-U $\rightarrow$\\ Karolinska-U, 0.369438) +- (0, 0.00683355)
      (Radboud-L, 0.514023) +- (0, 0.0456437)
      (Radboud-L $\rightarrow$\\ Radboud-R, 0.425298) +- (0, 0.00646147)};
    \addplot [fill=darkpastelpurple2, draw=darkpastelpurple2, error bars/.cd, y dir=both, y explicit, error bar style={thick}, error bar style=black] coordinates {
      (PANDA, 0.613157) +- (0, 0.016445)
      (Karolinska, 0.461135) +- (0, 0.0302627)
      (Radboud, 0.630168) +- (0, 0.0422488)
      (Radboud $\rightarrow$\\ Karolinska, 0.0705993) +- (0, 0.0134034)
      (Radboud-U, 0.594313) +- (0, 0.0487418)
      (Radboud-U $\rightarrow$\\ Karolinska-U, 0.0980739) +- (0, 0.0298326)
      (Radboud-L, 0.502797) +- (0, 0.0414738)
      (Radboud-L $\rightarrow$\\ Radboud-R, 0.282891) +- (0, 0.00887223)};
  \end{axis}
  \end{tikzpicture}\vspace{-2.5mm}
    \begin{tikzpicture}
  \centering
  \begin{axis}[
        ybar, axis on top,
        title={},
        height=4.525cm, width=18.0cm, 
        bar width=0.35cm,
        ymajorgrids, tick align=inside,
        enlarge y limits={value=.1,upper},
        ymin=0, ymax=0.9,
        set layers,
        ytick={0.2, 0.4, 0.6, 0.8},
        axis x line*=bottom,
        axis y line*=left,
        y axis line style={opacity=0},
        tickwidth=0pt,
        enlarge x limits=true,
        extra y tick style={
            grid style={
                solid,
                line width = 2.5pt,
                /pgfplots/on layer=axis background,
            },
        },
        tick style={
            grid style={
                dashed,
                /pgfplots/on layer=axis background,
            },
        },
        legend style={
            at={(0.5,-0.2)},
            anchor=north,
            legend columns=-1,
            /tikz/every even column/.append style={column sep=0.325cm},
            font=\footnotesize
        },
        ylabel={Kappa ($\uparrow$)},
        ylabel style={font=\footnotesize},
        symbolic x coords={
           PANDA,
           Karolinska,
           Radboud, 
           Radboud $\rightarrow$\\ Karolinska,
           Radboud-U, 
           Radboud-U $\rightarrow$\\ Karolinska-U,
           Radboud-L, 
           Radboud-L $\rightarrow$\\ Radboud-R},
       xtick=data,
       xticklabel=\empty,
       x tick label style={text width = 2.25cm, align=center, font=\footnotesize},
    ]
    \addplot [fill=skyblue2, draw=skyblue2, line width=0mm, error bars/.cd, y dir=both, y explicit, error bar style={thick}, error bar style=black] coordinates { 
      (PANDA, 0.8884) +- (0, 0.013359)
      (Karolinska, 0.886496 ) +- (0, 0.00971835)
      (Radboud, 0.853788) +- (0, 0.0262305)
      (Radboud $\rightarrow$\\ Karolinska, 0.24719) +- (0, 0.138006)
      (Radboud-U, 0.843966) +- (0, 0.0237577)
      (Radboud-U $\rightarrow$\\ Karolinska-U, 0.459637) +- (0, 0.112533)
      (Radboud-L, 0.826279) +- (0, 0.0361249)
      (Radboud-L $\rightarrow$\\ Radboud-R, 0.739375) +- (0, 0.0142154)};
    \addplot [fill=caribbeangreen2, draw=caribbeangreen2, error bars/.cd, y dir=both, y explicit, error bar style={thick}, error bar style=black] coordinates {
      (PANDA, 0.84559) +- (0, 0.0208975)
      (Karolinska, 0.779771) +- (0, 0.0216045)
      (Radboud, 0.846202) +- (0, 0.015932)
      (Radboud $\rightarrow$\\ Karolinska, 0.489885) +- (0, 0.0596735)
      (Radboud-U, 0.828197) +- (0, 0.0287752)
      (Radboud-U $\rightarrow$\\ Karolinska-U, 0.516602) +- (0, 0.018329)
      (Radboud-L, 0.806884) +- (0, 0.0328776)
      (Radboud-L $\rightarrow$\\ Radboud-R, 0.686432) +- (0, 0.0206927)};
    \addplot [fill=darkpastelpurple2, draw=darkpastelpurple2, error bars/.cd, y dir=both, y explicit, error bar style={thick}, error bar style=black] coordinates {
      (PANDA, 0.727169) +- (0, 0.0182163)
      (Karolinska, 0.582271) +- (0, 0.0332399)
      (Radboud, 0.747012) +- (0, 0.0270309)
      (Radboud $\rightarrow$\\ Karolinska, 0.214366) +- (0, 0.0376718)
      (Radboud-U, 0.711353) +- (0, 0.0360002)
      (Radboud-U $\rightarrow$\\ Karolinska-U, 0.283195) +- (0, 0.0345969)
      (Radboud-L, 0.66672) +- (0, 0.0305624)
      (Radboud-L $\rightarrow$\\ Radboud-R, 0.397871) +- (0, 0.00518936)};
  \end{axis}
  \end{tikzpicture}\vspace{-2.5mm}
    \begin{tikzpicture}
  \centering
  \begin{axis}[
        ybar, axis on top,
        title={},
        height=4.525cm, width=18.0cm, 
        bar width=0.35cm,
        ymajorgrids, tick align=inside,
        enlarge y limits={value=.1,upper},
        ymin=0, ymax=0.9,
        set layers,
        ytick={0.2, 0.4, 0.6, 0.8},
        axis x line*=bottom,
        axis y line*=left,
        y axis line style={opacity=0},
        tickwidth=0pt,
        enlarge x limits=true,
        extra y tick style={
            grid style={
                solid,
                line width = 2.5pt,
                /pgfplots/on layer=axis background,
            },
        },
        tick style={
            grid style={
                dashed,
                /pgfplots/on layer=axis background,
            },
        },
        legend style={
            at={(0.5,-0.2)},
            anchor=north,
            legend columns=-1,
            /tikz/every even column/.append style={column sep=0.325cm},
            font=\footnotesize
        },
        ylabel={Kappa ($\uparrow$)},
        ylabel style={font=\footnotesize},
        symbolic x coords={
           PANDA,
           Karolinska,
           Radboud, 
           Radboud $\rightarrow$\\ Karolinska,
           Radboud-U, 
           Radboud-U $\rightarrow$\\ Karolinska-U,
           Radboud-L, 
           Radboud-L $\rightarrow$\\ Radboud-R},
       xtick=data,
       xticklabel=\empty,
       x tick label style={text width = 2.25cm, align=center, font=\footnotesize},
    ]
    \addplot [fill=skyblue2, draw=skyblue2, line width=0mm, error bars/.cd, y dir=both, y explicit, error bar style={thick}, error bar style=black] coordinates { 
      (PANDA, 0.866334) +- (0, 0.00907365)
      (Karolinska, 0.86257) +- (0, 0.015705)
      (Radboud, 0.818953) +- (0, 0.0213635)
      (Radboud $\rightarrow$\\ Karolinska, 0.0245775) +- (0, 0.0183994)
      (Radboud-U, 0.81895) +- (0, 0.0287215)
      (Radboud-U $\rightarrow$\\ Karolinska-U, 0.206042) +- (0, 0.0755954)
      (Radboud-L, 0.805051) +- (0, 0.0483934)
      (Radboud-L $\rightarrow$\\ Radboud-R, 0.712784) +- (0, 0.0147616)};
    \addplot [fill=caribbeangreen2, draw=caribbeangreen2, error bars/.cd, y dir=both, y explicit, error bar style={thick}, error bar style=black] coordinates {
      (PANDA, 0.777674) +- (0, 0.0198599)
      (Karolinska, 0.714824) +- (0, 0.0230472)
      (Radboud, 0.743565) +- (0, 0.0255462)
      (Radboud $\rightarrow$\\ Karolinska, 0.0261816 ) +- (0, 0.0150573)
      (Radboud-U, 0.743879) +- (0, 0.0386241)
      (Radboud-U $\rightarrow$\\ Karolinska-U, 0.165819) +- (0, 0.0488119)
      (Radboud-L, 0.73137) +- (0, 0.055067)
      (Radboud-L $\rightarrow$\\ Radboud-R, 0.59083) +- (0, 0.0195734)};
    \addplot [fill=darkpastelpurple2, draw=darkpastelpurple2, error bars/.cd, y dir=both, y explicit, error bar style={thick}, error bar style=black] coordinates {
      (PANDA, 0.690939) +- (0, 0.0214606)
      (Karolinska, 0.590366) +- (0, 0.0260291)
      (Radboud, 0.687728) +- (0, 0.0414731)
      (Radboud $\rightarrow$\\ Karolinska, 0.208055) +- (0, 0.0128485)
      (Radboud-U, 0.665943) +- (0, 0.0300574)
      (Radboud-U $\rightarrow$\\ Karolinska-U, 0.320718) +- (0, 0.0234927)
      (Radboud-L, 0.630585) +- (0, 0.0435856)
      (Radboud-L $\rightarrow$\\ Radboud-R, 0.38761) +- (0, 0.00848339)};
  \end{axis}
  \end{tikzpicture}\vspace{-2.5mm}
    \begin{tikzpicture}
  \centering
  \begin{axis}[
        ybar, axis on top,
        title={},
        height=4.525cm, width=18.0cm, 
        bar width=0.35cm,
        ymajorgrids, tick align=inside,
        enlarge y limits={value=.1,upper},
        ymin=0, ymax=0.9,
        set layers,
        ytick={0.2, 0.4, 0.6, 0.8},
        axis x line*=bottom,
        axis y line*=left,
        y axis line style={opacity=0},
        tickwidth=0pt,
        enlarge x limits=true,
        extra y tick style={
            grid style={
                solid,
                line width = 2.5pt,
                /pgfplots/on layer=axis background,
            },
        },
        tick style={
            grid style={
                dashed,
                /pgfplots/on layer=axis background,
            },
        },
        legend style={
            at={(0.5,-0.2)},
            anchor=north,
            legend columns=-1,
            /tikz/every even column/.append style={column sep=0.325cm},
            font=\footnotesize
        },
        ylabel={Kappa ($\uparrow$)},
        ylabel style={font=\footnotesize},
        symbolic x coords={
           PANDA,
           Karolinska,
           Radboud, 
           Radboud $\rightarrow$\\ Karolinska,
           Radboud-U, 
           Radboud-U $\rightarrow$\\ Karolinska-U,
           Radboud-L, 
           Radboud-L $\rightarrow$\\ Radboud-R},
       xtick=data,
       xticklabel=\empty,
       x tick label style={text width = 2.25cm, align=center, font=\footnotesize},
    ]
    \addplot [fill=skyblue2, draw=skyblue2, line width=0mm, error bars/.cd, y dir=both, y explicit, error bar style={thick}, error bar style=black] coordinates { 
      (PANDA, 0.858159) +- (0, 0.00987344)
      (Karolinska, 0.86872) +- (0, 0.0181902)
      (Radboud, 0.815367) +- (0, 0.02357)
      (Radboud $\rightarrow$\\ Karolinska, 0.723546) +- (0, 0.037291)
      (Radboud-U, 0.809397) +- (0, 0.032308)
      (Radboud-U $\rightarrow$\\ Karolinska-U, 0.69802) +- (0, 0.0174124)
      (Radboud-L, 0.809747) +- (0, 0.0349779)
      (Radboud-L $\rightarrow$\\ Radboud-R, 0.689074) +- (0, 0.0186569)};
    \addplot [fill=caribbeangreen2, draw=caribbeangreen2, error bars/.cd, y dir=both, y explicit, error bar style={thick}, error bar style=black] coordinates {
      (PANDA, 0.768022) +- (0, 0.0174022)
      (Karolinska, 0.704212) +- (0, 0.0273398)
      (Radboud, 0.727299) +- (0, 0.0206647)
      (Radboud $\rightarrow$\\ Karolinska, 0.556405) +- (0, 0.0145983)
      (Radboud-U, 0.72103) +- (0, 0.0362594)
      (Radboud-U $\rightarrow$\\ Karolinska-U, 0.522532) +- (0, 0.0149313)
      (Radboud-L, 0.729058) +- (0, 0.033372)
      (Radboud-L $\rightarrow$\\ Radboud-R, 0.584524) +- (0, 0.0211163)};
    \addplot [fill=darkpastelpurple2, draw=darkpastelpurple2, error bars/.cd, y dir=both, y explicit, error bar style={thick}, error bar style=black] coordinates {
      (PANDA, 0.703755) +- (0, 0.0167515)
      (Karolinska, 0.612679) +- (0, 0.0267452)
      (Radboud, 0.695315) +- (0, 0.0419093)
      (Radboud $\rightarrow$\\ Karolinska, 0.344483) +- (0, 0.0133134)
      (Radboud-U, 0.665767) +- (0, 0.0482797)
      (Radboud-U $\rightarrow$\\ Karolinska-U, 0.400034) +- (0, 0.0166285)
      (Radboud-L, 0.635305) +- (0, 0.0365897)
      (Radboud-L $\rightarrow$\\ Radboud-R, 0.397994) +- (0, 0.00457531)};
  \end{axis}
  \end{tikzpicture}\vspace{-2.5mm}
    \begin{tikzpicture}
  \centering
  \begin{axis}[
        ybar, axis on top,
        title={},
        height=4.525cm, width=18.0cm, 
        bar width=0.35cm,
        ymajorgrids, tick align=inside,
        enlarge y limits={value=.1,upper},
        ymin=0, ymax=0.9,
        set layers,
        ytick={0.2, 0.4, 0.6, 0.8},
        axis x line*=bottom,
        axis y line*=left,
        y axis line style={opacity=0},
        tickwidth=0pt,
        enlarge x limits=true,
        extra y tick style={
            grid style={
                solid,
                line width = 2.5pt,
                /pgfplots/on layer=axis background,
            },
        },
        tick style={
            grid style={
                dashed,
                /pgfplots/on layer=axis background,
            },
        },
        legend style={
            at={(0.5,-0.2)},
            anchor=north,
            legend columns=-1,
            /tikz/every even column/.append style={column sep=0.325cm},
            font=\footnotesize
        },
        ylabel={Kappa ($\uparrow$)},
        ylabel style={font=\footnotesize},
        symbolic x coords={
           PANDA,
           Karolinska,
           Radboud, 
           Radboud $\rightarrow$\\ Karolinska,
           Radboud-U, 
           Radboud-U $\rightarrow$\\ Karolinska-U,
           Radboud-L, 
           Radboud-L $\rightarrow$\\ Radboud-R},
       xtick=data,
       xticklabel=\empty,
       x tick label style={text width = 2.25cm, align=center, font=\footnotesize},
    ]
    \addplot [fill=skyblue2, draw=skyblue2, line width=0mm, error bars/.cd, y dir=both, y explicit, error bar style={thick}, error bar style=black] coordinates { 
      (PANDA, 0.886692) +- (0, 0.0169589)
      (Karolinska, 0.882394) +- (0, 0.014148)
      (Radboud, 0.863841) +- (0, 0.0261018)
      (Radboud $\rightarrow$\\ Karolinska, 0.428641) +- (0, 0.0988533)
      (Radboud-U, 0.850576) +- (0, 0.0132259)
      (Radboud-U $\rightarrow$\\ Karolinska-U, 0.643878) +- (0, 0.0598084)
      (Radboud-L, 0.838932) +- (0, 0.0284458)
      (Radboud-L $\rightarrow$\\ Radboud-R, 0.734561) +- (0, 0.0187071)};
    \addplot [fill=caribbeangreen2, draw=caribbeangreen2, error bars/.cd, y dir=both, y explicit, error bar style={thick}, error bar style=black] coordinates {
      (PANDA, 0.839958) +- (0, 0.0160809)
      (Karolinska, 0.780891) +- (0, 0.0263926)
      (Radboud, 0.829961) +- (0, 0.0204068)
      (Radboud $\rightarrow$\\ Karolinska, 0.515981) +- (0, 0.0245259)
      (Radboud-U, 0.822937) +- (0, 0.0290758)
      (Radboud-U $\rightarrow$\\ Karolinska-U, 0.493689) +- (0, 0.0187603)
      (Radboud-L, 0.812939) +- (0, 0.0248034)
      (Radboud-L $\rightarrow$\\ Radboud-R, 0.685654) +- (0, 0.00678213)};
    \addplot [fill=darkpastelpurple2, draw=darkpastelpurple2, error bars/.cd, y dir=both, y explicit, error bar style={thick}, error bar style=black] coordinates {
      (PANDA, 0.734141) +- (0, 0.0190904)
      (Karolinska, 0.60592) +- (0, 0.0270284)
      (Radboud, 0.754972) +- (0, 0.0333241)
      (Radboud $\rightarrow$\\ Karolinska, 0.374623) +- (0, 0.0199929)
      (Radboud-U, 0.717829) +- (0, 0.0459769)
      (Radboud-U $\rightarrow$\\ Karolinska-U, 0.354394) +- (0, 0.00950279)
      (Radboud-L, 0.720773) +- (0, 0.0254158)
      (Radboud-L $\rightarrow$\\ Radboud-R, 0.44841) +- (0, 0.0081652)};
  \end{axis}
  \end{tikzpicture}\vspace{-2.5mm}
    \begin{tikzpicture}
  \centering
  \begin{axis}[
        ybar, axis on top,
        title={},
        height=4.525cm, width=18.0cm, 
        bar width=0.35cm,
        ymajorgrids, tick align=inside,
        enlarge y limits={value=.1,upper},
        ymin=0, ymax=0.9,
        set layers,
        ytick={0.2, 0.4, 0.6, 0.8},
        axis x line*=bottom,
        axis y line*=left,
        y axis line style={opacity=0},
        tickwidth=0pt,
        enlarge x limits=true,
        extra y tick style={
            grid style={
                solid,
                line width = 2.5pt,
                /pgfplots/on layer=axis background,
            },
        },
        tick style={
            grid style={
                dashed,
                /pgfplots/on layer=axis background,
            },
        },
        legend style={
            at={(0.5,-0.30)},
            anchor=north,
            legend columns=-1,
            /tikz/every even column/.append style={column sep=0.325cm},
            font=\footnotesize
        },
        ylabel={Kappa ($\uparrow$)},
        ylabel style={font=\footnotesize},
        symbolic x coords={
           PANDA,
           Karolinska,
           Radboud, 
           Radboud $\rightarrow$\\ Karolinska,
           Radboud-U, 
           Radboud-U $\rightarrow$\\ Karolinska-U,
           Radboud-L, 
           Radboud-L $\rightarrow$\\ Radboud-R},
       xtick=data,
       x tick label style={text width = 2.25cm, align=center, font=\footnotesize},
    ]
    \addplot [fill=skyblue2, draw=skyblue2, line width=0mm, error bars/.cd, y dir=both, y explicit, error bar style={thick}, error bar style=black] coordinates { 
      (PANDA, 0.876571) +- (0, 0.0117124)
      (Karolinska, 0.878736) +- (0, 0.013313)
      (Radboud, 0.847051) +- (0, 0.0178447)
      (Radboud $\rightarrow$\\ Karolinska, 0.504072) +- (0, 0.146641)
      (Radboud-U, 0.833371) +- (0, 0.0284642)
      (Radboud-U $\rightarrow$\\ Karolinska-U, 0.660097) +- (0, 0.0460499)
      (Radboud-L, 0.818717) +- (0, 0.0199882)
      (Radboud-L $\rightarrow$\\ Radboud-R, 0.719605) +- (0, 0.0213754)};
    \addplot [fill=caribbeangreen2, draw=caribbeangreen2, error bars/.cd, y dir=both, y explicit, error bar style={thick}, error bar style=black] coordinates {
      (PANDA, 0.827421) +- (0, 0.0133668)
      (Karolinska, 0.774161) +- (0, 0.0232635)
      (Radboud, 0.809273) +- (0, 0.0220188)
      (Radboud $\rightarrow$\\ Karolinska, 0.222778) +- (0, 0.0738892)
      (Radboud-U, 0.805385) +- (0, 0.0375729)
      (Radboud-U $\rightarrow$\\ Karolinska-U, 0.358065) +- (0, 0.0535879)
      (Radboud-L, 0.791257) +- (0, 0.0210861)
      (Radboud-L $\rightarrow$\\ Radboud-R, 0.65632) +- (0, 0.0127742)};
    \addplot [fill=darkpastelpurple2, draw=darkpastelpurple2, error bars/.cd, y dir=both, y explicit, error bar style={thick}, error bar style=black] coordinates {
      (PANDA, 0.694544) +- (0, 0.0116528)
      (Karolinska, 0.594539) +- (0, 0.0296609)
      (Radboud, 0.696469) +- (0, 0.0268229)
      (Radboud $\rightarrow$\\ Karolinska, 0.378458) +- (0, 0.0143491)
      (Radboud-U, 0.672094) +- (0, 0.0407603)
      (Radboud-U $\rightarrow$\\ Karolinska-U, 0.368012) +- (0, 0.0126055)
      (Radboud-L, 0.653078) +- (0, 0.0454299)
      (Radboud-L $\rightarrow$\\ Radboud-R, 0.386323) +- (0, 0.00664311)};
    \legend{
            \textit{ABMIL},
            \textit{Mean Feature},
            \textit{kNN}
            }
  \end{axis}
  \end{tikzpicture}\vspace{-1.0mm}
    \caption{\textbf{Comparison of downstream ISUP grade classification models.} Performance of the three ISUP grade models \textit{ABMIL}, \textit{Mean Feature}, and \textit{kNN} when using different feature extractors: Resnet-IN \textbf{(top row)}, UNI, CONCH, CONCHv1.5, H-optimus-1, and Virchow2 \textbf{(bottom row)}. Each row shows results across all PANDA subsets, corresponding to the same experiments as in Figure~\ref{fig:main_results3a_12fms_kappa}~\&~\ref{fig:main_results3b_12fms_kappa}, but reorganized to enable direct comparison of the downstream ISUP grade models. All results are reported as mean $\pm$ standard deviation over $10$ cross-validation folds. Results for the remaining six feature extractors are shown in Figure~\ref{fig:main_results_attmil_mean-pool_knn_kappa_remaining6}.}
    \label{fig:main_results_attmil_mean-pool_knn_kappa}
\end{figure*}

\input{figures/fig_detailed_comparison}

\begin{figure*}[h]
\centering
    \begin{subfigure}[t]{0.245\textwidth}
        \centering%
        \includegraphics[clip, trim=1.5cm 0.95cm 1.85cm 0.95cm, width=1.0\linewidth]{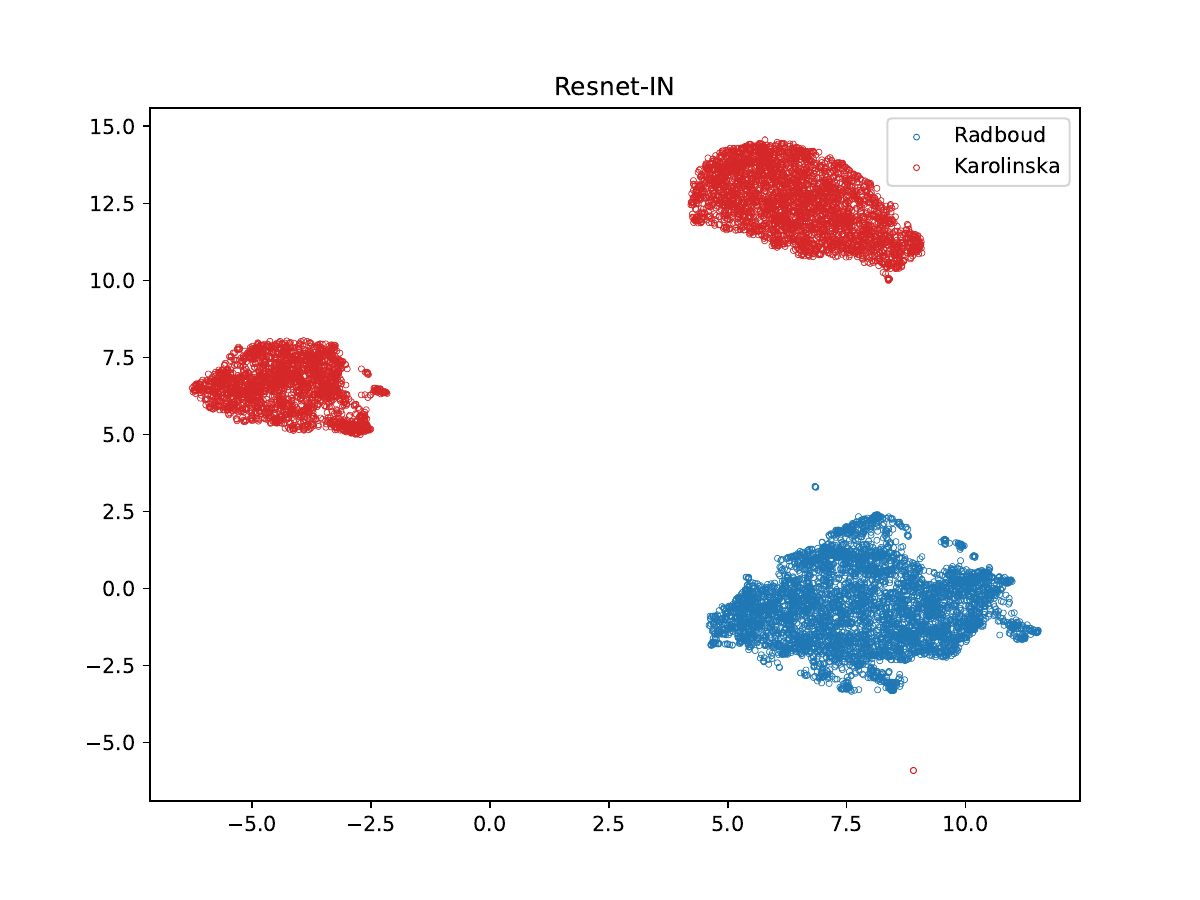}
    \end{subfigure}
    \begin{subfigure}[t]{0.245\textwidth}
        \centering%
        \includegraphics[clip, trim=1.5cm 0.95cm 1.85cm 0.95cm, width=1.0\linewidth]{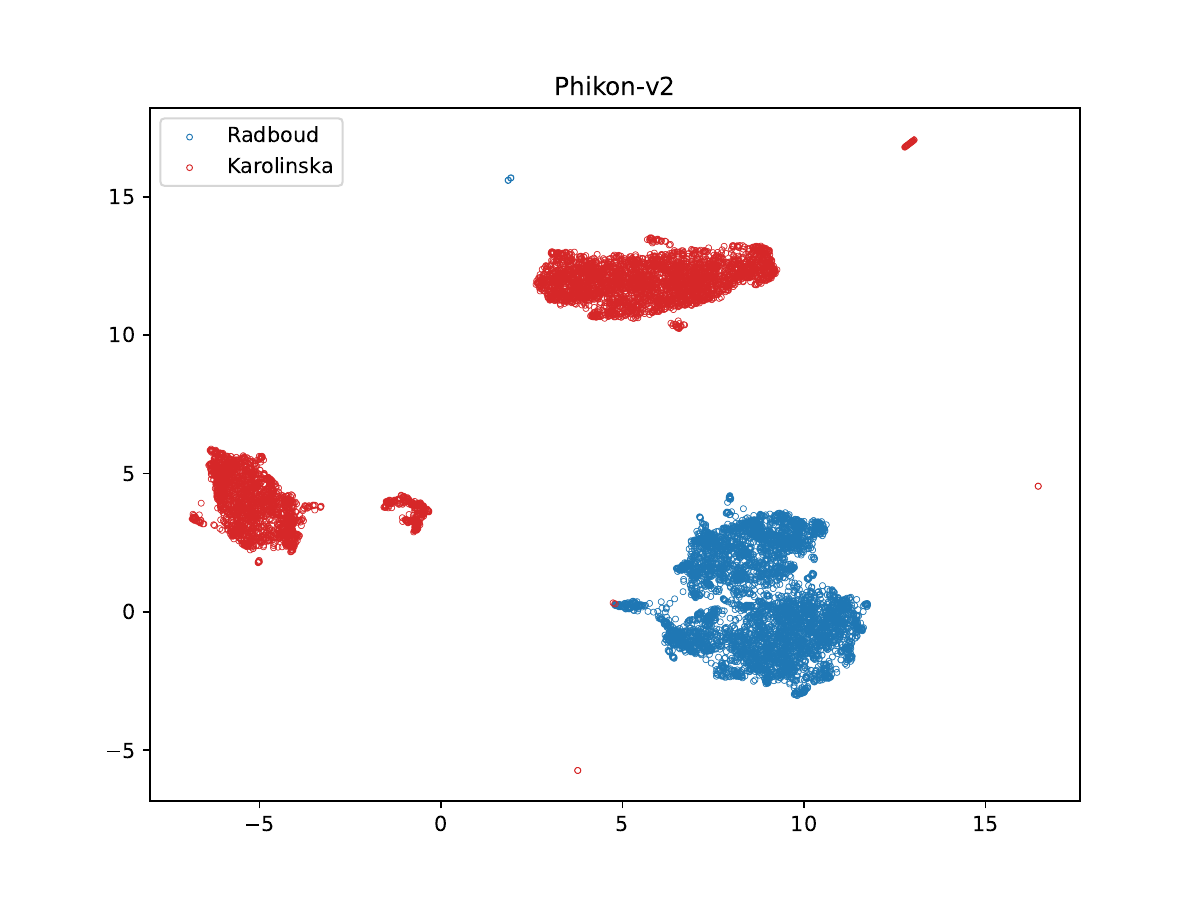}
    \end{subfigure}
    \begin{subfigure}[t]{0.245\textwidth}
        \centering%
        \includegraphics[clip, trim=1.5cm 0.95cm 1.85cm 0.95cm, width=1.0\linewidth]{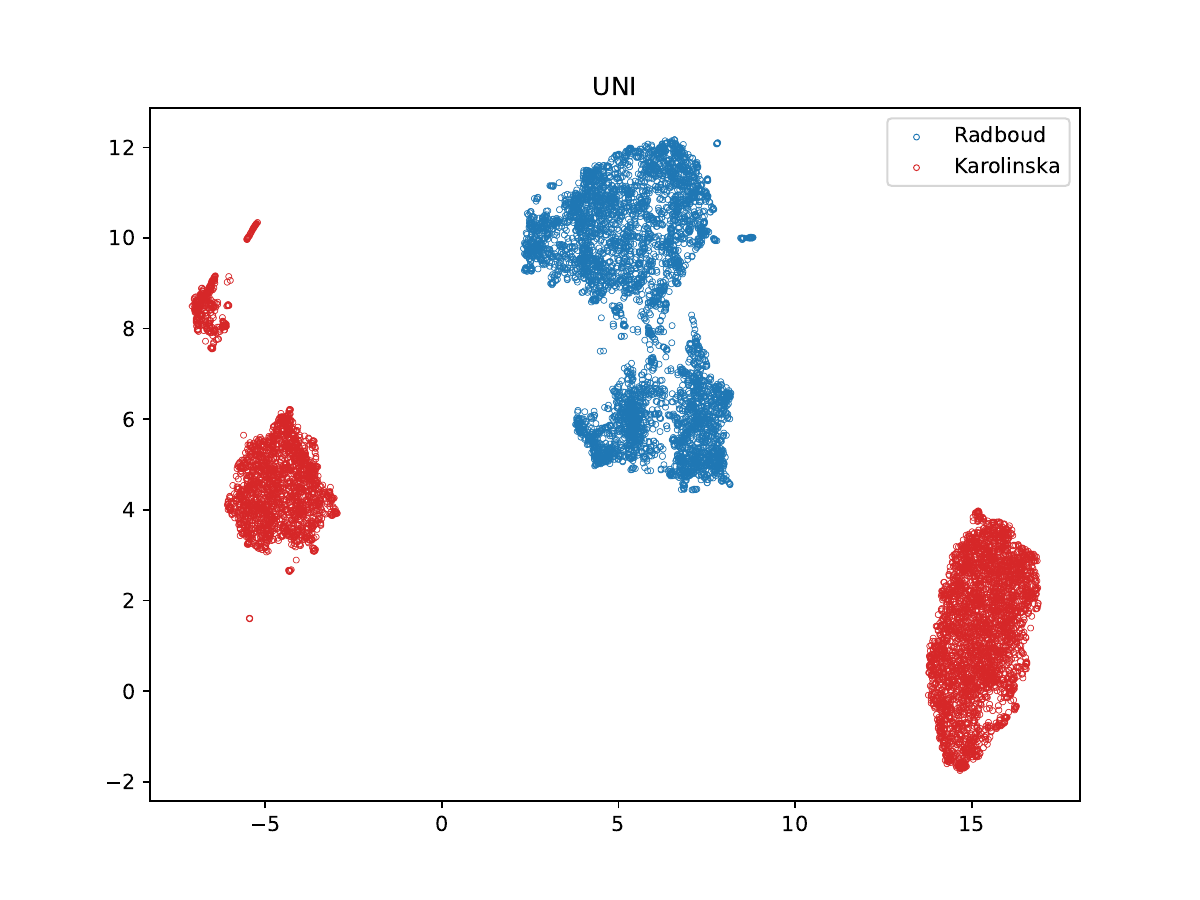}
    \end{subfigure}
    \begin{subfigure}[t]{0.245\textwidth}
        \centering%
        \includegraphics[clip, trim=1.5cm 0.95cm 1.85cm 0.95cm, width=1.0\linewidth]{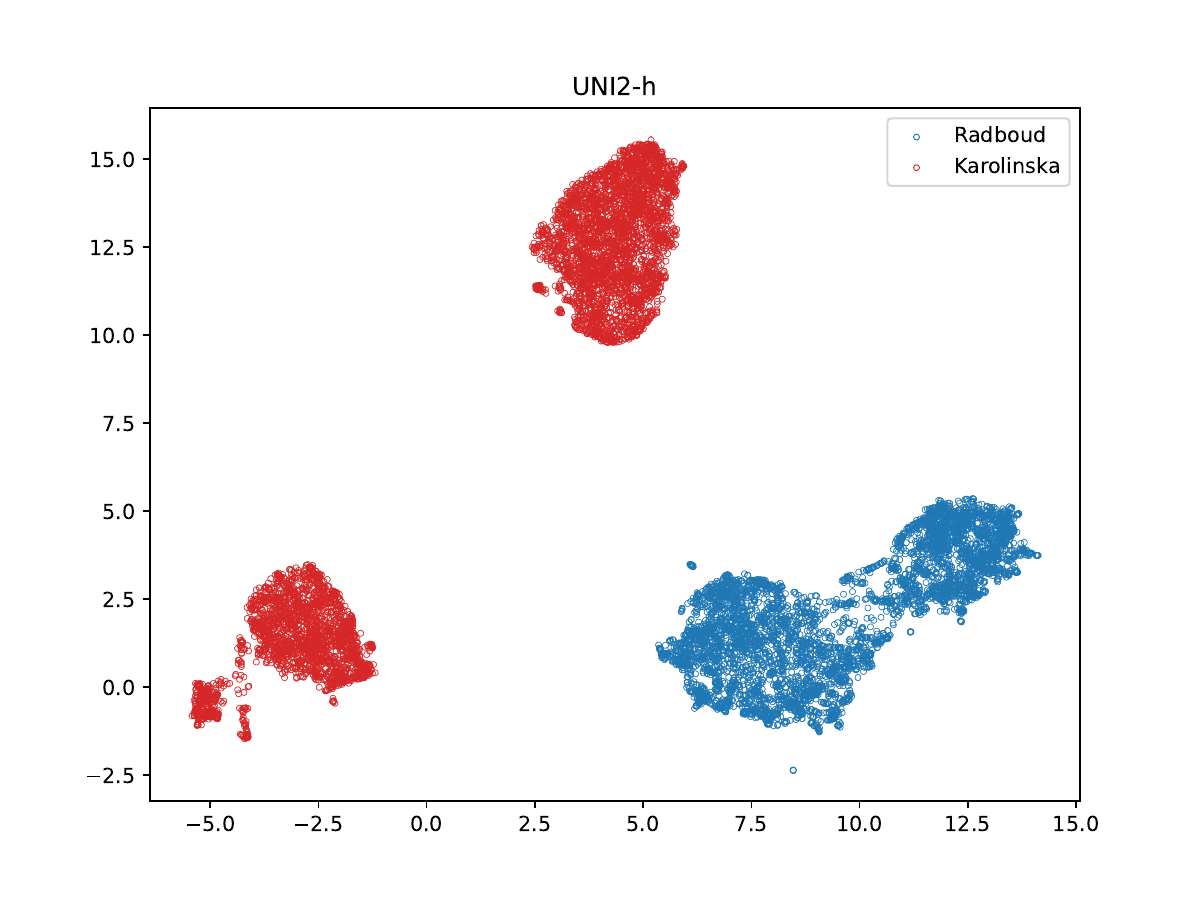}
    \end{subfigure}
    
    \begin{subfigure}[t]{0.245\textwidth}
        \centering%
        \includegraphics[clip, trim=1.5cm 0.95cm 1.85cm 0.95cm, width=1.0\linewidth]{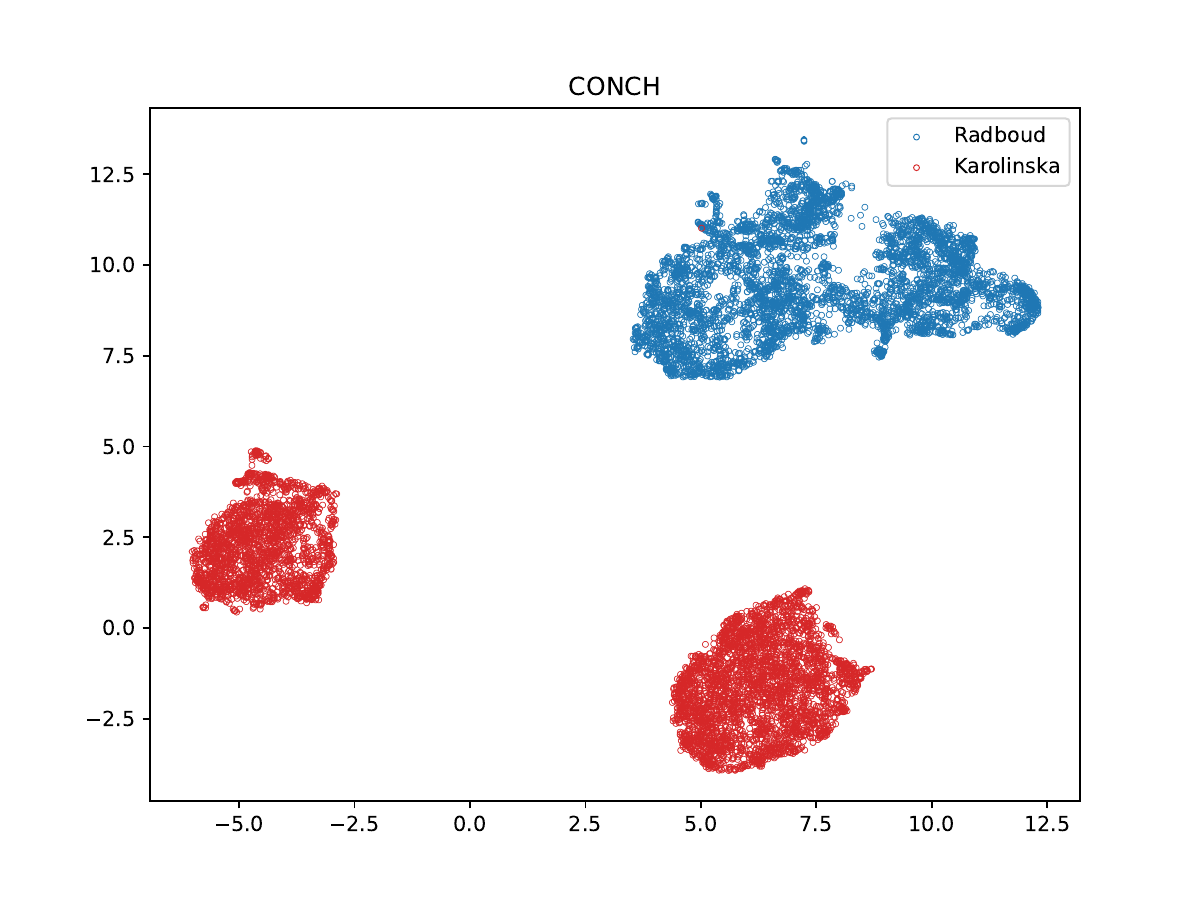}
    \end{subfigure}
    \begin{subfigure}[t]{0.245\textwidth}
        \centering%
        \includegraphics[clip, trim=1.5cm 0.95cm 1.85cm 0.95cm, width=1.0\linewidth]{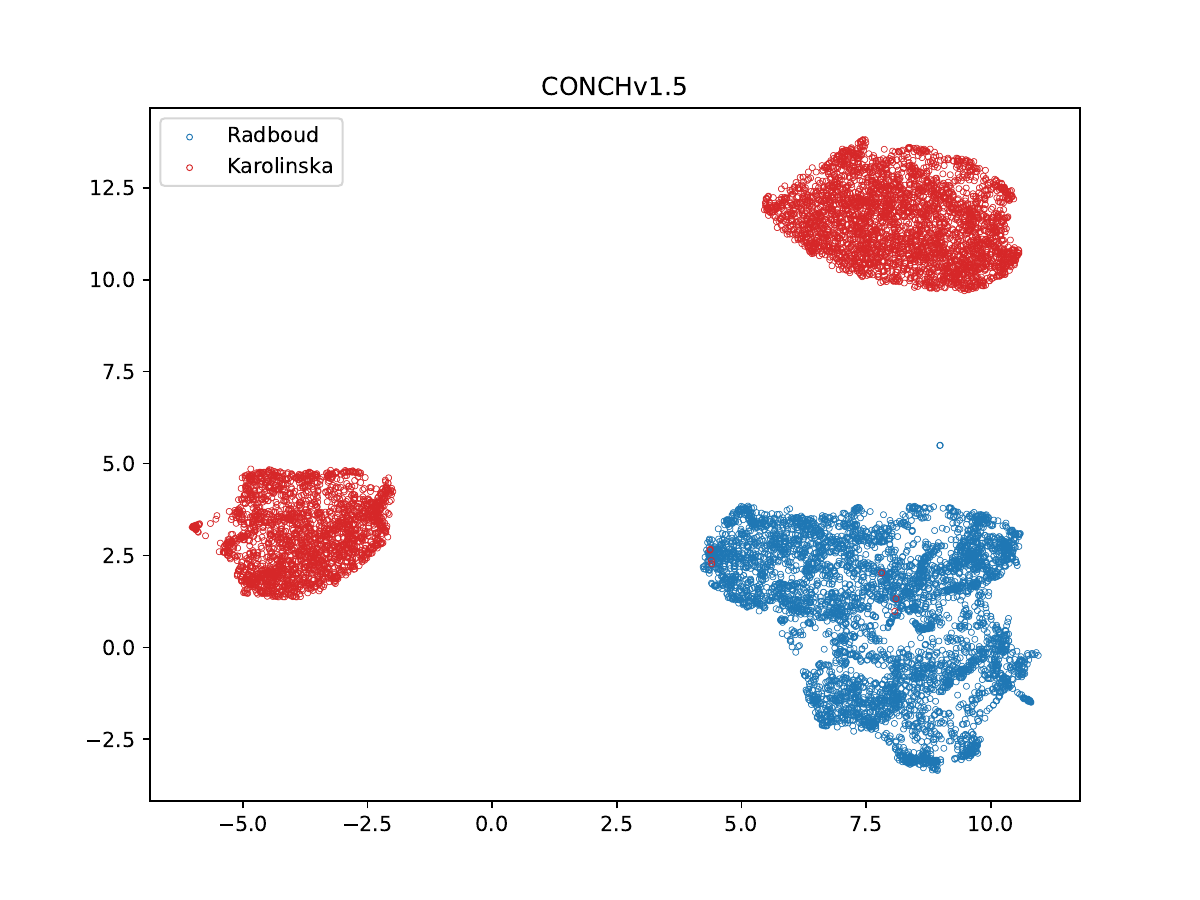}
    \end{subfigure}
    \begin{subfigure}[t]{0.245\textwidth}
        \centering%
        \includegraphics[clip, trim=1.5cm 0.95cm 1.85cm 0.95cm, width=1.0\linewidth]{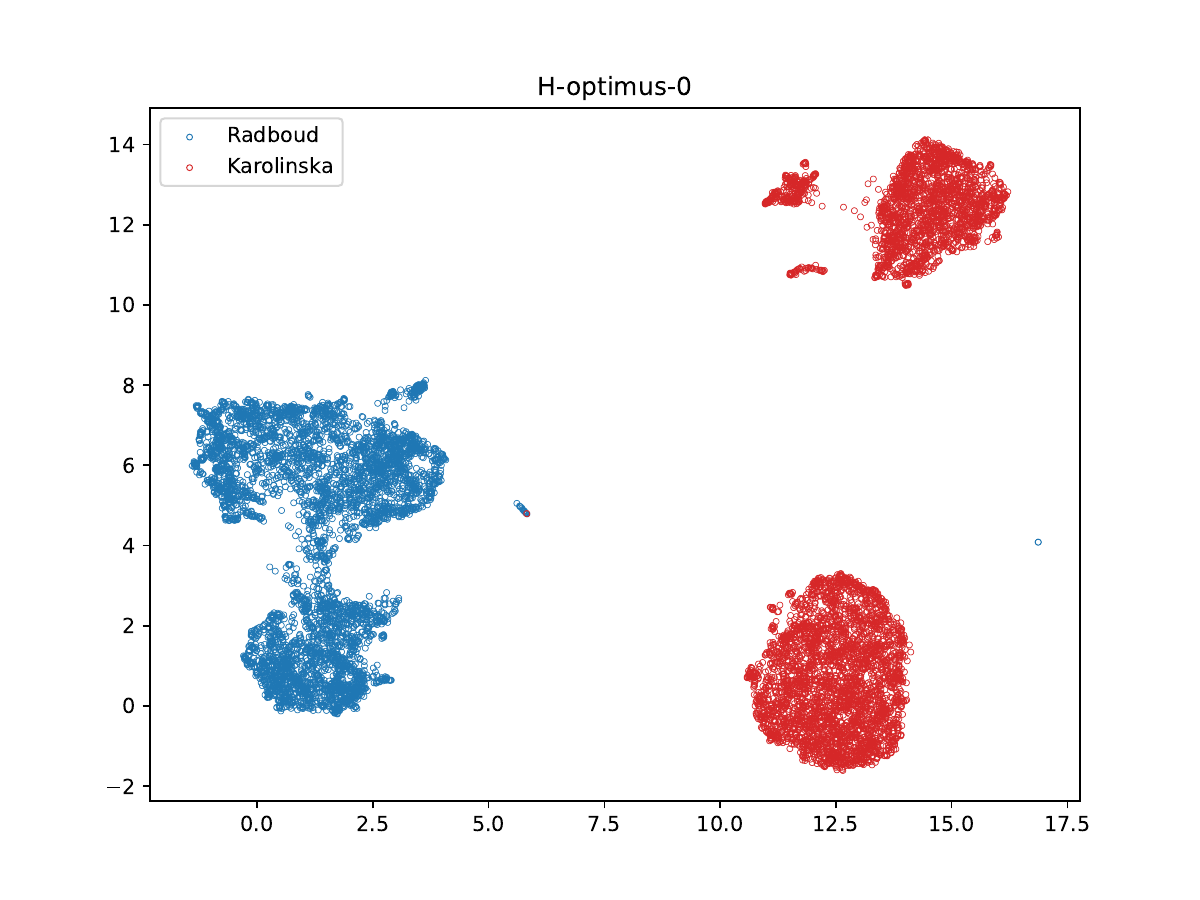}
    \end{subfigure}
    \begin{subfigure}[t]{0.245\textwidth}
        \centering%
        \includegraphics[clip, trim=1.5cm 0.95cm 1.85cm 0.95cm, width=1.0\linewidth]{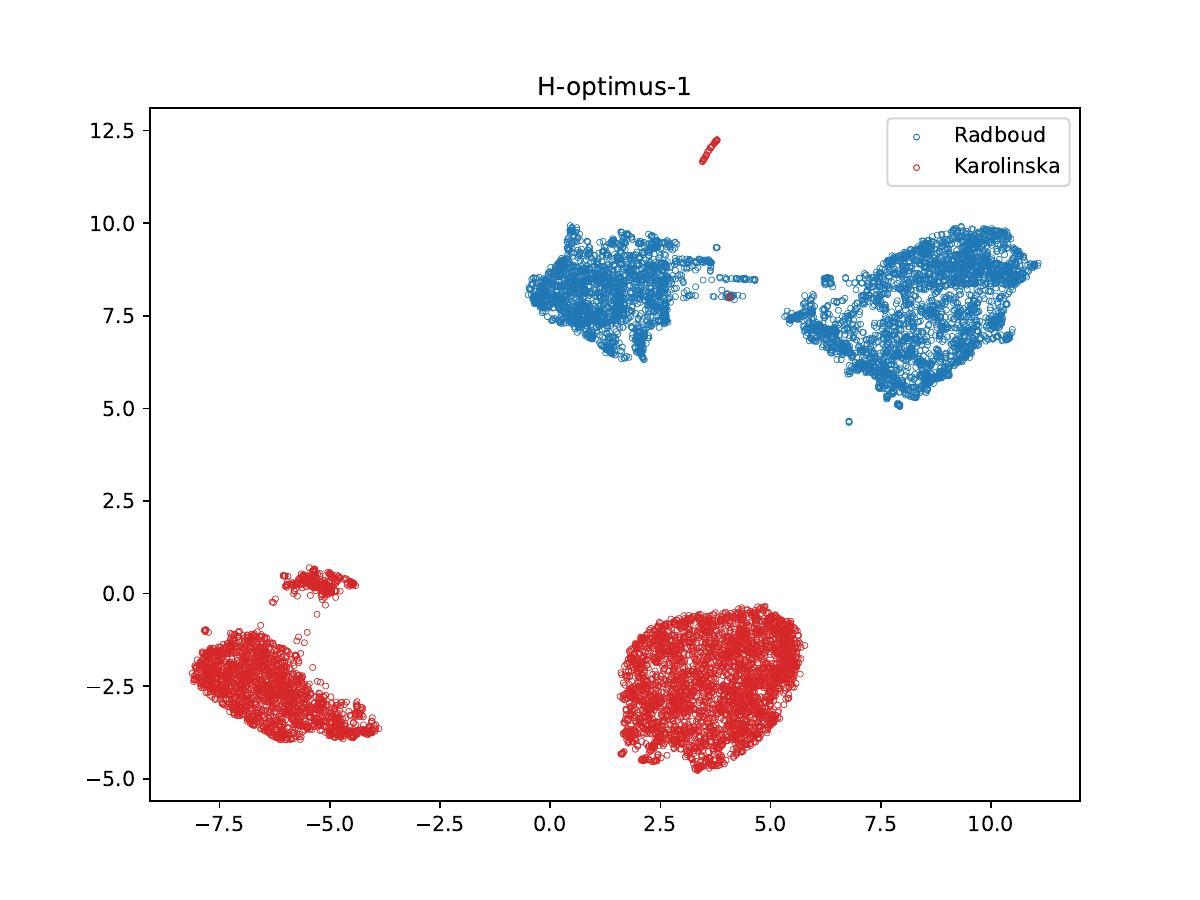}
    \end{subfigure}
    
    \begin{subfigure}[t]{0.245\textwidth}
        \centering%
        \includegraphics[clip, trim=1.5cm 0.95cm 1.85cm 0.95cm, width=1.0\linewidth]{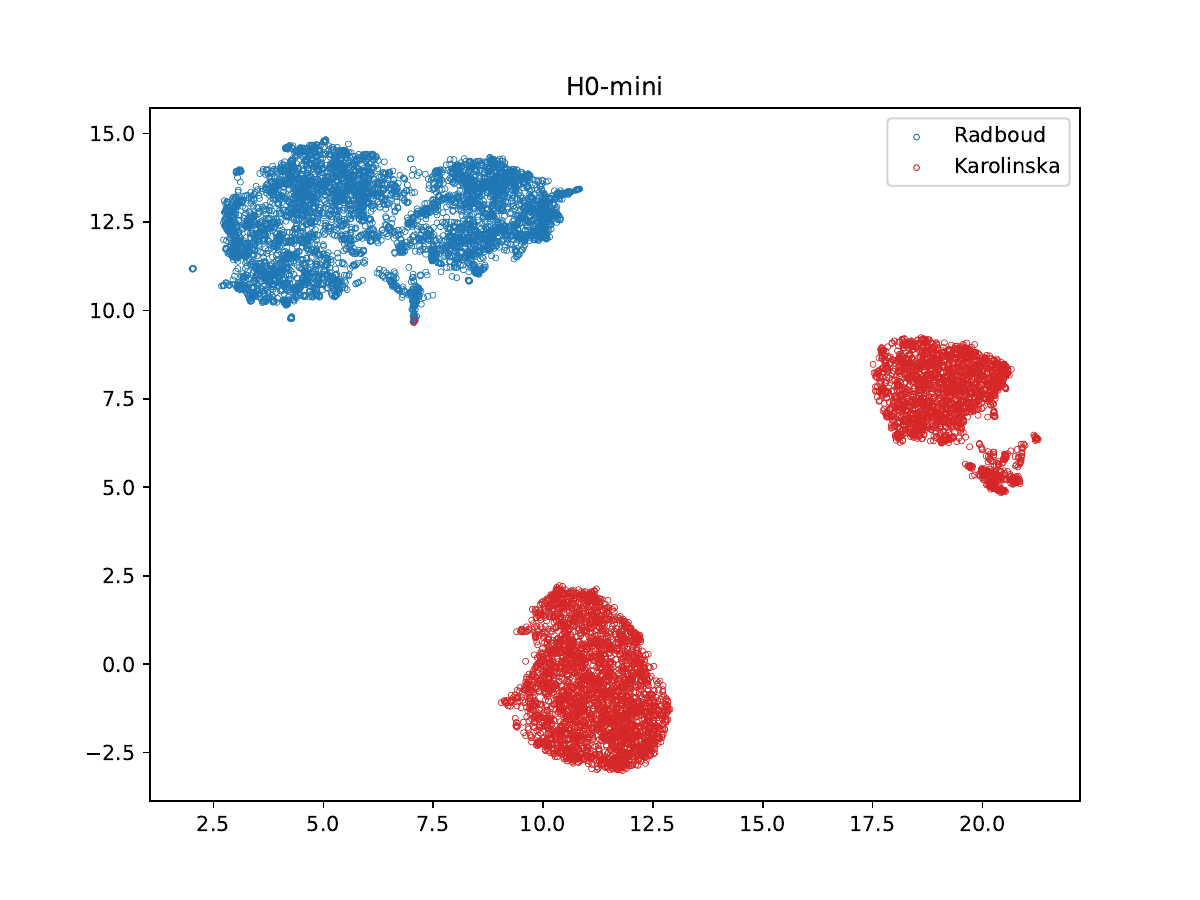}
    \end{subfigure}
    \begin{subfigure}[t]{0.245\textwidth}
        \centering%
        \includegraphics[clip, trim=1.5cm 0.95cm 1.85cm 0.95cm, width=1.0\linewidth]{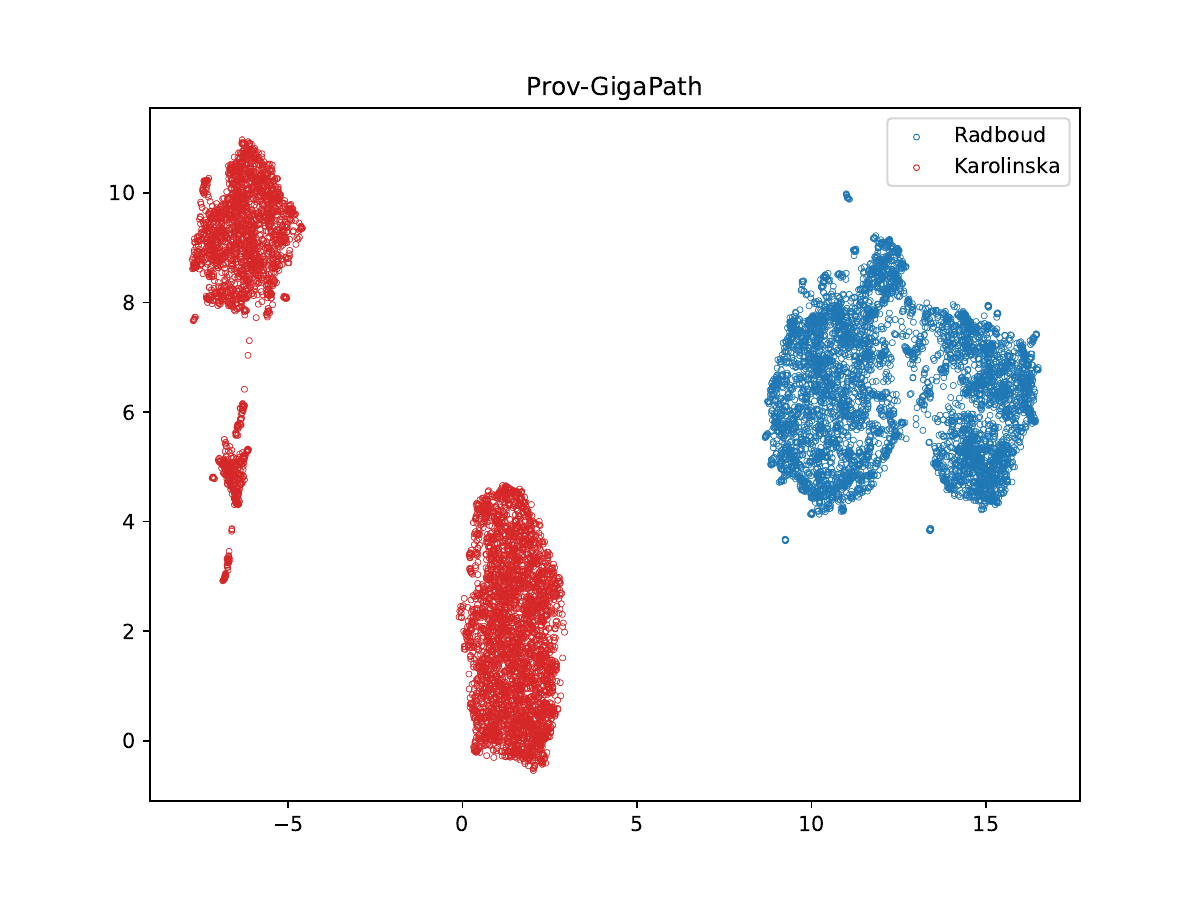}
    \end{subfigure}
    \begin{subfigure}[t]{0.245\textwidth}
        \centering%
        \includegraphics[clip, trim=1.5cm 0.95cm 1.85cm 0.95cm, width=1.0\linewidth]{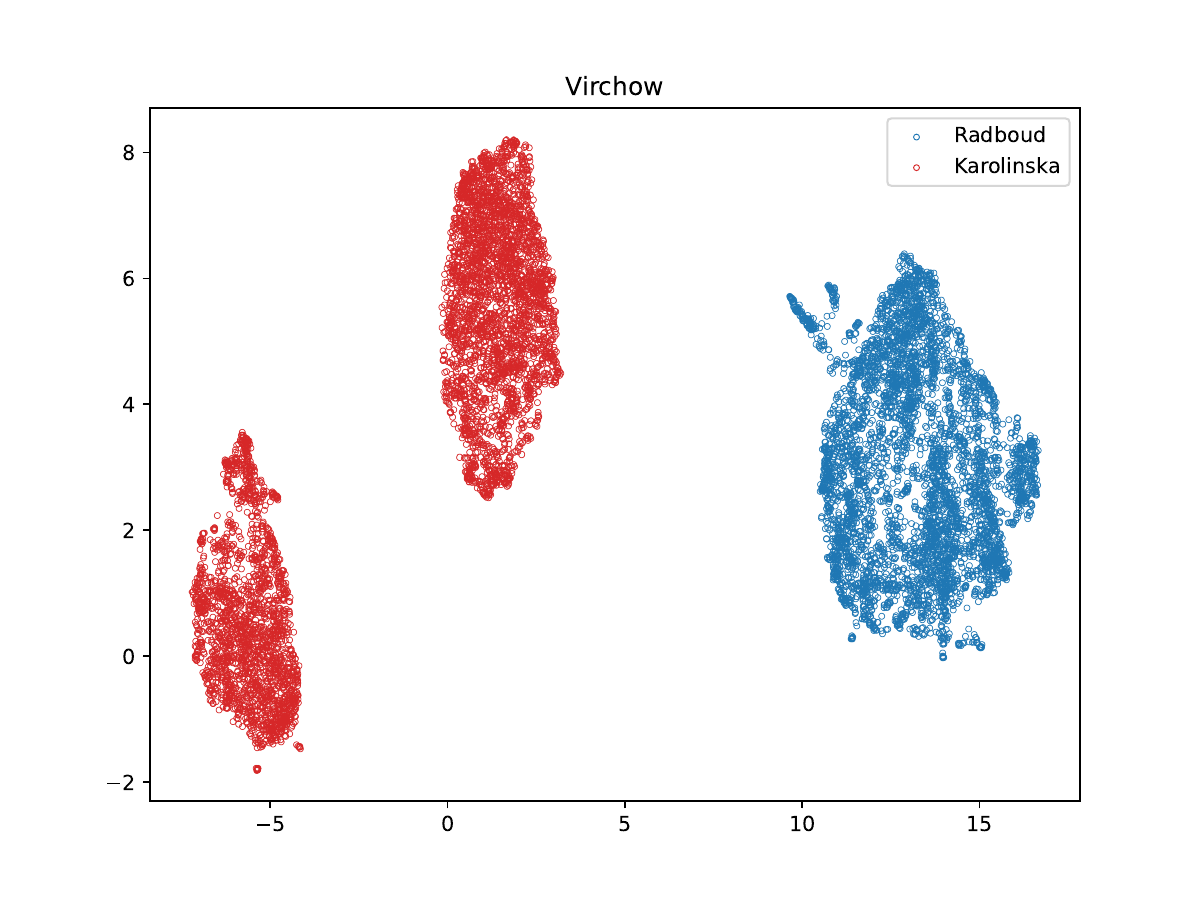}
    \end{subfigure}
    \begin{subfigure}[t]{0.245\textwidth}
        \centering%
        \includegraphics[clip, trim=1.5cm 0.95cm 1.85cm 0.95cm, width=1.0\linewidth]{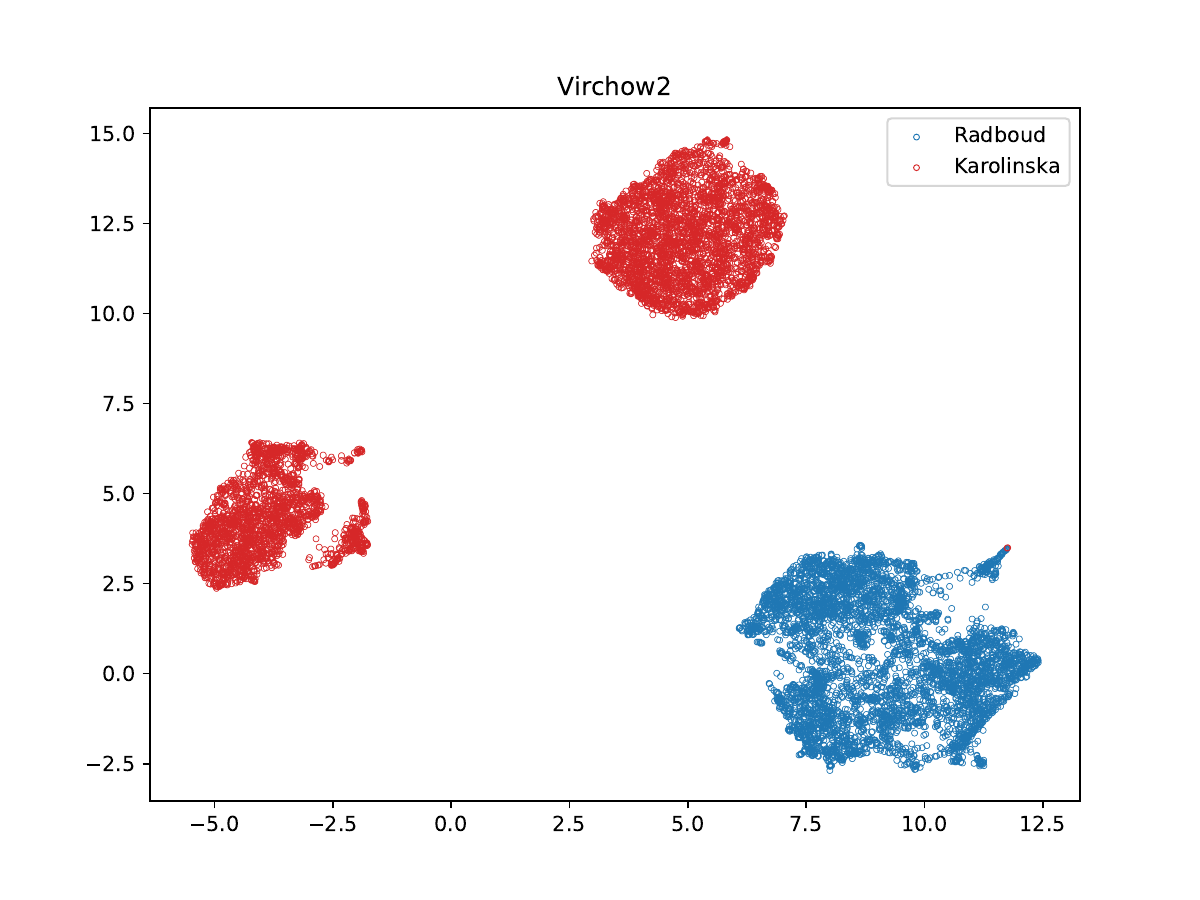}
    \end{subfigure}
    \caption{\textbf{UMAP visualizations of WSI-level feature representations across sites.} UMAP projections of the mean patch-level feature vectors for all twelve evaluated models on the full PANDA dataset, with one point per WSI. Points are colored by data collection site: \textcolor{Cerulean}{\textbf{blue}} for Radboud and \textcolor{OrangeRed2}{\textbf{red}} for Karolinska. Across all models, WSIs from Radboud and Karolinska form clearly separated clusters. In addition, the Karolinska WSIs are further divided into two sub-clusters, likely reflecting scanner-specific variation within that site.}
  \label{fig:umap_plots_radboud_vs_karolinska}
\end{figure*}

\begin{figure*}[h]
\centering
    \begin{subfigure}[t]{0.325\textwidth}
        \centering%
        \includegraphics[clip, trim=1.5cm 0.95cm 1.85cm 0.95cm, width=0.925\linewidth]{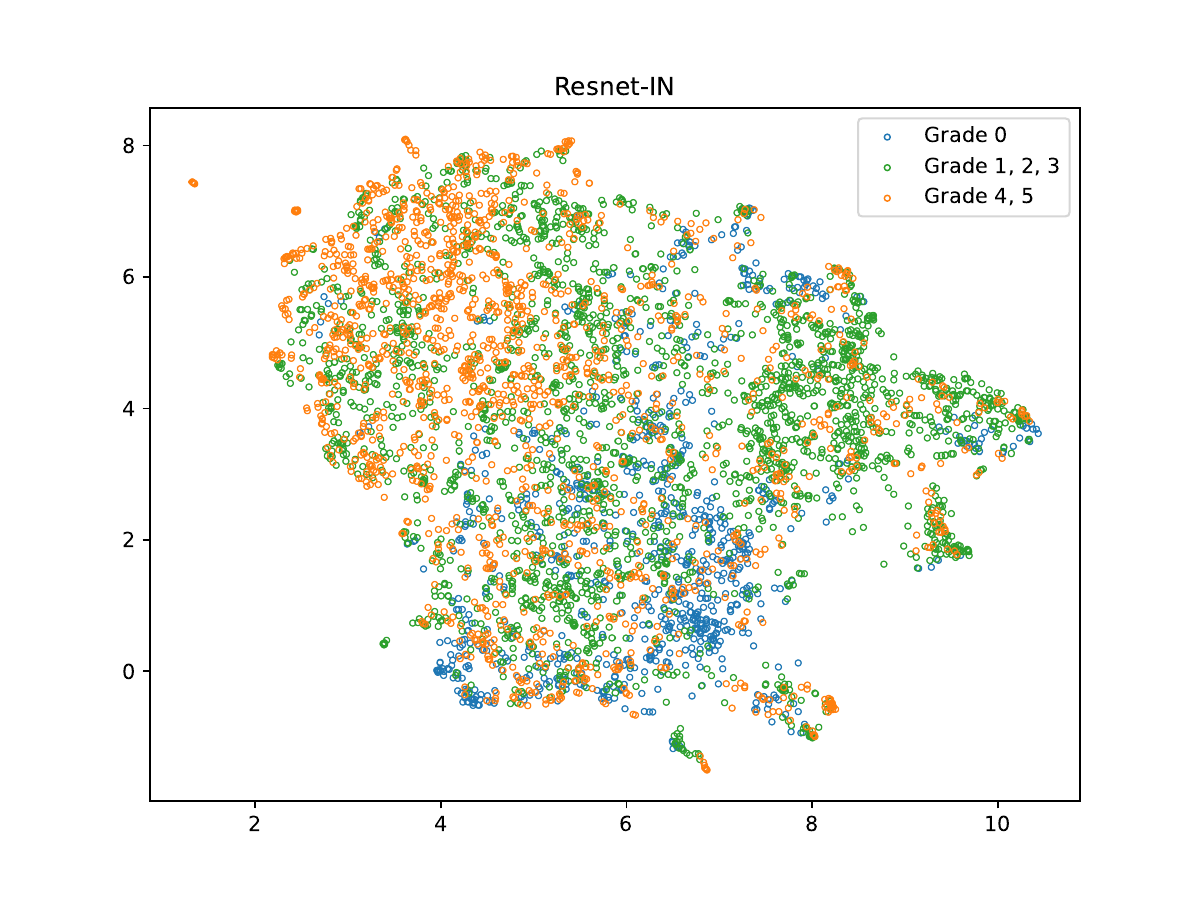}
    \end{subfigure}
    \begin{subfigure}[t]{0.325\textwidth}
        \centering%
        \includegraphics[clip, trim=1.5cm 0.95cm 1.85cm 0.95cm, width=0.925\linewidth]{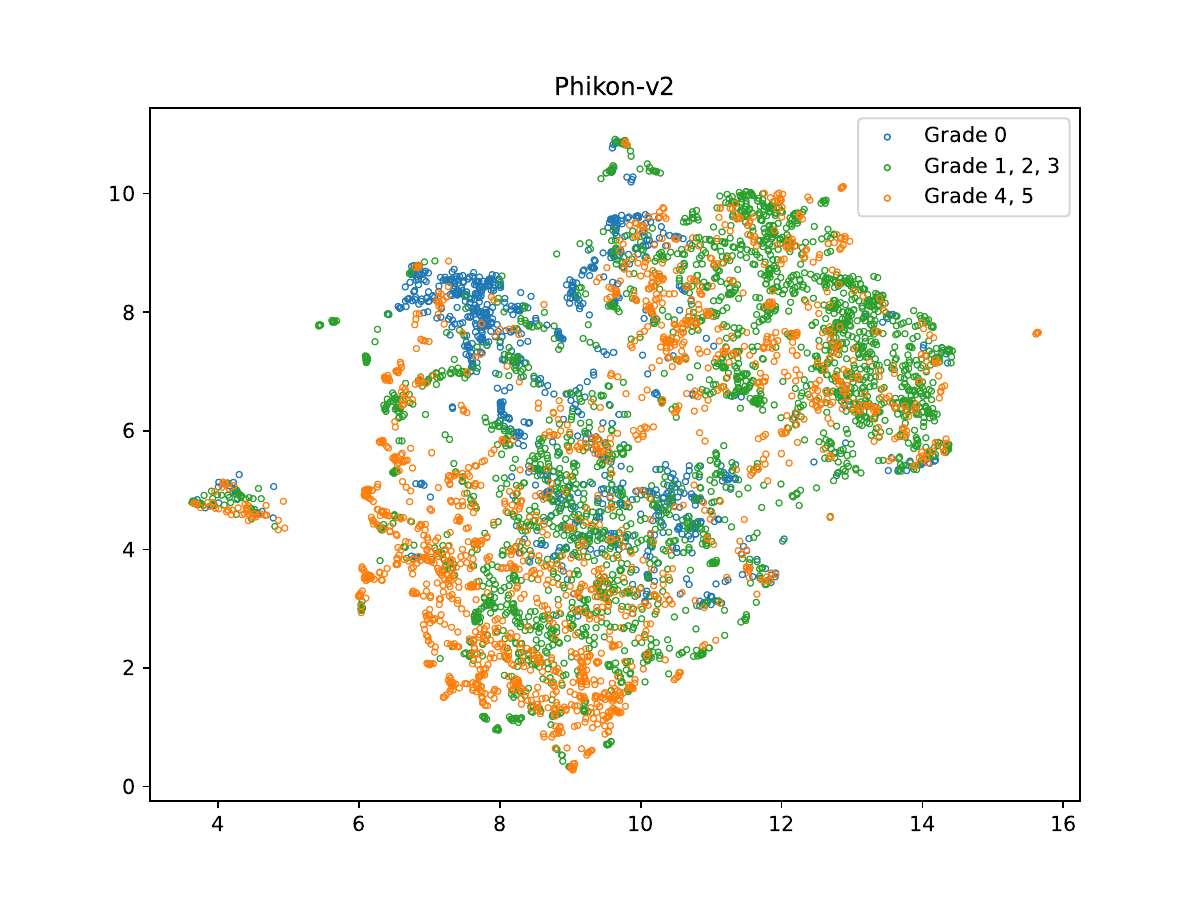}
    \end{subfigure}
    \begin{subfigure}[t]{0.325\textwidth}
        \centering%
        \includegraphics[clip, trim=1.5cm 0.95cm 1.85cm 0.95cm, width=0.925\linewidth]{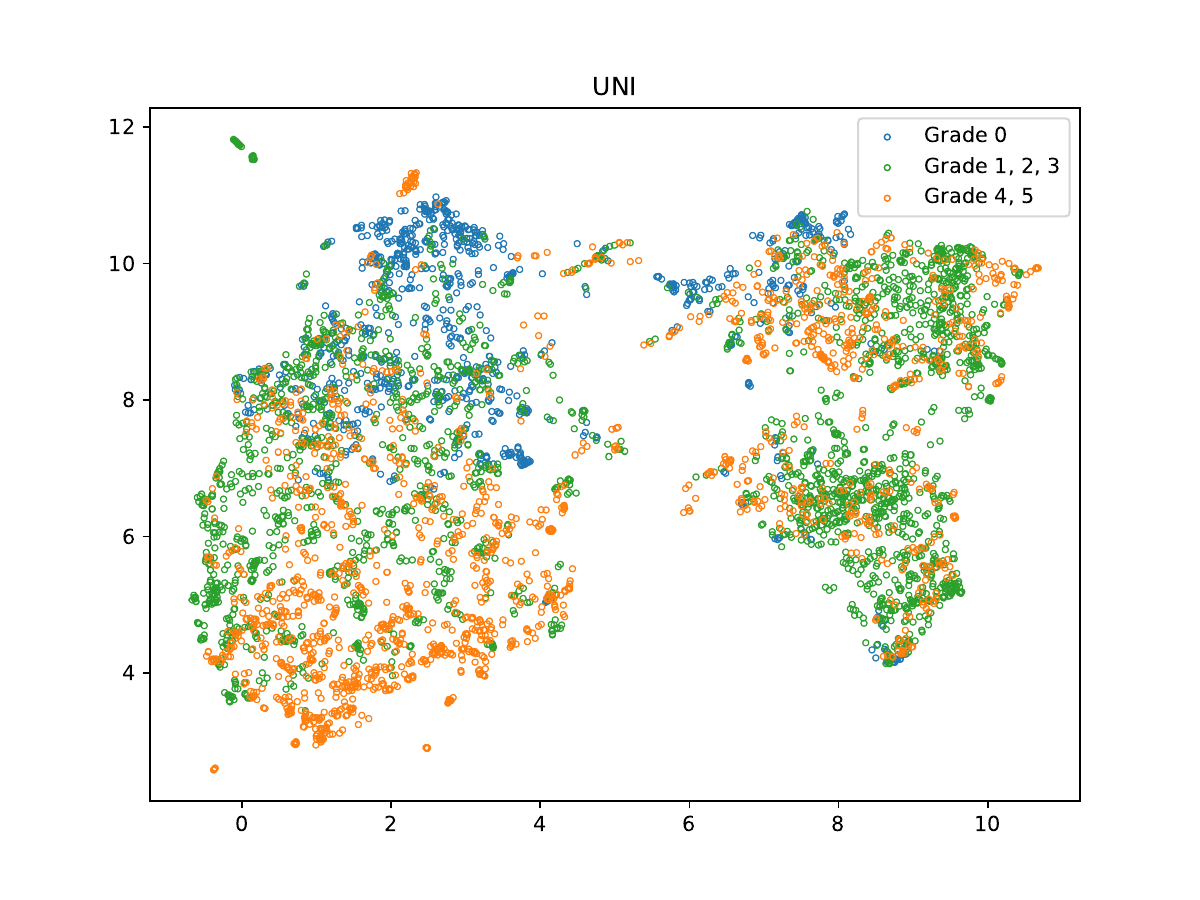}
    \end{subfigure}\vspace{-1.0mm}
    
    \begin{subfigure}[t]{0.325\textwidth}
        \centering%
        \includegraphics[clip, trim=1.5cm 0.95cm 1.85cm 0.95cm, width=0.925\linewidth]{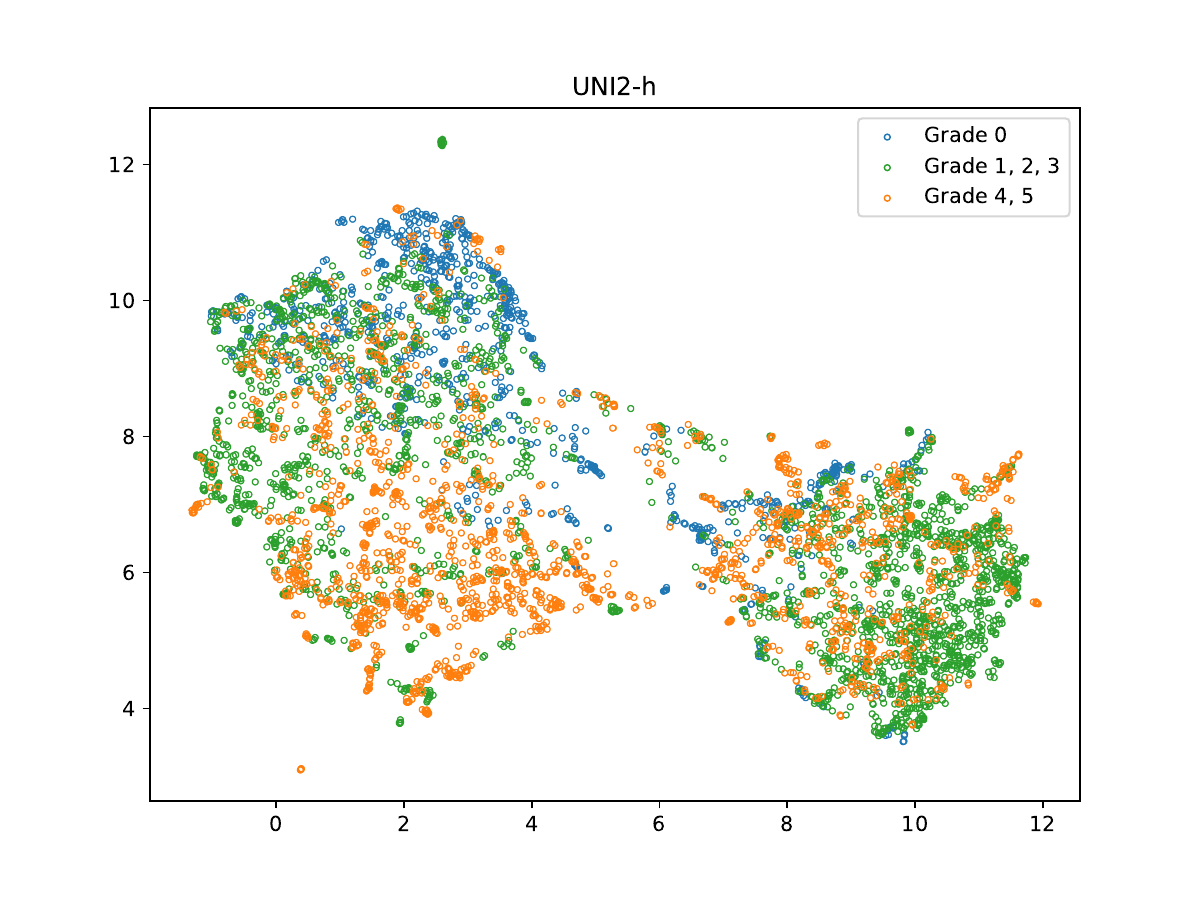}
    \end{subfigure}
    \begin{subfigure}[t]{0.325\textwidth}
        \centering%
        \includegraphics[clip, trim=1.5cm 0.95cm 1.85cm 0.95cm, width=0.925\linewidth]{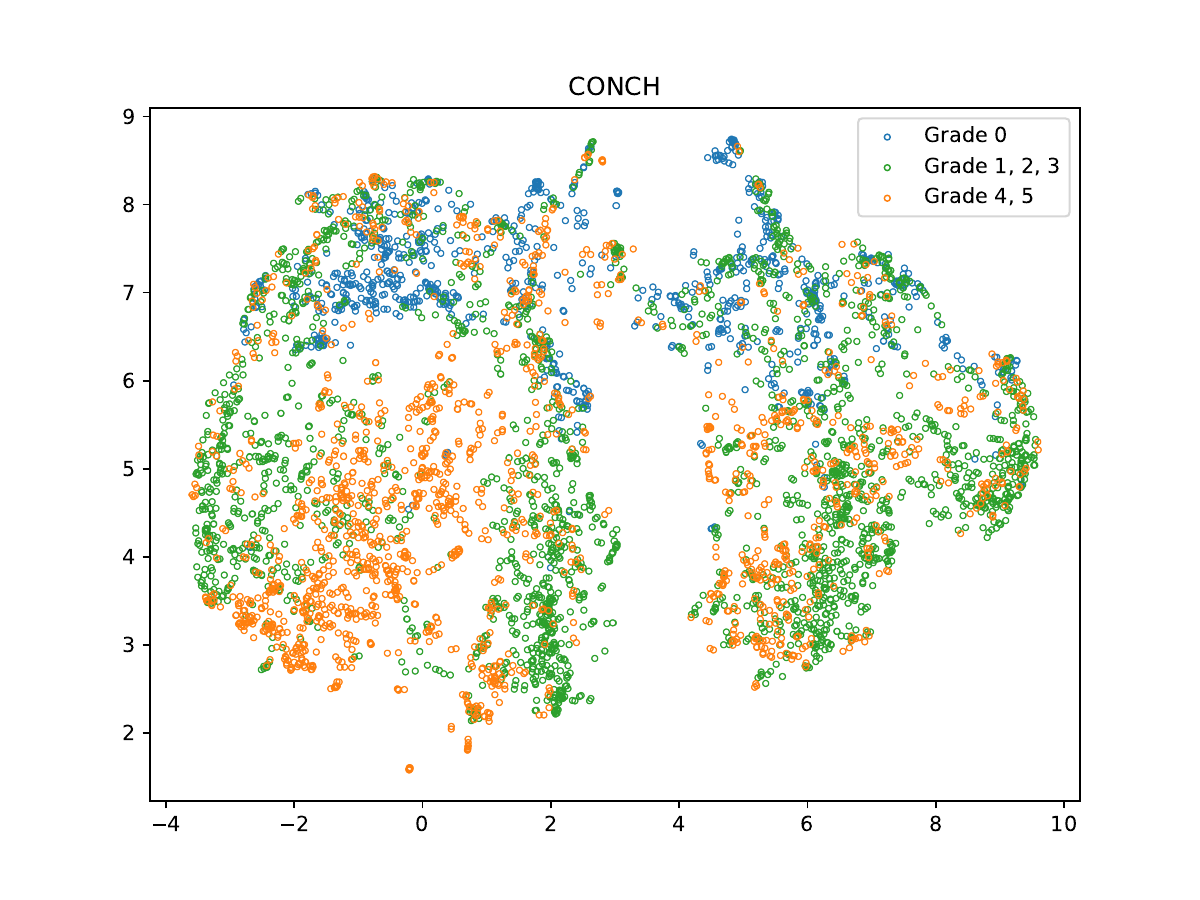}
    \end{subfigure}
    \begin{subfigure}[t]{0.325\textwidth}
        \centering%
        \includegraphics[clip, trim=1.5cm 0.95cm 1.85cm 0.95cm, width=0.925\linewidth]{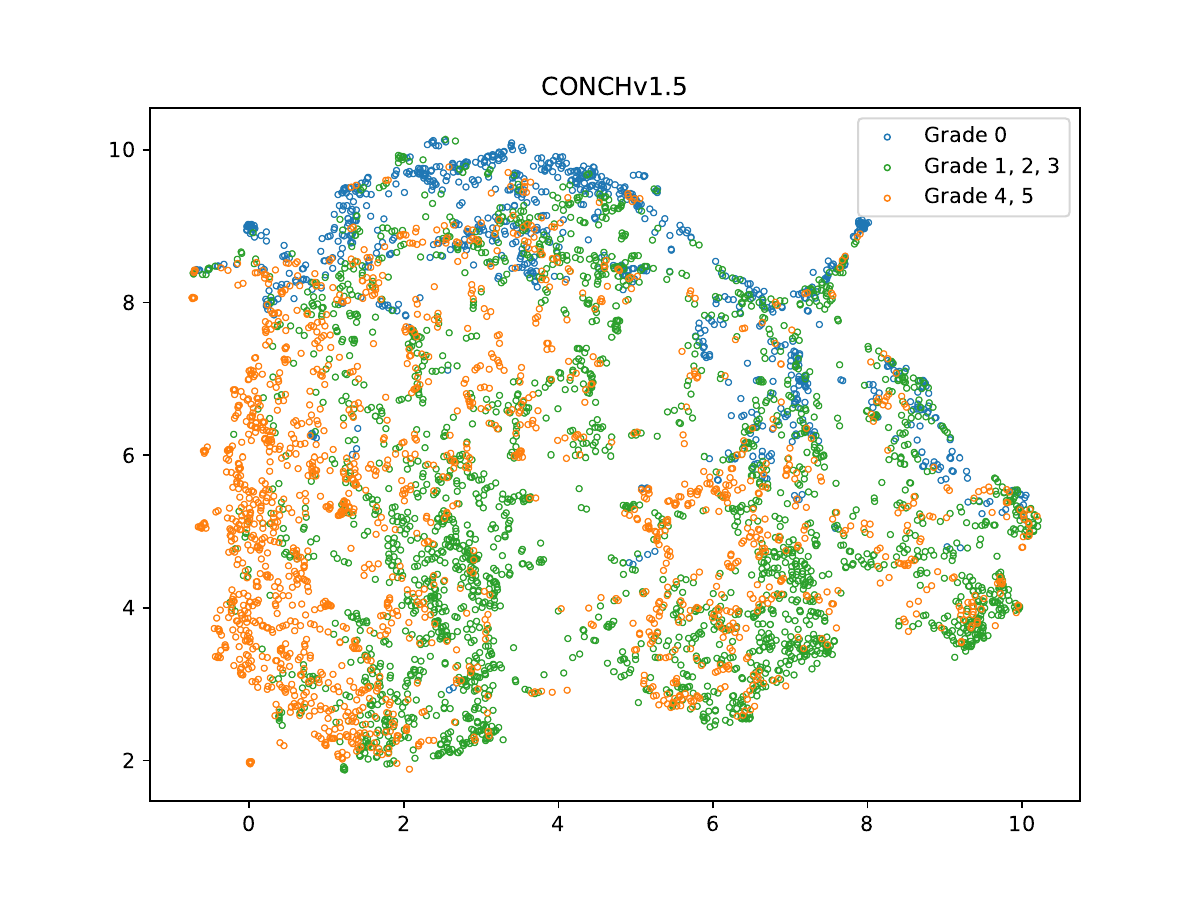}
    \end{subfigure}\vspace{-1.0mm}

    \begin{subfigure}[t]{0.325\textwidth}
        \centering%
        \includegraphics[clip, trim=1.5cm 0.95cm 1.85cm 0.95cm, width=0.925\linewidth]{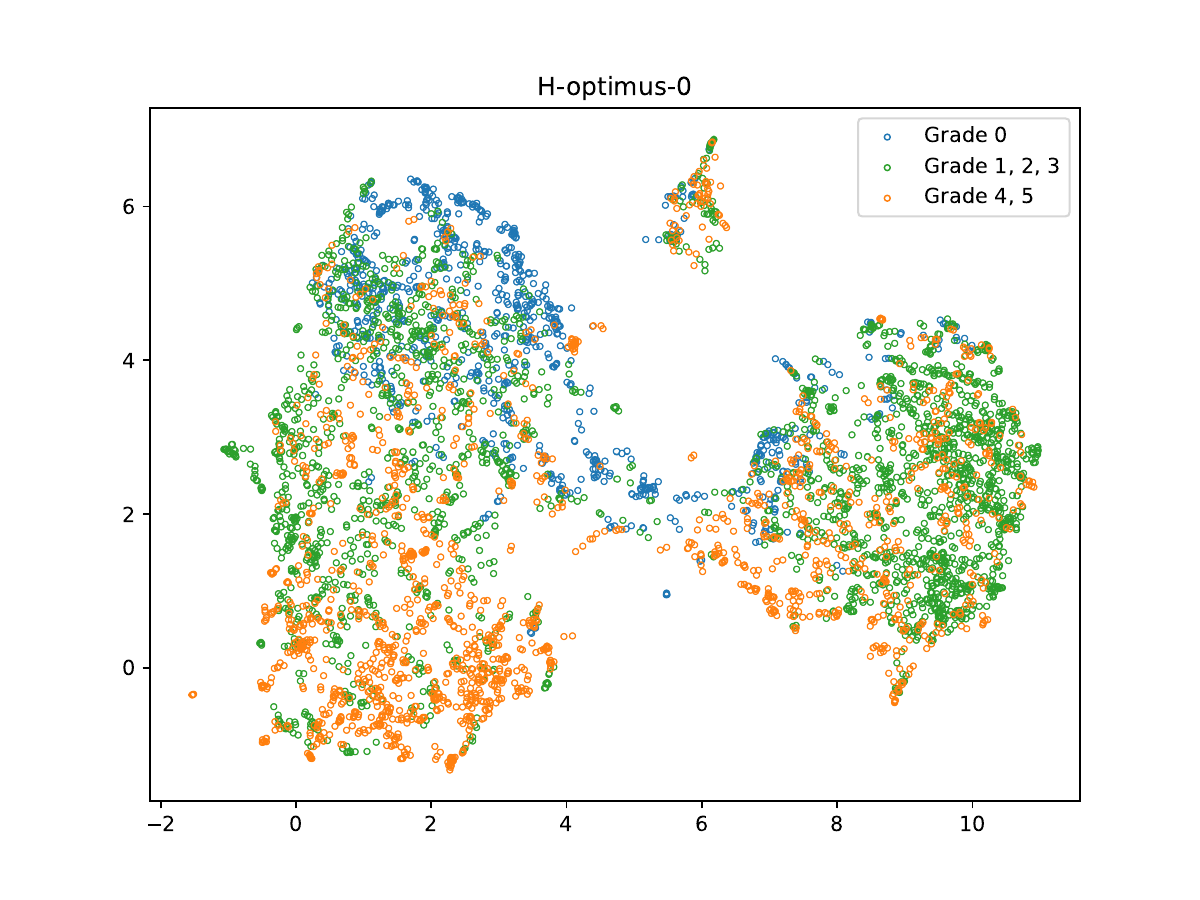}
    \end{subfigure}
    \begin{subfigure}[t]{0.325\textwidth}
        \centering%
        \includegraphics[clip, trim=1.5cm 0.95cm 1.85cm 0.95cm, width=0.925\linewidth]{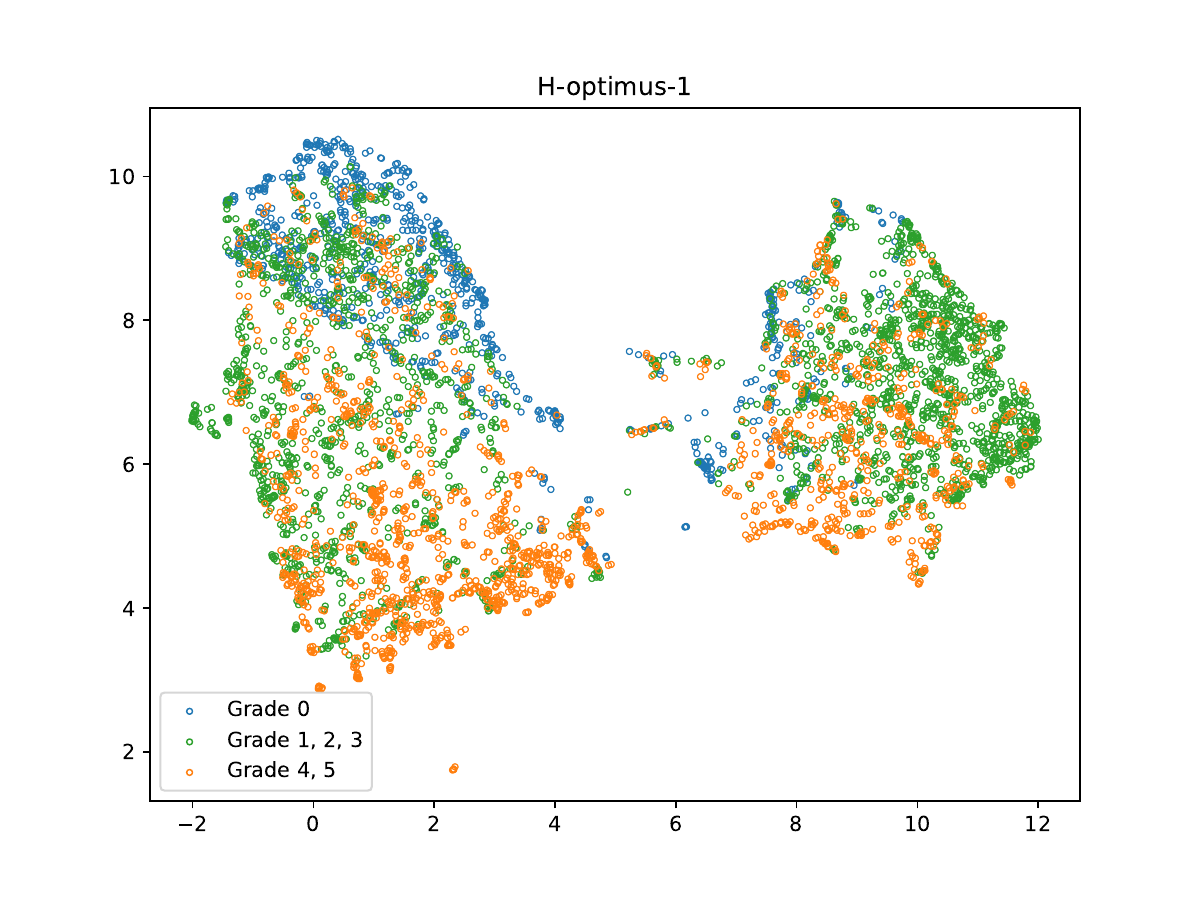}
    \end{subfigure}
    \begin{subfigure}[t]{0.325\textwidth}
        \centering%
        \includegraphics[clip, trim=1.5cm 0.95cm 1.85cm 0.95cm, width=0.925\linewidth]{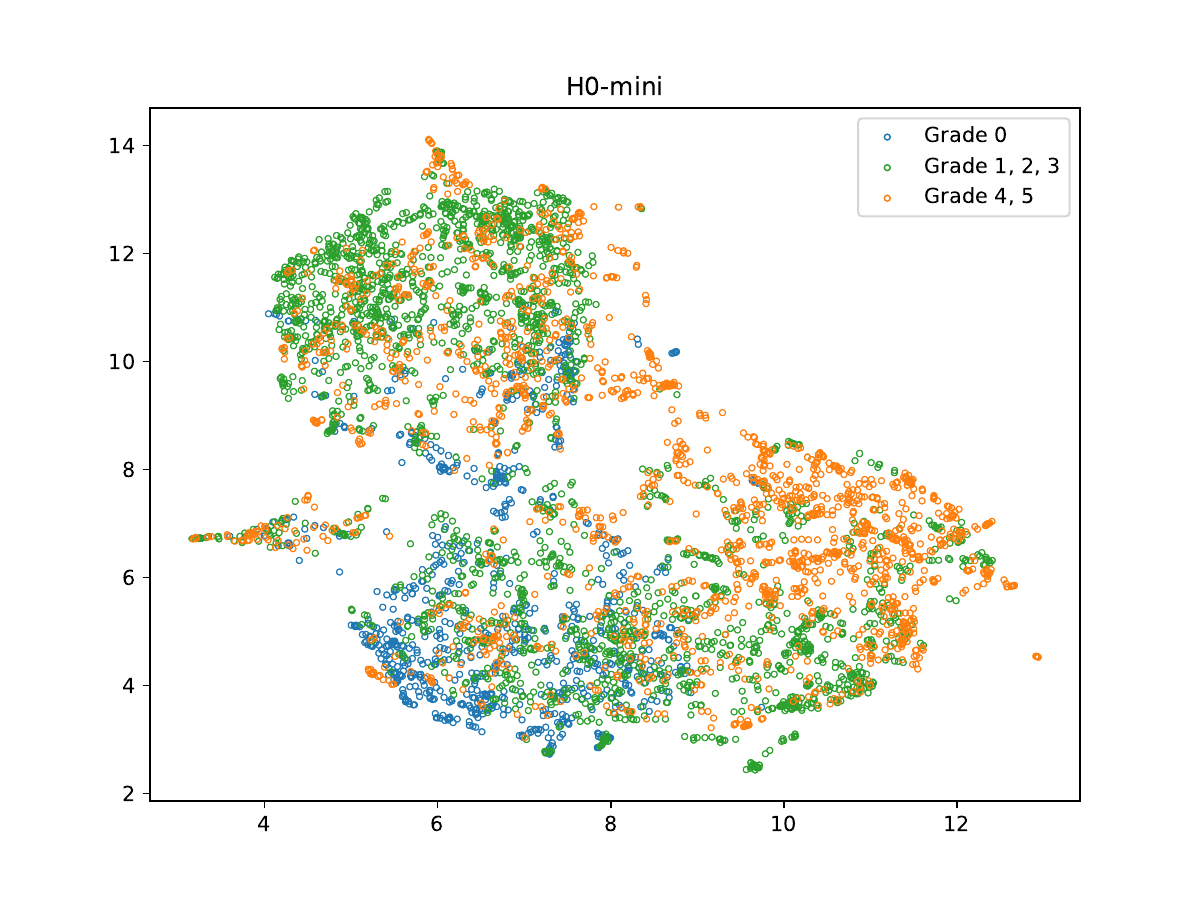}
    \end{subfigure}\vspace{-1.0mm}
    
    \begin{subfigure}[t]{0.325\textwidth}
        \centering%
        \includegraphics[clip, trim=1.5cm 0.95cm 1.85cm 0.95cm, width=0.925\linewidth]{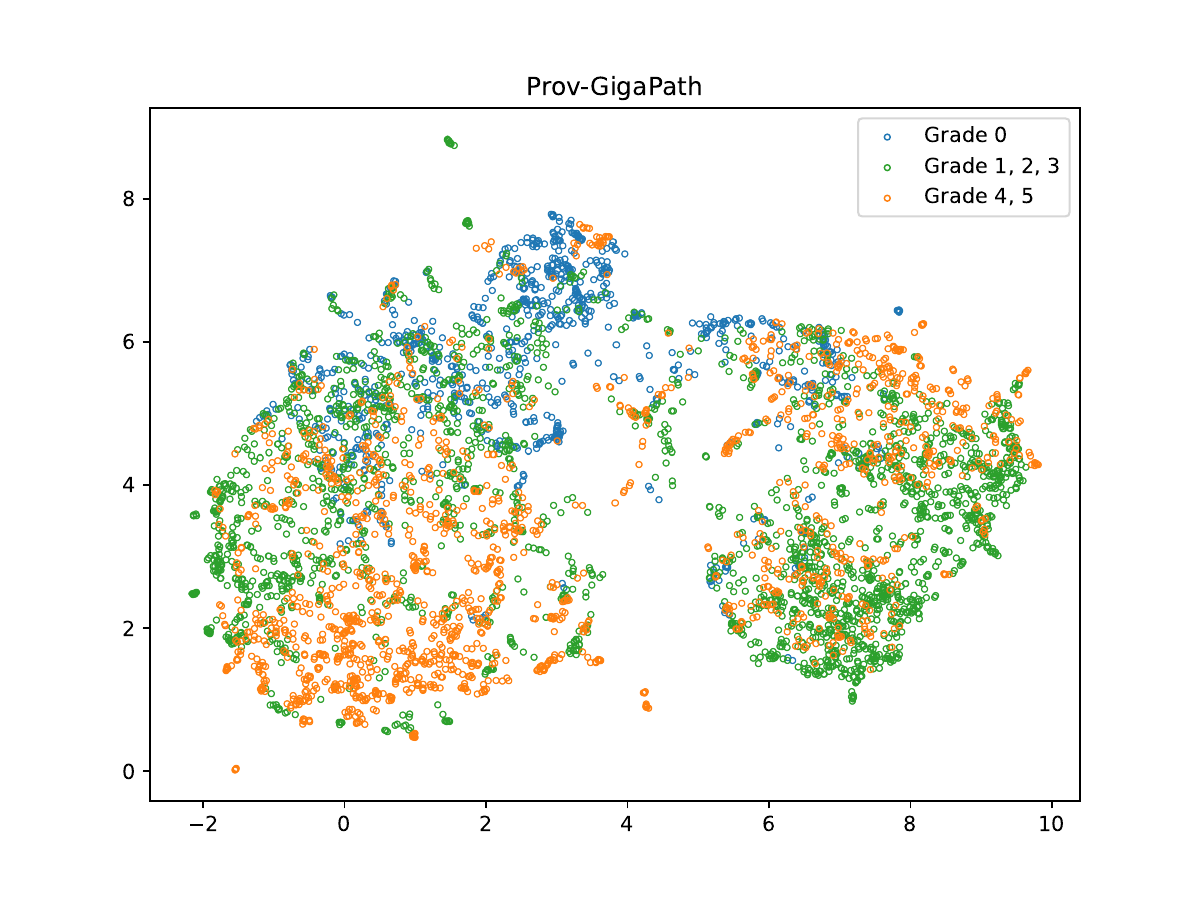}
    \end{subfigure}
    \begin{subfigure}[t]{0.325\textwidth}
        \centering%
        \includegraphics[clip, trim=1.5cm 0.95cm 1.85cm 0.95cm, width=0.925\linewidth]{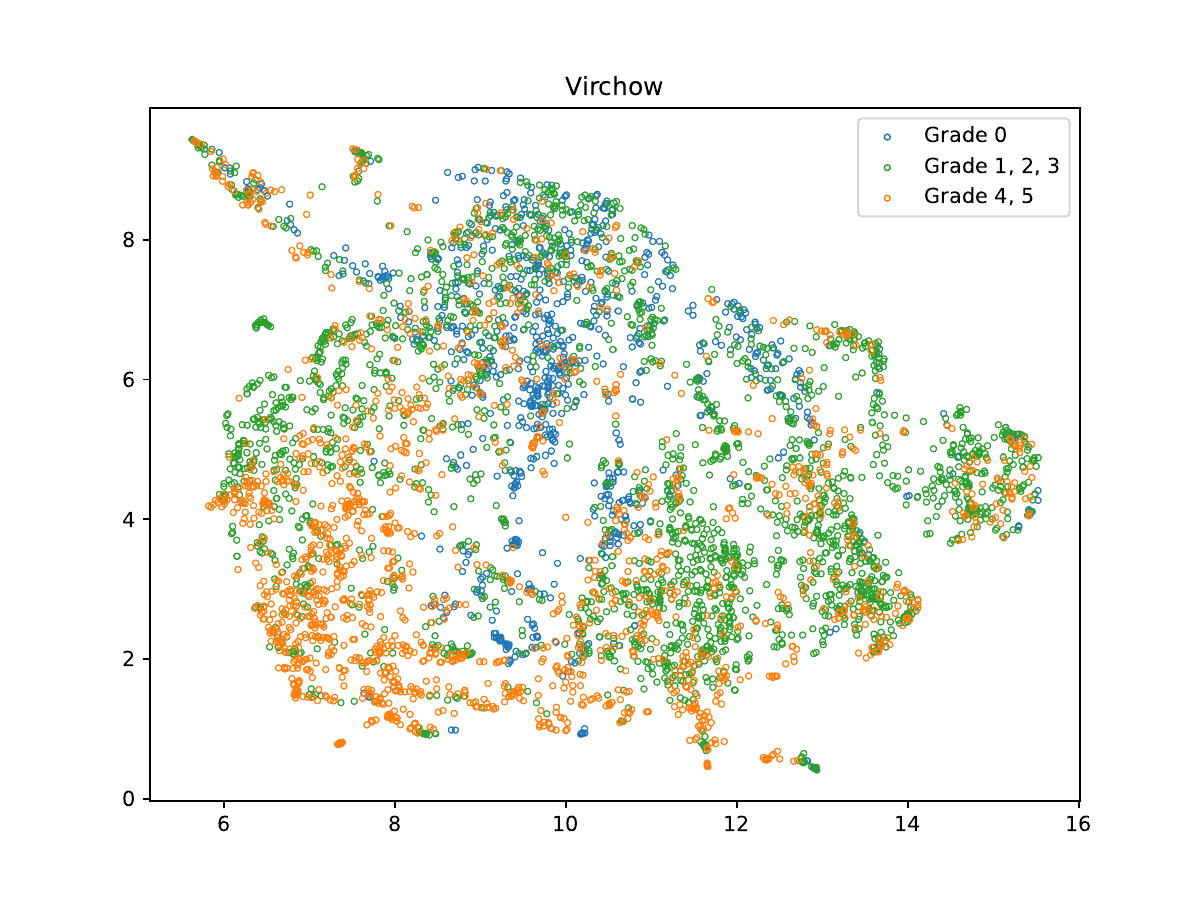}
    \end{subfigure}
    \begin{subfigure}[t]{0.325\textwidth}
        \centering%
        \includegraphics[clip, trim=1.5cm 0.95cm 1.85cm 0.95cm, width=0.925\linewidth]{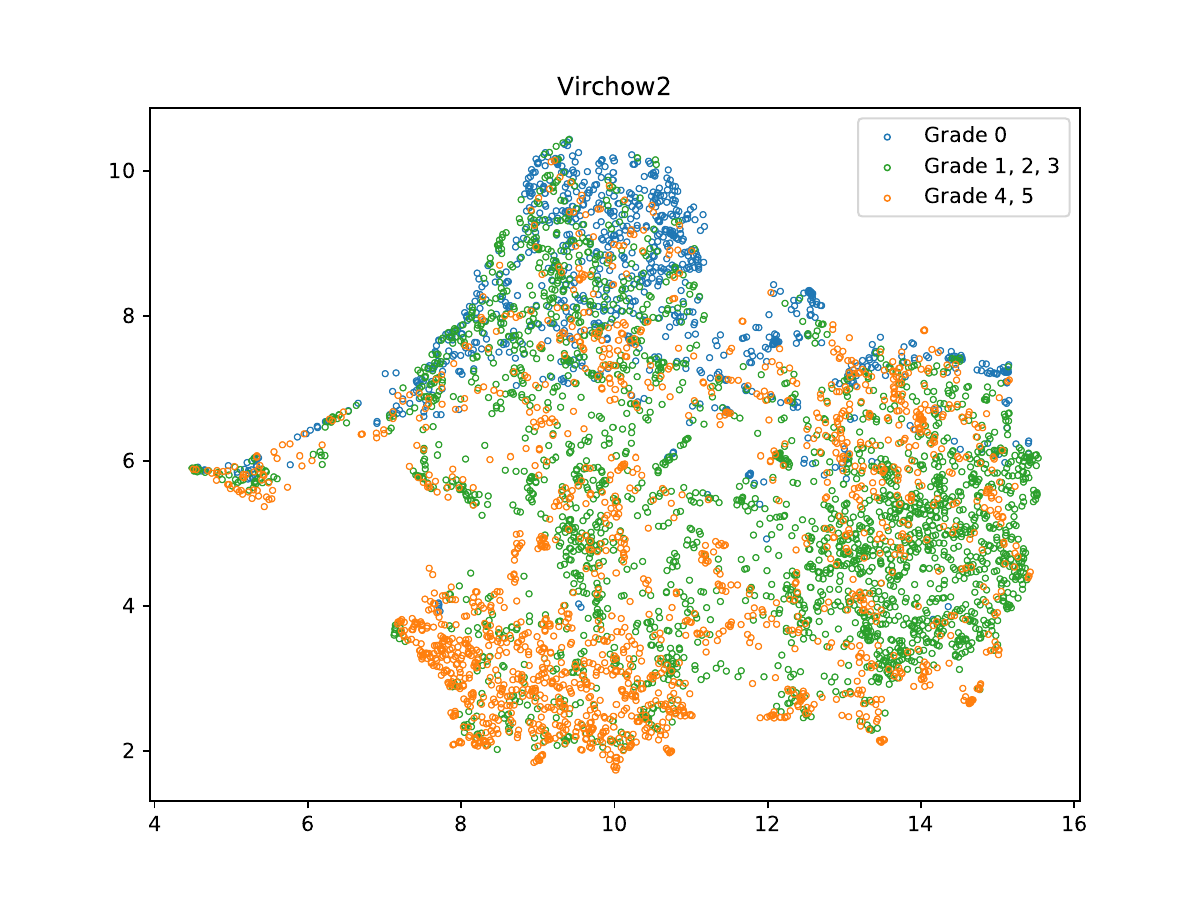}
    \end{subfigure}\vspace{-1.0mm}
    \caption{\textbf{UMAP visualizations of WSI-level feature representations within a single site.} UMAP projections of the mean patch-level feature vectors for all twelve evaluated models, restricted to Radboud WSIs. Points are colored by coarse ISUP grade groups: \textcolor{Cerulean}{\textbf{blue}} for grade $0$ (non-tumor), \textcolor{ForestGreen}{\textbf{green}} for grade $1$--$3$, and \textcolor{goldenpoppy3}{\textbf{orange}} for grade $4$--$5$. Across all models, partial clustering by grade is observable, but the separation is relatively weak and exhibits substantial overlap, especially compared to the strong site-driven clustering observed in Figure~\ref{fig:umap_plots_radboud_vs_karolinska}.}  
    \label{fig:umap_plots_coarse_isup_grades}
\end{figure*}

\section*{Results}

We evaluate eight recent vision-only PFMs (Phikon-v2~\citep{filiot2024phikonv2}, Prov-GigaPath~\citep{gigapath2024}, UNI~\citep{chen2024uni}, UNI2-h~\citep{UNI2h2024}, Virchow~\citep{virchow2024}, Virchow2~\citep{virchow22024}, H-optimus-0~\citep{hoptimus0}, H-optimus-1~\citep{hoptimus1}), two vision-language PFMs (CONCH~\citep{conch2024}, CONCHv1.5~\citep{ding2025multimodal}), and a compact distilled PFM (H0-mini~\citep{filiot2025h0mini}, distilled from H-optimus-0), alongside a natural-image baseline (Resnet-IN).

For this evaluation, we utilize histological grading of prostate cancer biopsy WSIs~\cite{strom2020artificial, bulten2022artificial} as our specific practical application. Histological grading is used to provide important prognostic information for patients, categorizing biopsies into five different International Society of Urological Pathology (ISUP) grade groups~\cite{epstein2010update, epstein2016contemporary}. With non-tumor biopsies denoted as grade $0$, each biopsy WSI is thus assigned an ISUP grade $0$--$5$.

Following recent benchmarking studies~\cite{wolflein2024benchmarking, campanella2025clinical, neidlinger2025benchmarking}, PFMs are evaluated by utilizing them as \emph{frozen} patch-level feature extractors in weakly supervised WSI-level prediction models. Specifically, we utilize the feature extractors in three different ISUP grade classification models (see \textit{Methods} for details): \textit{ABMIL} uses attention-based multiple instance learning (ABMIL)~\cite{ilse2018attention, laleh2022benchmarking} to process the patch-level feature vectors and output a predicted ISUP grade $\hat{y}(x) \in \{0, \dots, 5\}$ for each biopsy WSI $x$. \textit{Mean Feature} removes the trainable ABMIL aggregator and directly feeds the mean patch-level feature vector as input to a small classification network. Lastly, \textit{kNN} removes also this trainable classification network, instead utilizing the k-nearest neighbors algorithm. \textit{kNN} therefore contains no trainable components, enabling a direct evaluation of the underlying PFM. 

\begin{table}[t]
	\caption{\textbf{Overview of all evaluated models}, listing their architectures, number of model parameters, feature dimensions, and the scale of their pretraining data. Evaluated models include a natural-image baseline (Resnet-IN), eight vision-only PFMs, a compact distilled PFM (H0-mini), and two vision-language PFMs.}\vspace{-2.25mm}	
    \label{table:fms}
    \centering
	\resizebox{1.0\linewidth}{!}{%
        \begin{tabular}{lcccc}
\toprule
Model Name &Architecture &Size &\makecell{Feature\\ Dimension} &Pretraining Data\\
\midrule
Resnet-IN~\citep{he2016deep}          &\small{ResNet-50}     &25M      &1024 &1.3M natural images\\
\midrule
Phikon-v2~\citep{filiot2024phikonv2}  &ViT-L         &307M     &1024 &58K WSIs\\
UNI~\citep{chen2024uni}               &ViT-L         &307M     &1024 &100K WSIs\\
UNI2-h~\citep{UNI2h2024}              &ViT-H         &682M     &1536 &350K WSIs\\
H-optimus-0~\citep{hoptimus0}         &ViT-G         &1.1B &1536 &500K WSIs\\
H-optimus-1~\citep{hoptimus1}         &ViT-G         &1.1B &1536 &1M WSIs\\
Prov-GigaPath~\citep{gigapath2024}    &ViT-G         &1.1B &1536 &170K WSIs\\
Virchow~\citep{virchow2024}           &ViT-H         &632M     &2560 &1.5M WSIs\\
Virchow2~\citep{virchow22024}         &ViT-H         &632M     &2560 &3.1M WSIs\\
\midrule
H0-mini~\citep{filiot2025h0mini}      &ViT-B         &86M      &768  &500K + 6K WSIs\\
\midrule
CONCH~\citep{conch2024}               &ViT-B         &86M      &512  &\small{21K WSIs + 1.1M image-text pairs}\\
CONCHv1.5~\citep{ding2025multimodal}  &ViT-L         &307M     &768  &N/A\\
\bottomrule
\end{tabular}

	}
\end{table}

Experiments are conducted on the PANDA dataset~\cite{bulten2022artificial}, containing $10{,}616$ prostate biopsy WSIs with corresponding ISUP grade labels. The data was collected from two different sites: Radboud University Medical Center (Radboud) in the Netherlands, and Karolinska Institutet (Karolinska) in Sweden. The two sites differ in terms of both pathology lab procedures and utilized scanners, creating a clear distribution shift for the WSI image data (Figure~\ref{fig:shifts_covariate_radboud_karolinska}). By training ISUP grade models exclusively on Radboud data and then evaluating on Karolinska data, we are thus able to study robustness to a commonly encountered cross-site image shift. By creating further PANDA subsets, we also evaluate robustness to shifts in the label distribution over ISUP grades (Figure~\ref{fig:shifts_label_radboud_karolinska}~\&~\ref{fig:shifts_label_radboud-l_radboud-r}). Performance is measured using quadratically weighted Cohen's kappa as the main metric.

\subsubsection*{Main Model Comparison}
Figure~\ref{fig:main_results3a_12fms_kappa}~\&~\ref{fig:main_results3b_12fms_kappa} show the performance of all evaluated PFMs across different PANDA subsets, when utilized as patch-level feature extractors in the three ISUP grade models.

First, we observe that when models are trained and evaluated on the full PANDA dataset (first column of Figure~\ref{fig:main_results3a_12fms_kappa}), all PFMs perform well (e.g., $0.888\pm0.013$ kappa for UNI with \textit{ABMIL}) and clearly outperform Resnet-IN. Similar results are achieved also when ISUP grade models are both trained and evaluated exclusively on data from either Karolinska or Radboud (second and third column of Figure~\ref{fig:main_results3a_12fms_kappa}, respectively). However, when models are trained on Radboud data and evaluated on Karolinska data (\textit{Radboud $\rightarrow$ Karolinska} in Figure~\ref{fig:main_results3a_12fms_kappa}), the performance drops drastically (e.g., $0.247\pm0.138$ kappa for UNI with \textit{ABMIL}). This clear performance drop is observed for all three ISUP grade models, i.e.\ even when applying kNN directly on top of the PFM patch-level features.

When repeating this experiment on subsets with perfectly uniform ISUP grade label distributions (\textit{Radboud-U} and \textit{Radboud-U $\rightarrow$ Karolinska-U} in Figure~\ref{fig:main_results3b_12fms_kappa}), the performance drop from Radboud to Karolinska data is slightly reduced but remains substantial. When models instead are trained and evaluated on Radboud data with left- or right-skewed label distributions (\textit{Radboud-L} and \textit{Radboud-L $\rightarrow$ Radboud-R} in Figure~\ref{fig:main_results3b_12fms_kappa}), there is a consistent but relatively small drop in performance, at least for the \textit{ABMIL} and \textit{Mean Feature} ISUP grade models. This indicates that visual domain shift, rather than label imbalance, is the dominant challenge.

When comparing the second-generation PFMs (UNI2-h, CONCHv1.5, H-optimus-1, Virchow2) with their corresponding first-generation counterparts (UNI, CONCH, H-optimus-0, Virchow), the performance is highly similar across all in-distribution settings. For the two distribution shift settings with clear performance drops (\textit{Radboud $\rightarrow$ Karolinska}, \textit{Radboud-U $\rightarrow$ Karolinska-U}), improvements are modest and inconsistent. While CONCHv1.5 outperforms CONCH in all evaluated cases, UNI2-h, H-optimus-1, and Virchow2 show only partial and setting-dependent improvements over their predecessors. Similarly, the compact distilled model H0-mini performs comparably to its substantially larger teacher model H-optimus-0, but does not show a clear advantage in the distribution shift settings.

Figure~\ref{fig:main_results_attmil_mean-pool_knn_kappa}~\&~~\ref{fig:main_results_attmil_mean-pool_knn_kappa_remaining6} contain the same results as Figure~\ref{fig:main_results3a_12fms_kappa}~\&~\ref{fig:main_results3b_12fms_kappa}, but presented in order to enable a direct performance comparison of the three ISUP grade models. We observe that \textit{ABMIL} achieves the best performance in most cases across the different PANDA subsets and PFMs, followed by \textit{Mean Feature} and \textit{kNN}. The ranking of the three models is more varied for \textit{Radboud $\rightarrow$ Karolinska} and \textit{Radboud-U $\rightarrow$ Karolinska-U}, but for these cases relatively poor performance is achieved by all three models.

\subsubsection*{Extended Model Analysis}
To better understand the observed robustness limitations, we perform additional analyses of model behavior. First, a more detailed performance comparison of three representative models (UNI, CONCH, Resnet-IN) is given in Figure~\ref{fig:detailed_comparison} (top), when utilized in the \textit{ABMIL} ISUP grade model. 

The left and middle plots show how the performance on Radboud and Karolinska test data, respectively, is affected when training models on varying amounts of Radboud data. While the performance on Radboud (in-distribution, ID) data consistently improves with more training data, this is not the case on Karolinska (out-of-distribution, OOD) data, suggesting overfitting to site-specific characteristics. We also observe that UNI and CONCH perform especially well relative to Resnet-IN when the amount of training data is limited. 

The right plot of Figure~\ref{fig:detailed_comparison} (top) shows how the performance on Karolinska test data is affected when varying the proportion of training data sampled from Karolinska. There, we observe that when increasing from $0\%$ Karolinska training data, the performance quickly improves to a level close to that achieved with $100\%$ Karolinska training data.

Figure~\ref{fig:detailed_comparison} (bottom) shows the same performance comparison but for the three ISUP grade models instead, when using UNI as the patch-level feature extractor. We observe that, in virtually all cases, \textit{ABMIL} achieves the best performance followed by \textit{Mean Feature} and \textit{kNN}.

Figure~\ref{fig:umap_plots_radboud_vs_karolinska} shows UMAP~\citep{mcinnes2018umap} projections of the mean patch-level feature vectors for all PFMs on the full PANDA dataset, with one point per WSI and points colored by site (Radboud vs Karolinska). As can be observed, Radboud and Karolinska WSIs form clearly separated clusters for all PFMs. In addition, the Karolinska WSIs are further divided into two sub-clusters.

Figure~\ref{fig:umap_plots_coarse_isup_grades} instead shows UMAP visualizations of just the Radboud WSIs for all PFMs, colored according to three ISUP grade groups (grade $0$, grade $1$--$3$, grade $4$--$5$). The grade groups show partial clustering for all models including Resnet-IN, indicating that the learned features do capture clinically relevant signal. However, this separation is substantially weaker and less well-defined than the site-driven clustering observed in Figure~\ref{fig:umap_plots_radboud_vs_karolinska}, with considerable overlap between grade groups.

\section*{Discussion}

PFMs clearly improve over natural-image features for prostate cancer grading, yet robust generalization under clinically realistic distribution shift remains far from solved. 

Across the benchmark, PFMs achieve strong performance in standard evaluation settings, but substantial degradation is consistently observed under cross-site transfer from Radboud to Karolinska. This indicates that sensitivity to domain shift is a general limitation of current PFMs, rather than a weakness of any individual model. Importantly, this limitation persists despite the scale and diversity of modern PFM pretraining. Models such as H-optimus-1 and Virchow have been trained on more than one million WSIs, yet this does not ensure that models built on top of these PFMs always will perform well in downstream prediction tasks. In other words, while PFMs are pretrained on increasingly large and diverse datasets, their downstream use remains sensitive to acquisition- and site-related variation.

A second important finding is that the dominant challenge in this benchmark is site-related image shift rather than label-distribution shift. Performance degradation under \textit{Radboud-L $\rightarrow$ Radboud-R} is modest compared with the large drop observed for \textit{Radboud $\rightarrow$ Karolinska}. A substantial degradation is also seen for \textit{Radboud-U $\rightarrow$ Karolinska-U}, where controlling for label distribution reduces one source of shift but still leaves a large performance gap. This suggests that the primary bottleneck is not class imbalance, but visually grounded domain shift induced by differences in pathology workflow, staining, and scanning. Methods that address only label imbalance are therefore unlikely to be sufficient for improving robustness in practice.

Moreover, the comparison across model families suggests that recent advances in model scale and training do not translate into consistent improvements in robustness. While UNI2-h, H-optimus-1, and Virchow2 were pretrained on at least twice as many WSIs as their corresponding predecessors UNI, H-optimus-0, and Virchow (Table~\ref{table:fms}), they show only incremental improvements in cross-site generalization at best. Scaling alone thus appears insufficient to resolve robustness challenges.

While H0-mini performs comparably to its teacher model H-optimus-0, despite using fewer than 8\% of the parameters (86 million vs 1.1 billion), it does not show a clear improvement in cross-site generalization, even though it is explicitly designed for improved robustness. This highlights a potential disconnect between representation-level robustness and downstream performance, suggesting that improving embedding invariance alone may be insufficient for robust generalization.

The two vision-language models CONCH and CONCHv1.5 show inconsistent cross-site generalization. While CONCHv1.5 still suffers substantial performance drops from Radboud to Karolinska, it is among the most robust of all evaluated PFMs in this setting. In contrast, CONCH shows some of the weakest cross-site performance across several evaluation settings. These mixed results make it difficult to draw a clear conclusion regarding the benefit of multimodal pretraining. Prior work has suggested that training on image-caption pairs from diverse sources may encourage more robust representations, but our findings do not provide consistent support for this hypothesis. More systematic comparisons between vision-only and vision-language PFMs, ideally in controlled settings, are therefore needed to better understand when and why multimodal pretraining might improve robustness.

Another observation is that the downstream learning setup still matters. The \textit{ABMIL} model consistently outperforms simpler aggregation strategies (Figure~\ref{fig:main_results_attmil_mean-pool_knn_kappa}~\&~\ref{fig:main_results_attmil_mean-pool_knn_kappa_remaining6}), demonstrating the value of trainable slide-level modeling. At the same time, the persistence of large performance drops even for \textit{kNN} indicates that the robustness limitation is not solely due to overfitting in the downstream model, but is already present in the learned feature representations.

The UMAP analysis (Figure~\ref{fig:umap_plots_radboud_vs_karolinska}~\&~\ref{fig:umap_plots_coarse_isup_grades}) provides further insight into this limitation. Across all evaluated PFMs, site-related structure is much more pronounced than grade-related structure: Radboud and Karolinska form clearly separated clusters, and the Karolinska WSIs further split into two sub-clusters, likely reflecting scanner-specific variation within that site (Karolinska WSIs were digitized using two different scanners). In contrast, when considering Radboud alone, the ISUP grade groups show only partial and comparatively subtle separation. These results suggest a clear hierarchy in the learned representations, where variation related to data acquisition (site and scanner) dominates over variation related to disease grade. While grading signal is present, it forms a weaker axis of variation. This imbalance provides a plausible explanation for the observed lack of robustness, as downstream models trained on a single site may rely on domain-specific structure that does not transfer across sites.

Our findings are consistent with recent work demonstrating that PFMs are sensitive to scanner-induced variability under controlled experimental conditions~\citep{thiringer2026scanner}. In that setting, identical tissue sections scanned across multiple devices revealed pronounced scanner-specific structure in the learned embedding spaces. The present work extends these findings to a concrete practical deployment setting. Rather than isolating scanner effects, the PANDA site split represents a compound real-world distribution shift, combining scanner differences with variation in staining, tissue processing, and patient cohorts. Importantly, we show that the representation-level sensitivities observed in controlled studies can translate directly into substantial degradation in downstream clinical prediction performance.

\begin{figure*}[t]
\newcommand{\wid}{0.33\textwidth}
     \centering
     \begin{subfigure}[b]{\wid}
         \centering
         \includegraphics[width=0.47\textwidth]{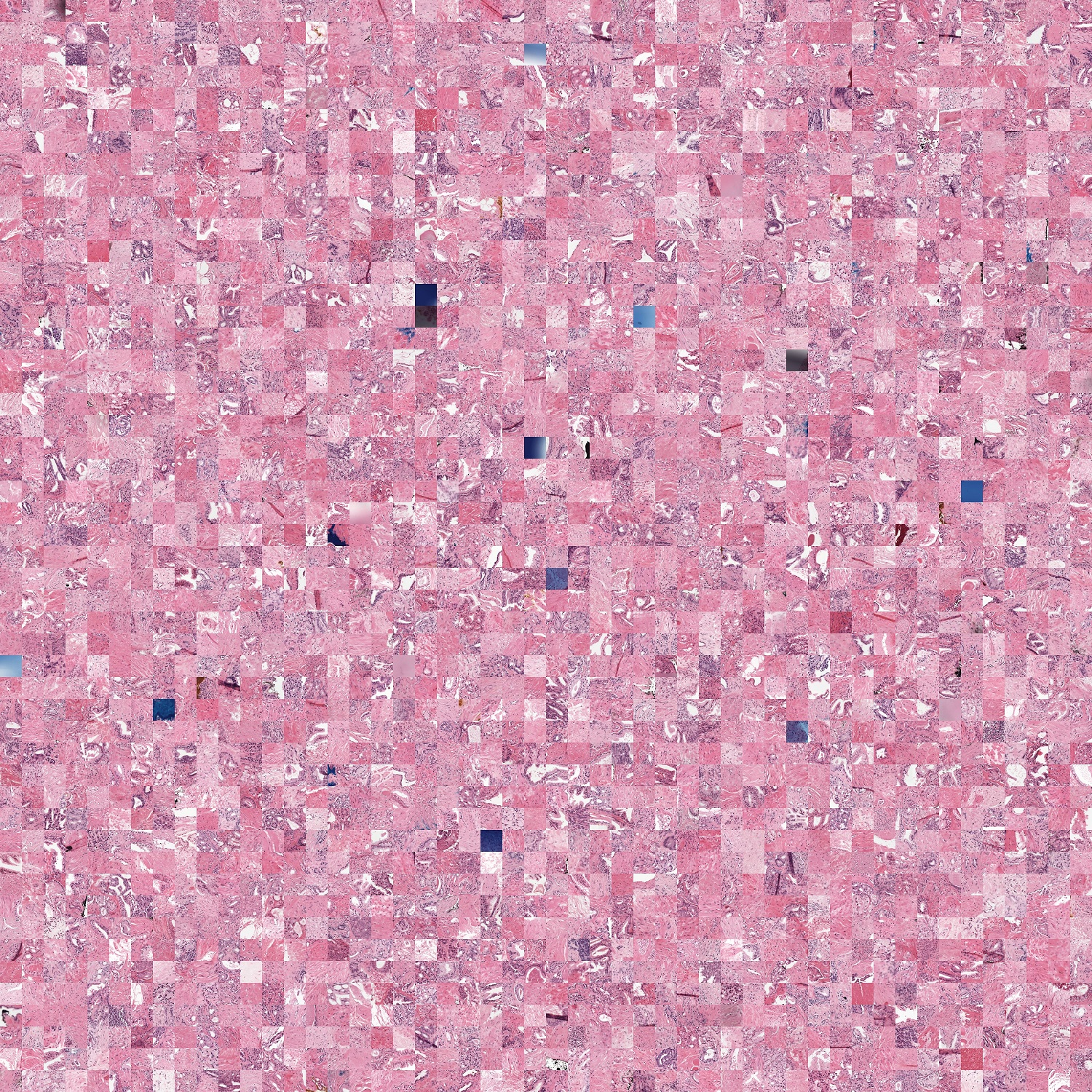}
         \includegraphics[width=0.47\textwidth]{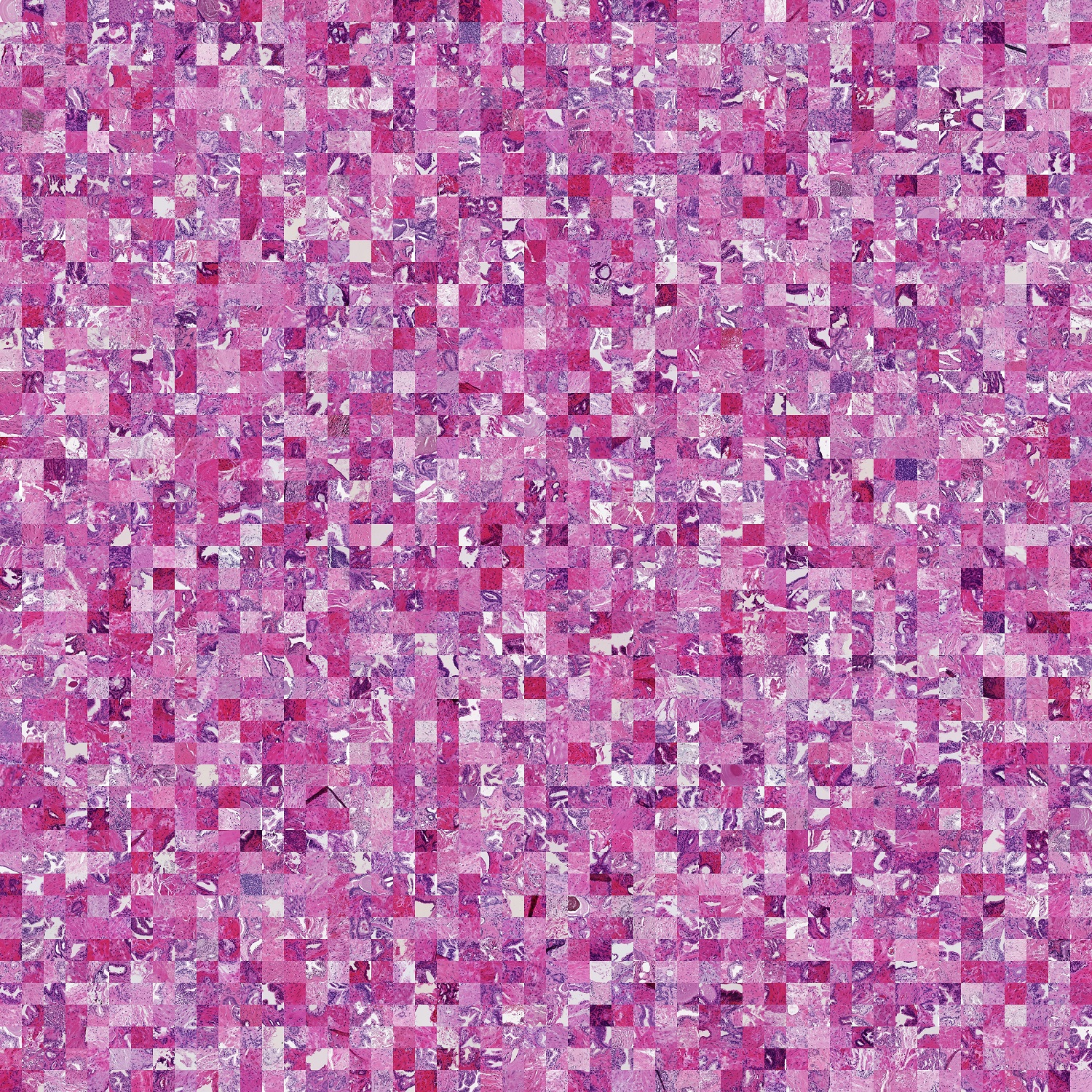}\vspace{1.0mm}
         \caption{WSI image data shift, \textit{Radboud} $\rightarrow$ \textit{Karolinska}.}
         \label{fig:shifts_covariate_radboud_karolinska}
     \end{subfigure}
     \begin{subfigure}[b]{\wid}
         \centering
         \includestandalone[width=0.47\textwidth]{figures/label_distr_radboud}\vspace{0.5mm}
         \includestandalone[width=0.47\textwidth]{figures/label_distr_karolinska}\vspace{1.0mm}
         \caption{Grade label shift, \textit{Radboud} $\rightarrow$ \textit{Karolinska}.}
         \label{fig:shifts_label_radboud_karolinska}
     \end{subfigure}
     \begin{subfigure}[b]{\wid}
         \centering
         \includestandalone[width=0.47\textwidth]{figures/label_distr_radboud-l}\vspace{0.5mm}
         \includestandalone[width=0.47\textwidth]{figures/label_distr_radboud-r}\vspace{1.0mm}
         \caption{Grade label shift, \textit{Radboud-L} $\rightarrow$ \textit{Radboud-R}.}
         \label{fig:shifts_label_radboud-l_radboud-r}
     \end{subfigure}
    \caption{\textbf{Illustration of the distribution shifts studied in this work.} We evaluate robustness under two common types of distribution shift in computational pathology. \textbf{(a) WSI image (domain) shift:} Visualization of $2{,}500$ randomly sampled patches from the Radboud and Karolinska subsets, illustrating differences in staining, scanning, and overall image appearance between sites. \textbf{(b) Label-distribution shift (Radboud $\rightarrow$ Karolinska):} Distribution of ISUP grades $0$--$5$ in the Radboud and Karolinska subsets, showing a substantial shift in class proportions between sites. \textbf{(c) Label-distribution shift (Radboud-L $\rightarrow$ Radboud-R):} Artificially constructed left- and right-skewed label distributions within the Radboud subset, used to isolate the effect of label-distribution shift independently of image-domain differences.}
    \label{fig:shifts}
\end{figure*}

The detailed data-scaling experiments (Figure~\ref{fig:detailed_comparison}) further emphasize the role of training data. Increasing the amount of single-site training data improves in-distribution performance but does not improve out-of-distribution performance and can even degrade it, suggesting overfitting to site-specific characteristics. In contrast, introducing even a small fraction of in-domain (Karolinska) data leads to rapid improvements in performance, indicating that exposure to target-domain variation, even at limited scale, can be highly effective for improving robustness.

Taken together, these results paint a consistent picture: current PFMs learn representations that are highly effective within domains but remain entangled with acquisition-specific variation, limiting their ability to generalize across sites. More broadly, these findings reinforce a key practical lesson for computational pathology: the quality and diversity of the labeled data used to train downstream models remain crucial, even in the foundation-model era. While PFMs provide strong and reusable feature representations, they do not remove the need for downstream training data that captures the variability encountered in real-world deployment. Incorporating data from multiple sites, or developing methods that explicitly promote domain-invariant representations, may therefore be at least as important as selecting the strongest PFM based on in-distribution benchmark performance alone.

This study has several limitations. First, it focuses on a single downstream task, prostate cancer grading on the PANDA dataset, and conclusions may not transfer directly to other pathology applications or cancer types. Second, we evaluate all PFMs as frozen patch-level feature extractors within a standardized weakly supervised pipeline in order to isolate representation quality and enable fair comparison. Relative rankings may differ under end-to-end fine-tuning, alternative aggregation strategies, or domain-adaptation methods. Third, the benchmark evaluates robustness using two sites, which provides a clean and practically relevant setting but does not fully capture the range of domain variability encountered in broader clinical deployment.

Future work should therefore extend this framework to multi-site training and evaluation across larger numbers of institutions, and systematically study how increasing site diversity during downstream training affects generalization. In particular, the results in Figure~\ref{fig:detailed_comparison} (right) suggest that introducing even small amounts of data from a second site can substantially improve performance on that site. However, the current experimental setup does not allow us to fully disentangle whether this reflects true improvements in cross-domain generalization or partial adaptation to shared domain characteristics, since training and test data originate from overlapping site distributions. A more rigorous evaluation would require data from three or more independent sites. For example, one could study whether a model trained on a mixture of data from sites $1$ and $2$ generalizes better to a previously unseen site $3$, compared to a model trained exclusively on site $1$. More generally, it would be valuable to investigate how performance scales as a function of the number of distinct sites included during training, and whether increasing site diversity leads to more domain-invariant representations. Finally, it would also be valuable to investigate whether the UMAP-observed differences in site separation can be translated into quantitative measures of feature-space invariance that better predict downstream robustness.

\emph{The main takeaways from our study are:}
(1) PFMs substantially outperform a natural-image baseline for prostate cancer grading, but absolute performance under cross-site shifts can still deteriorate severely.
(2) Large-scale pretraining of PFMs on diverse datasets does not guarantee downstream cross-site generalization, and increasing model size or pretraining data alone does not reliably improve robustness.
(3) PFMs are significantly more sensitive to image-domain shifts than to label distribution shifts, indicating that acquisition-related variability is the dominant challenge.
(4) Representation-level analysis reveals persistent domain separation, suggesting that current PFMs encode site- and scanner-specific variation more strongly than pathology-relevant signals.
(5) Strong PFMs and high-quality downstream data complement rather than replace each other: even in the foundation-model era, the quality and diversity of the data used to train downstream prediction models remains critical for reliable deployment.

\begin{figure*}[t]
    \centering%
    \includegraphics[width=0.925\linewidth]{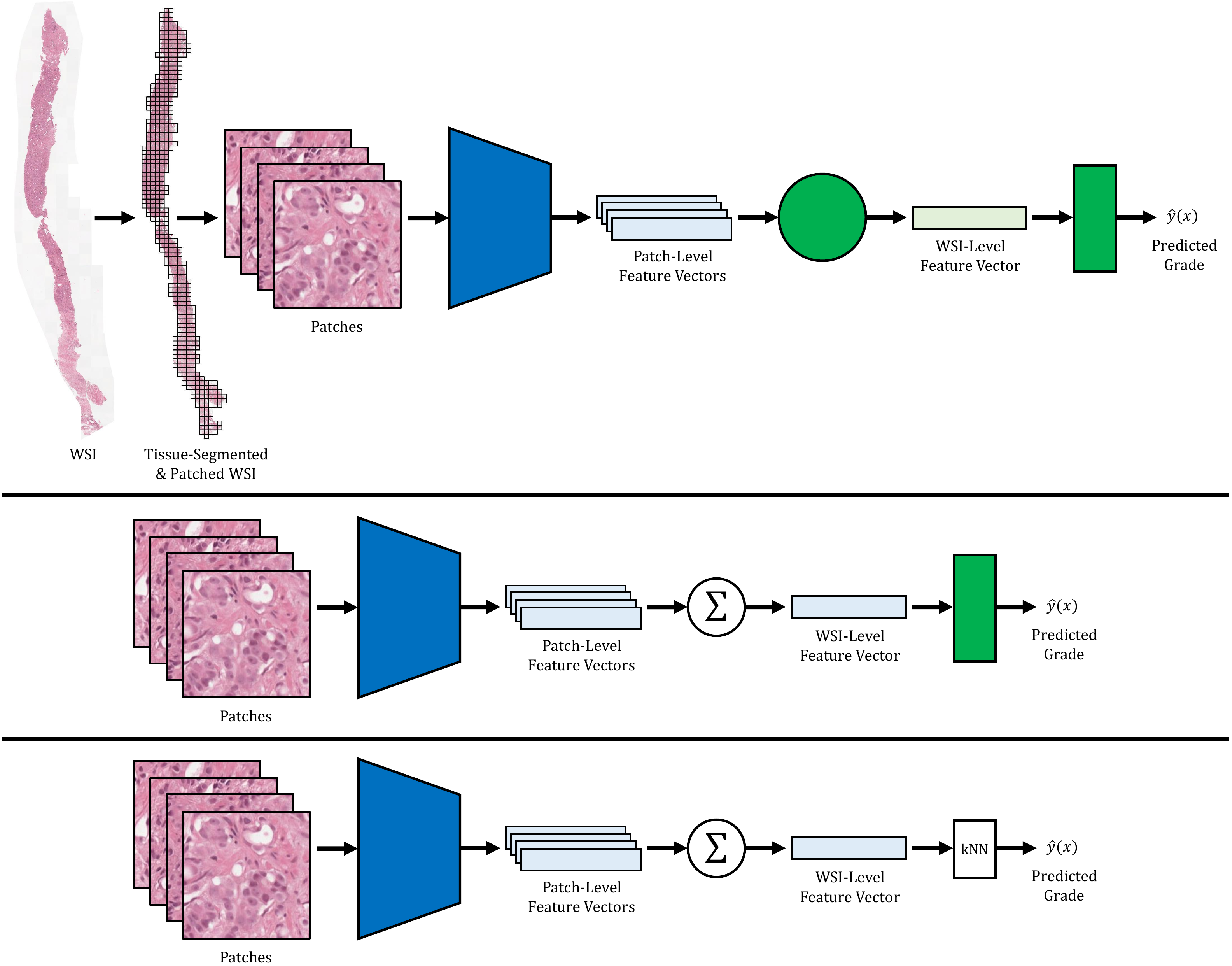}
    \caption{\textbf{Overview of the three evaluated ISUP grade classification models.} \textbf{Top:} \textit{ABMIL}. \textbf{Middle:} \textit{Mean Feature}. \textbf{Bottom:} \textit{kNN}. All models use the same WSI preprocessing pipeline. First, the prostate biopsy WSI $x$ is tissue-segmented and divided into non-overlapping patches $\tilde{x}_i$ of size $256 \times 256$ using CLAM~\citep{lu2021data}. Next, a feature vector $p(\tilde{x}_i)$ is extracted for each patch using a pretrained and frozen feature extractor. The models then process these patch-level feature vectors $p(\tilde{x}_i)$ further, outputting a predicted ISUP grade $\hat{y}(x) \in \{0, \dots, 5\}$. In the figure, \textcolor{Cerulean}{\textbf{blue}} denotes the pretrained and frozen feature extractor, whereas \textcolor{ForestGreen}{\textbf{green}} denotes trainable model components.}    
    \label{fig:model_overview}
\end{figure*}

\section*{Methods}

We next describe the experimental setup, datasets, and modeling framework used to evaluate the robustness of PFMs under distribution shift. Our evaluation follows a standardized pipeline in which pretrained PFMs are used as frozen patch-level feature extractors within weakly supervised slide-level classification models, enabling direct comparison of their representation quality. We construct multiple dataset splits to isolate two clinically relevant types of distribution shift, cross-site variation in WSI image appearance and shifts in the ISUP grade label distribution, and evaluate model performance under each setting using consistent training and evaluation protocols. This setup enables controlled evaluation of how different PFMs behave under distribution shift while minimizing confounding factors from downstream model design.

\subsubsection*{Experimental Setup}
We conduct all experiments using the publicly available development set of the PANDA dataset~\citep{bulten2022artificial}. It contains $10{,}616$ prostate biopsy WSIs with corresponding ISUP grade labels (grade $0$--$5$). The $10{,}616$ biopsy WSIs were collected from two different sites: Radboud University Medical Center (Nijmegen, the Netherlands) and Karolinska Institutet (Stockholm, Sweden), and originate from a total of $2{,}113$ patients. The data collection site (Radboud or Karolinska) is known for each WSI in the dataset, while there is no available information on which patient each WSI originates from.

The WSIs from the two data collection sites were digitized using different scanners (Radboud: 3DHistech, Karolinska: Leica or Hamamatsu). Together with other potential differences in staining and processing procedures, this creates a clear shift in appearance of the corresponding WSI image data, as illustrated in Figure~\ref{fig:shifts_covariate_radboud_karolinska} (visualization of randomly sampled tissue patches from Radboud and Karolinska WSIs). We do not apply stain normalization or color augmentation, in order to preserve the natural domain shift between sites.

Following the PANDA challenge~\citep{bulten2022artificial}, model performance is evaluated using quadratically weighted Cohen's kappa (`kappa' for short) as the primary metric, which measures the level of agreement between the predicted and ground truth ISUP grades. As a secondary metric, we also compute the mean absolute error (MAE) between grade predictions and labels. All results are reported as mean $\pm$ standard deviation across the 10 cross-validation folds. All models are trained and evaluated using identical data splits to ensure fair comparison across feature extractors.

\paragraph{Main Evaluation}
For the main model comparison in Figure~\ref{fig:main_results3a_12fms_kappa}~-~\ref{fig:main_results_attmil_mean-pool_knn_kappa}, we create different subsets of the original PANDA development set, resulting in seven different datasets in total. \textbf{1. \textit{PANDA}} is the full original dataset, containing $10{,}433$ WSIs after pre-processing (removal of WSIs with no extracted tissue patches). \textbf{2. \textit{Karolinska}} is a subset of \textit{PANDA}, containing all WSIs collected at Karolinska, $5{,}434$ WSIs. Similarly, \textbf{3. \textit{Radboud}} contains all WSIs collected at Radboud, $4{,}999$ WSIs. The label distribution for \textit{Radboud} and \textit{Karolinska}, as shown in Figure~\ref{fig:shifts_label_radboud_karolinska}, are quite different: nearly a uniform distribution over grade $0$--$5$ for \textit{Radboud}, whereas the distribution is heavily left-skewed for \textit{Karolinska}. In order to study the effect of this label shift, we create further subsets with a perfectly uniform label distribution: \textbf{4. \textit{Radboud-U}} containing $3{,}996$ WSIs, and \textbf{5. \textit{Karolinska-U}} containing $1{,}506$ WSIs. We also create subsets of \textit{Radboud} with a left-skewed or right-skewed label distribution (as shown in Figure~\ref{fig:shifts_label_radboud-l_radboud-r}): \textbf{6. \textit{Radboud-L}} containing $2{,}484$ WSIs, and \textbf{7. \textit{Radboud-R}} containing $2{,}515$ WSIs.

For in-distribution evaluations (\textit{PANDA}, \textit{Karolinska}, \textit{Radboud}, \textit{Radboud-U}, \textit{Radboud-L}), models are trained and evaluated using random 10-fold cross-validation with $80\%/10\%/10\%$ train/val/test splits. For the \textit{Radboud $\rightarrow$ Karolinska}, \textit{Radboud-U $\rightarrow$ Karolinska-U} and \textit{Radboud-L $\rightarrow$ Radboud-R} results, the $10$ trained cross-validation models are all evaluated on the full corresponding test set (e.g. for \textit{Radboud $\rightarrow$ Karolinska}, the $10$ models trained on \textit{Radboud} are all evaluated on \textit{Karolinska}).

\paragraph{Detailed Evaluation}
For the detailed model comparison in Figure~\ref{fig:detailed_comparison}, we construct additional subsets of \textit{Radboud} and \textit{Karolinska}. For the left and middle plots of Figure~\ref{fig:detailed_comparison}, we set aside $1{,}000$ randomly sampled WSIs from \textit{Radboud} to form \textit{Radboud-Subsets-Test}, while the remaining $3{,}999$ WSIs form \textit{Radboud-Subsets-Dev-100}. We then also create \textit{Radboud-Subsets-Dev-50}, \textit{Radboud-Subsets-Dev-25}, \textit{Radboud-Subsets-Dev-10}, \textit{Radboud-Subsets-Dev-5} and \textit{Radboud-Subsets-Dev-2}, by randomly sampling $50\%$, $25\%$, $10\%$, $5\%$ or $2\%$ of the WSIs in \textit{Radboud-Subsets-Dev-100}. The \textit{Radboud-Subsets-Dev} datasets are used for training models (random 10-fold cross-validation), which then are evaluated on both \textit{Radboud-Subsets-Test} and \textit{Karolinska}.\looseness=-1

For the right plots of Figure~\ref{fig:detailed_comparison}, we set aside $1{,}000$ randomly sampled WSIs from \textit{Karolinska} to form \textit{Karolinska-1k-Test}. We then randomly sample the remaining $4{,}434$ WSIs from \textit{Karolinska} and all $4{,}999$ WSIs of \textit{Radboud}, to form a series of datasets all containing $4{,}000$ WSIs, with $0\%$, $1\%$, $5\%$, $10\%$, $25\%$, $50\%$ or $100\%$ of the WSIs being sampled from \textit{Karolinska}. These mixed Radboud/Karolinska datasets are used for training models (random 10-fold cross-validation), which then are evaluated on \textit{Karolinska-1k-Test}.

\subsubsection*{ISUP Grade Classification Models}
We evaluate three different models, which all take a prostate biopsy WSI $x$ as input and output a predicted ISUP grade $\hat{y}(x) \in \{0, \dots, 5\}$. An overview of the models is shown in Figure~\ref{fig:model_overview}. Although there is an ordinal relationship among the $6$ ISUP grades $\{0, \dots, 5\}$, we treat this as a regular multi-class classification problem, with $C=6$ classes.

All three models utilize the same initial WSI processing steps. First, the input biopsy WSI $x$ is tissue-segmented and divided into $P$ non-overlapping patches $\{\tilde{x}_i\}_{i=1}^P$ using CLAM~\citep{lu2021data}. Patches are of size $256 \times 256$ at $20\times$ magnification, and the number of extracted tissue patches $P$ varies for different WSIs. Next, a feature vector $p(\tilde{x}_i)$ is extracted for each patch $\tilde{x}_i$, using a pretrained and frozen feature extractor. The three models then process these \emph{patch-level} feature vectors $\{p(\tilde{x}_i)\}_{i=1}^P$ further, finally outputting a predicted ISUP grade $\hat{y}(x) \in \{0, \dots, 5\}$ for the WSI $x$.

\paragraph{ABMIL}
The first ISUP grade classification model, \textit{ABMIL} in Figure~\ref{fig:model_overview} (top), utilizes an ABMIL model~\citep{ilse2018attention} to aggregate the set of patch-level feature vectors $\{p(\tilde{x}_i)\}_{i=1}^P$ into a single WSI-level feature vector $w(x)$. This feature vector $w(x)$ is then fed as input to a linear classification layer, outputting logits for the $C\!=\!6$ classes/grades. Finally, $\hat{y}(x) \in \{0, \dots, 5\}$ is computed as the argmax over the logits. Both the ABMIL aggregator and the linear classification layer are trained using the standard cross-entropy loss. Our ABMIL implementation is based on CLAM~\citep{lu2021data} (without the instance-level clustering), and we set the model and training hyperparameters according to UNI (see \textit{Methods - Weakly supervised slide classification} in \cite{chen2024uni}). Specifically, models are trained using the AdamW optimizer~\citep{loshchilov2018decoupled} with a cosine learning rate schedule, for a maximum of 20 epochs. Early stopping is performed based on the val loss for each of the 10 random train/val/test cross-validation folds.

\paragraph{Mean Feature}
The second model, \textit{Mean Feature} in Figure~\ref{fig:model_overview} (middle), simplifies \textit{ABMIL} by removing the trainable ABMIL WSI-level aggregator. Instead, a single WSI-level feature vector $w(x)$ is extracted by directly computing the mean over all patch-level feature vectors $\{p(\tilde{x}_i)\}_{i=1}^P$. The WSI-level feature vector $w(x)$ is then fed as input to a network head consisting of a fully-connected layer (with dropout and ReLU activation) and a linear classification layer, outputting logits for the $C\!=\!6$ classes/grades. The network head (which is the only trainable model component left) is trained using the cross-entropy loss, with the same hyperparameters as used for \textit{ABMIL}.

\paragraph{kNN}
The third ISUP grade model, \textit{kNN} in Figure~\ref{fig:model_overview} (bottom), in turn simplifies \textit{Mean Feature} by removing its trainable network head. As for \textit{Mean Feature}, a WSI-level feature vector $w(x)$ is directly computed as the mean over the patch-level feature vectors $\{p(\tilde{x}_i)\}_{i=1}^P$. To output a predicted ISUP grade $\hat{y}(x) \in \{0, \dots, 5\}$, we then instead utilize \texttt{KNeighborsClassifier} from scikit-learn~\citep{scikit-learn} with $k = 5$. We include \textit{kNN} as a simplest possible baseline, without any trainable model parameters.

\subsubsection*{Patch-Level Feature Extractors}
We evaluate eight recent vision-only PFMs (Phikon-v2~\citep{filiot2024phikonv2}, Prov-GigaPath~\citep{gigapath2024}, UNI~\citep{chen2024uni}, UNI2-h~\citep{UNI2h2024}, Virchow~\citep{virchow2024}, Virchow2~\citep{virchow22024}, H-optimus-0~\citep{hoptimus0}, H-optimus-1~\citep{hoptimus1}), two vision-language PFMs (CONCH~\citep{conch2024}, CONCHv1.5~\citep{ding2025multimodal}), a compact distilled PFM (H0-mini~\citep{filiot2025h0mini}), and a natural-image baseline (Resnet-IN). An overview of all evaluated models, including architecture, parameter count, feature dimensionality, and pretraining data, is provided in Table~\ref{table:fms}. All models are used as frozen patch-level feature extractors, and are not updated during the training of the downstream ISUP grade classification models.

The evaluated PFMs span a range of architectures and pretraining regimes. All models are based on vision transformer (ViT) architectures~\citep{dosovitskiy2021an}, with sizes ranging from ViT-Base ($86$ million parameters) to ViT-Giant ($1.1$ billion). The vision-only PFMs are pretrained on large collections of WSIs using self-supervised learning objectives, with dataset sizes ranging from approximately $58{,}000$ WSIs (Phikon-v2) to $3.1$ million WSIs (Virchow2). For example, UNI is a widely used vision-only PFM based on ViT-Large, pretrained using DINOv2~\citep{oquab2024dinov2} on a pan-cancer dataset (20 major tissue types) comprising approximately $100$ million tissue patches from more than $100{,}000$ WSIs. 

In contrast, CONCH and CONCHv1.5 incorporate multimodal pretraining using image-text pairs in addition to WSIs, representing an alternative training paradigm based on vision-language supervision. The CONCH image encoder is a ViT-Base model, first pretrained on an in-house dataset of $16$ million tissue patches from more than $21{,}000$ WSIs using self-supervised learning (iBOT~\citep{zhou2022image}), and subsequently trained with a vision-language objective (CoCa~\citep{yu2022coca}) on more than $1.1$ million pathology-specific image-caption pairs. This image-caption dataset was curated via automatic processing of figures extracted from publicly available articles from PubMed.

H0-mini is a compact distilled variant of H-optimus-0, designed to reduce model size and computational cost while preserving representation quality. It is based on a ViT-Base model with $86$ million parameters, corresponding to less than $8\%$ of the parameters of the original ViT-Giant H-optimus-0. The distillation is performed using a dataset of $43$ million image patches extracted from $6{,}093$ WSIs from the TCGA dataset, covering $16$ cancer types.

Finally, Resnet-IN is a Resnet-50 model~\citep{he2016deep} pretrained on the ImageNet dataset~\citep{russakovsky2015imagenet} of natural images. It is included as a simple reference baseline to quantify the benefit of pathology-specific pretraining.

\subsubsection*{Ethics Statement}
This study exclusively utilizes publicly available, de-identified data from the PANDA dataset~\citep{bulten2022artificial}.

\section*{Data Availability}

The biopsy whole-slide images and corresponding ISUP grade labels for the development set of the PANDA dataset are available at \url{https://www.kaggle.com/c/prostate-cancer-grade-assessment/data}.
\section*{Code Availability}

The code for this study is based on CLAM which is available at \url{https://github.com/mahmoodlab/CLAM}. Further implementation details are available from FKG upon reasonable request. 
\section*{Acknowledgments}

The project was supported by funding from the Swedish Cancer Society (23 2905 Pj 01 H), VINNOVA (SwAIPP2), and the Swedish e-science Research Centre (SeRC) (eMPHasis project).
\section*{Author Contributions}

FKG was responsible for project conceptualization, software implementation, preparation of figures and tables, and manuscript drafting. MR supervised the project and acquired funding. Both authors contributed to the design of experiments, interpretation of results, and manuscript revision.
\section*{Competing Interests}

MR is a co-founder and shareholder of Stratipath AB. FKG declares no competing interests.
\section*{Declaration of Generative AI Use}

During the preparation of this manuscript, the authors used ChatGPT 5.3 to assist with language editing and drafting. All content produced using this tool was critically reviewed, edited and validated by the authors, who take full responsibility for the final content of the manuscript.

{
    \small
    \bibliographystyle{ieeenat_fullname}
    \bibliography{references}
}

\clearpage
\appendix
\onecolumn

\renewcommand{\thefigure}{S\arabic{figure}}
\setcounter{figure}{0}

\renewcommand{\thetable}{S\arabic{table}}
\setcounter{table}{0}

\renewcommand{\theequation}{S\arabic{equation}}
\setcounter{equation}{0}

\subsection*{\centering{Evaluating Computational Pathology Foundation Models\\ for Prostate Cancer Grading under Distribution Shifts}}
\section*{\centering{Supplementary Material}}

\vspace{6.0mm}

\section{Supplementary Figures}
\label{appendix:figures}

This section contains Figure \ref{fig:main_results_attmil_mean-pool_knn_kappa_remaining6} - \ref{fig:main_results_attmil_mean-pool_knn_mae_remaining6}.

\vspace{5.0mm}

\begin{figure*}[h]
    \centering
    \begin{tikzpicture}
  \centering
  \begin{axis}[
        ybar, axis on top,
        title={},
        height=4.525cm, width=18.0cm, 
        bar width=0.35cm,
        ymajorgrids, tick align=inside,
        enlarge y limits={value=.1,upper},
        ymin=0, ymax=0.9,
        set layers,
        ytick={0.2, 0.4, 0.6, 0.8},
        axis x line*=bottom,
        axis y line*=left,
        y axis line style={opacity=0},
        tickwidth=0pt,
        enlarge x limits=true,
        extra y tick style={
            grid style={
                solid,
                line width = 2.5pt,
                /pgfplots/on layer=axis background,
            },
        },
        tick style={
            grid style={
                dashed,
                /pgfplots/on layer=axis background,
            },
        },
        legend style={
            at={(0.5,-0.2)},
            anchor=north,
            legend columns=-1,
            /tikz/every even column/.append style={column sep=0.325cm},
            font=\footnotesize
        },
        ylabel={Kappa ($\uparrow$)},
        ylabel style={font=\footnotesize},
        symbolic x coords={
           PANDA,
           Karolinska,
           Radboud, 
           Radboud $\rightarrow$\\ Karolinska,
           Radboud-U, 
           Radboud-U $\rightarrow$\\ Karolinska-U,
           Radboud-L, 
           Radboud-L $\rightarrow$\\ Radboud-R},
       xtick=data,
       xticklabel=\empty,
       x tick label style={text width = 2.25cm, align=center, font=\footnotesize},
    ]
    \addplot [fill=skyblue2, draw=skyblue2, line width=0mm, error bars/.cd, y dir=both, y explicit, error bar style={thick}, error bar style=black] coordinates { 
      (PANDA, 0.874011) +- (0, 0.0113699)
      (Karolinska, 0.874871) +- (0, 0.0112448)
      (Radboud, 0.837963) +- (0, 0.00868408)
      (Radboud $\rightarrow$\\ Karolinska, 0.168457) +- (0, 0.0897553)
      (Radboud-U, 0.839083) +- (0, 0.0319653)
      (Radboud-U $\rightarrow$\\ Karolinska-U, 0.507484) +- (0, 0.0579279)
      (Radboud-L, 0.817192) +- (0, 0.0346085)
      (Radboud-L $\rightarrow$\\ Radboud-R, 0.714851) +- (0, 0.0166124)};
    \addplot [fill=caribbeangreen2, draw=caribbeangreen2, error bars/.cd, y dir=both, y explicit, error bar style={thick}, error bar style=black] coordinates {
      (PANDA, 0.789349) +- (0, 0.0215116)
      (Karolinska, 0.744444) +- (0, 0.0361009)
      (Radboud, 0.7775) +- (0, 0.0219794)
      (Radboud $\rightarrow$\\ Karolinska, 0.212961) +- (0, 0.0533835)
      (Radboud-U, 0.760223) +- (0, 0.0327461)
      (Radboud-U $\rightarrow$\\ Karolinska-U, 0.350377) +- (0, 0.0322908)
      (Radboud-L, 0.760892) +- (0, 0.0253868)
      (Radboud-L $\rightarrow$\\ Radboud-R, 0.629822) +- (0, 0.00907769)};
    \addplot [fill=darkpastelpurple2, draw=darkpastelpurple2, error bars/.cd, y dir=both, y explicit, error bar style={thick}, error bar style=black] coordinates {
      (PANDA, 0.687072) +- (0, 0.0202816)
      (Karolinska, 0.556356) +- (0, 0.0316019)
      (Radboud, 0.727437) +- (0, 0.0197546)
      (Radboud $\rightarrow$\\ Karolinska, 0.126199) +- (0, 0.0605454)
      (Radboud-U, 0.670146) +- (0, 0.0246172)
      (Radboud-U $\rightarrow$\\ Karolinska-U, 0.10539) +- (0, 0.0530505)
      (Radboud-L, 0.600378) +- (0, 0.0572836)
      (Radboud-L $\rightarrow$\\ Radboud-R, 0.350412) +- (0, 0.00958643)};
  \end{axis}
  \end{tikzpicture}\vspace{-2.5mm}
    \begin{tikzpicture}
  \centering
  \begin{axis}[
        ybar, axis on top,
        title={},
        height=4.525cm, width=18.0cm, 
        bar width=0.35cm,
        ymajorgrids, tick align=inside,
        enlarge y limits={value=.1,upper},
        ymin=0, ymax=0.9,
        set layers,
        ytick={0.2, 0.4, 0.6, 0.8},
        axis x line*=bottom,
        axis y line*=left,
        y axis line style={opacity=0},
        tickwidth=0pt,
        enlarge x limits=true,
        extra y tick style={
            grid style={
                solid,
                line width = 2.5pt,
                /pgfplots/on layer=axis background,
            },
        },
        tick style={
            grid style={
                dashed,
                /pgfplots/on layer=axis background,
            },
        },
        legend style={
            at={(0.5,-0.2)},
            anchor=north,
            legend columns=-1,
            /tikz/every even column/.append style={column sep=0.325cm},
            font=\footnotesize
        },
        ylabel={Kappa ($\uparrow$)},
        ylabel style={font=\footnotesize},
        symbolic x coords={
           PANDA,
           Karolinska,
           Radboud, 
           Radboud $\rightarrow$\\ Karolinska,
           Radboud-U, 
           Radboud-U $\rightarrow$\\ Karolinska-U,
           Radboud-L, 
           Radboud-L $\rightarrow$\\ Radboud-R},
       xtick=data,
       xticklabel=\empty,
       x tick label style={text width = 2.25cm, align=center, font=\footnotesize},
    ]
    \addplot [fill=skyblue2, draw=skyblue2, line width=0mm, error bars/.cd, y dir=both, y explicit, error bar style={thick}, error bar style=black] coordinates { 
      (PANDA, 0.885085) +- (0, 0.0123753)
      (Karolinska, 0.885949) +- (0, 0.0159321)
      (Radboud, 0.85256) +- (0, 0.0152281)
      (Radboud $\rightarrow$\\ Karolinska, 0.270728) +- (0, 0.143959)
      (Radboud-U, 0.855195) +- (0, 0.0186979)
      (Radboud-U $\rightarrow$\\ Karolinska-U, 0.536572) +- (0, 0.0657804)
      (Radboud-L, 0.840353) +- (0, 0.0261901)
      (Radboud-L $\rightarrow$\\ Radboud-R, 0.730126) +- (0, 0.0199094)};
    \addplot [fill=caribbeangreen2, draw=caribbeangreen2, error bars/.cd, y dir=both, y explicit, error bar style={thick}, error bar style=black] coordinates {
      (PANDA, 0.835556) +- (0, 0.0115376)
      (Karolinska, 0.764095) +- (0, 0.0362231)
      (Radboud, 0.82717) +- (0, 0.025873)
      (Radboud $\rightarrow$\\ Karolinska, 0.335084) +- (0, 0.0431102)
      (Radboud-U, 0.815075) +- (0, 0.0347275)
      (Radboud-U $\rightarrow$\\ Karolinska-U, 0.370097) +- (0, 0.0361245)
      (Radboud-L, 0.802894) +- (0, 0.0221661)
      (Radboud-L $\rightarrow$\\ Radboud-R, 0.677703) +- (0, 0.0047424)};
    \addplot [fill=darkpastelpurple2, draw=darkpastelpurple2, error bars/.cd, y dir=both, y explicit, error bar style={thick}, error bar style=black] coordinates {
      (PANDA, 0.719051) +- (0, 0.0217944)
      (Karolinska, 0.560601) +- (0, 0.0475562)
      (Radboud, 0.757702) +- (0, 0.0320785)
      (Radboud $\rightarrow$\\ Karolinska, 0.21015) +- (0, 0.0267073)
      (Radboud-U, 0.705477) +- (0, 0.0383834)
      (Radboud-U $\rightarrow$\\ Karolinska-U, 0.358436) +- (0, 0.0180455)
      (Radboud-L, 0.673361) +- (0, 0.0298184)
      (Radboud-L $\rightarrow$\\ Radboud-R, 0.4149) +- (0, 0.00668088)};
  \end{axis}
  \end{tikzpicture}\vspace{-2.5mm}
    \begin{tikzpicture}
  \centering
  \begin{axis}[
        ybar, axis on top,
        title={},
        height=4.525cm, width=18.0cm, 
        bar width=0.35cm,
        ymajorgrids, tick align=inside,
        enlarge y limits={value=.1,upper},
        ymin=0, ymax=0.9,
        set layers,
        ytick={0.2, 0.4, 0.6, 0.8},
        axis x line*=bottom,
        axis y line*=left,
        y axis line style={opacity=0},
        tickwidth=0pt,
        enlarge x limits=true,
        extra y tick style={
            grid style={
                solid,
                line width = 2.5pt,
                /pgfplots/on layer=axis background,
            },
        },
        tick style={
            grid style={
                dashed,
                /pgfplots/on layer=axis background,
            },
        },
        legend style={
            at={(0.5,-0.2)},
            anchor=north,
            legend columns=-1,
            /tikz/every even column/.append style={column sep=0.325cm},
            font=\footnotesize
        },
        ylabel={Kappa ($\uparrow$)},
        ylabel style={font=\footnotesize},
        symbolic x coords={
           PANDA,
           Karolinska,
           Radboud, 
           Radboud $\rightarrow$\\ Karolinska,
           Radboud-U, 
           Radboud-U $\rightarrow$\\ Karolinska-U,
           Radboud-L, 
           Radboud-L $\rightarrow$\\ Radboud-R},
       xtick=data,
       xticklabel=\empty,
       x tick label style={text width = 2.25cm, align=center, font=\footnotesize},
    ]
    \addplot [fill=skyblue2, draw=skyblue2, line width=0mm, error bars/.cd, y dir=both, y explicit, error bar style={thick}, error bar style=black] coordinates { 
      (PANDA, 0.889037) +- (0, 0.01254)
      (Karolinska, 0.89135) +- (0, 0.0195419)
      (Radboud, 0.860933) +- (0, 0.0227627)
      (Radboud $\rightarrow$\\ Karolinska, 0.472049) +- (0, 0.105997)
      (Radboud-U, 0.849584) +- (0, 0.0248498)
      (Radboud-U $\rightarrow$\\ Karolinska-U, 0.656305) +- (0, 0.0318687)
      (Radboud-L, 0.833236) +- (0, 0.0307769)
      (Radboud-L $\rightarrow$\\ Radboud-R, 0.729401) +- (0, 0.0238521)};
    \addplot [fill=caribbeangreen2, draw=caribbeangreen2, error bars/.cd, y dir=both, y explicit, error bar style={thick}, error bar style=black] coordinates {
      (PANDA, 0.841891) +- (0, 0.0163428)
      (Karolinska, 0.782086) +- (0, 0.0264358)
      (Radboud, 0.832701) +- (0, 0.0163559)
      (Radboud $\rightarrow$\\ Karolinska, 0.359176) +- (0, 0.0426587)
      (Radboud-U, 0.823268) +- (0, 0.0363036)
      (Radboud-U $\rightarrow$\\ Karolinska-U, 0.380124) +- (0, 0.0558697)
      (Radboud-L, 0.80787) +- (0, 0.0274059)
      (Radboud-L $\rightarrow$\\ Radboud-R, 0.681326) +- (0, 0.00623155)};
    \addplot [fill=darkpastelpurple2, draw=darkpastelpurple2, error bars/.cd, y dir=both, y explicit, error bar style={thick}, error bar style=black] coordinates {
      (PANDA, 0.724572) +- (0, 0.0149668)
      (Karolinska, 0.587677) +- (0, 0.037649)
      (Radboud, 0.751543) +- (0, 0.0341701)
      (Radboud $\rightarrow$\\ Karolinska, 0.294091) +- (0, 0.012755)
      (Radboud-U, 0.712543) +- (0, 0.040858)
      (Radboud-U $\rightarrow$\\ Karolinska-U, 0.30383) +- (0, 0.0111311)
      (Radboud-L, 0.682925) +- (0, 0.0170415)
      (Radboud-L $\rightarrow$\\ Radboud-R, 0.407121) +- (0, 0.00866511)};
  \end{axis}
  \end{tikzpicture}\vspace{-2.5mm}
    \begin{tikzpicture}
  \centering
  \begin{axis}[
        ybar, axis on top,
        title={},
        height=4.525cm, width=18.0cm, 
        bar width=0.35cm,
        ymajorgrids, tick align=inside,
        enlarge y limits={value=.1,upper},
        ymin=0, ymax=0.9,
        set layers,
        ytick={0.2, 0.4, 0.6, 0.8},
        axis x line*=bottom,
        axis y line*=left,
        y axis line style={opacity=0},
        tickwidth=0pt,
        enlarge x limits=true,
        extra y tick style={
            grid style={
                solid,
                line width = 2.5pt,
                /pgfplots/on layer=axis background,
            },
        },
        tick style={
            grid style={
                dashed,
                /pgfplots/on layer=axis background,
            },
        },
        legend style={
            at={(0.5,-0.2)},
            anchor=north,
            legend columns=-1,
            /tikz/every even column/.append style={column sep=0.325cm},
            font=\footnotesize
        },
        ylabel={Kappa ($\uparrow$)},
        ylabel style={font=\footnotesize},
        symbolic x coords={
           PANDA,
           Karolinska,
           Radboud, 
           Radboud $\rightarrow$\\ Karolinska,
           Radboud-U, 
           Radboud-U $\rightarrow$\\ Karolinska-U,
           Radboud-L, 
           Radboud-L $\rightarrow$\\ Radboud-R},
       xtick=data,
       xticklabel=\empty,
       x tick label style={text width = 2.25cm, align=center, font=\footnotesize},
    ]
    \addplot [fill=skyblue2, draw=skyblue2, line width=0mm, error bars/.cd, y dir=both, y explicit, error bar style={thick}, error bar style=black] coordinates { 
      (PANDA, 0.885995) +- (0, 0.0134458)
      (Karolinska, 0.884609) +- (0, 0.0129327)
      (Radboud, 0.859427) +- (0, 0.0152547)
      (Radboud $\rightarrow$\\ Karolinska, 0.405437) +- (0, 0.0840904)
      (Radboud-U, 0.852509) +- (0, 0.0202243)
      (Radboud-U $\rightarrow$\\ Karolinska-U, 0.64361) +- (0, 0.0204429)
      (Radboud-L, 0.827465) +- (0, 0.0269886)
      (Radboud-L $\rightarrow$\\ Radboud-R, 0.729103) +- (0, 0.0170106)};
    \addplot [fill=caribbeangreen2, draw=caribbeangreen2, error bars/.cd, y dir=both, y explicit, error bar style={thick}, error bar style=black] coordinates {
      (PANDA, 0.822077) +- (0, 0.018959)
      (Karolinska, 0.761503) +- (0, 0.0308394)
      (Radboud, 0.806233) +- (0, 0.0249663)
      (Radboud $\rightarrow$\\ Karolinska, 0.489018) +- (0, 0.0232659)
      (Radboud-U, 0.788607) +- (0, 0.0386399)
      (Radboud-U $\rightarrow$\\ Karolinska-U, 0.420712) +- (0, 0.0236664)
      (Radboud-L, 0.799799) +- (0, 0.0347504)
      (Radboud-L $\rightarrow$\\ Radboud-R, 0.655461) +- (0, 0.0135482)};
    \addplot [fill=darkpastelpurple2, draw=darkpastelpurple2, error bars/.cd, y dir=both, y explicit, error bar style={thick}, error bar style=black] coordinates {
      (PANDA, 0.687414) +- (0, 0.0244893)
      (Karolinska, 0.540177) +- (0, 0.0366539)
      (Radboud, 0.711327) +- (0, 0.0364132)
      (Radboud $\rightarrow$\\ Karolinska, 0.346756) +- (0, 0.00713873)
      (Radboud-U, 0.674127) +- (0, 0.0351979)
      (Radboud-U $\rightarrow$\\ Karolinska-U, 0.314316) +- (0, 0.0137609)
      (Radboud-L, 0.633288) +- (0, 0.0313997)
      (Radboud-L $\rightarrow$\\ Radboud-R, 0.375735) +- (0, 0.00683095)};
  \end{axis}
  \end{tikzpicture}\vspace{-2.5mm}
    \begin{tikzpicture}
  \centering
  \begin{axis}[
        ybar, axis on top,
        title={},
        height=4.525cm, width=18.0cm, 
        bar width=0.35cm,
        ymajorgrids, tick align=inside,
        enlarge y limits={value=.1,upper},
        ymin=0, ymax=0.9,
        set layers,
        ytick={0.2, 0.4, 0.6, 0.8},
        axis x line*=bottom,
        axis y line*=left,
        y axis line style={opacity=0},
        tickwidth=0pt,
        enlarge x limits=true,
        extra y tick style={
            grid style={
                solid,
                line width = 2.5pt,
                /pgfplots/on layer=axis background,
            },
        },
        tick style={
            grid style={
                dashed,
                /pgfplots/on layer=axis background,
            },
        },
        legend style={
            at={(0.5,-0.2)},
            anchor=north,
            legend columns=-1,
            /tikz/every even column/.append style={column sep=0.325cm},
            font=\footnotesize
        },
        ylabel={Kappa ($\uparrow$)},
        ylabel style={font=\footnotesize},
        symbolic x coords={
           PANDA,
           Karolinska,
           Radboud, 
           Radboud $\rightarrow$\\ Karolinska,
           Radboud-U, 
           Radboud-U $\rightarrow$\\ Karolinska-U,
           Radboud-L, 
           Radboud-L $\rightarrow$\\ Radboud-R},
       xtick=data,
       xticklabel=\empty,
       x tick label style={text width = 2.25cm, align=center, font=\footnotesize},
    ]
    \addplot [fill=skyblue2, draw=skyblue2, line width=0mm, error bars/.cd, y dir=both, y explicit, error bar style={thick}, error bar style=black] coordinates { 
      (PANDA, 0.8862) +- (0, 0.0116586)
      (Karolinska, 0.874592) +- (0, 0.017091)
      (Radboud, 0.857914) +- (0, 0.0195372)
      (Radboud $\rightarrow$\\ Karolinska, 0.459026) +- (0, 0.0749562)
      (Radboud-U, 0.842319) +- (0, 0.023979)
      (Radboud-U $\rightarrow$\\ Karolinska-U, 0.675557) +- (0, 0.0598927)
      (Radboud-L, 0.835485) +- (0, 0.0248449)
      (Radboud-L $\rightarrow$\\ Radboud-R, 0.72981) +- (0, 0.012295)};
    \addplot [fill=caribbeangreen2, draw=caribbeangreen2, error bars/.cd, y dir=both, y explicit, error bar style={thick}, error bar style=black] coordinates {
      (PANDA, 0.832176) +- (0, 0.0189604)
      (Karolinska, 0.750182) +- (0, 0.0345796)
      (Radboud, 0.83313) +- (0, 0.0140878)
      (Radboud $\rightarrow$\\ Karolinska, 0.416334) +- (0, 0.066083)
      (Radboud-U, 0.815392) +- (0, 0.0404605)
      (Radboud-U $\rightarrow$\\ Karolinska-U, 0.472304) +- (0, 0.0337215)
      (Radboud-L, 0.802815) +- (0, 0.023774)
      (Radboud-L $\rightarrow$\\ Radboud-R, 0.682357) +- (0, 0.00911505)};
    \addplot [fill=darkpastelpurple2, draw=darkpastelpurple2, error bars/.cd, y dir=both, y explicit, error bar style={thick}, error bar style=black] coordinates {
      (PANDA, 0.721069) +- (0, 0.0251075)
      (Karolinska, 0.567143) +- (0, 0.0311424)
      (Radboud, 0.744075) +- (0, 0.0198549)
      (Radboud $\rightarrow$\\ Karolinska, 0.228048) +- (0, 0.0367065)
      (Radboud-U, 0.711507) +- (0, 0.0339886)
      (Radboud-U $\rightarrow$\\ Karolinska-U, 0.328888) +- (0, 0.0249578)
      (Radboud-L, 0.652109) +- (0, 0.0390751)
      (Radboud-L $\rightarrow$\\ Radboud-R, 0.379005) +- (0, 0.00637403)};
  \end{axis}
  \end{tikzpicture}\vspace{-2.5mm}
    \begin{tikzpicture}
  \centering
  \begin{axis}[
        ybar, axis on top,
        title={},
        height=4.525cm, width=18.0cm, 
        bar width=0.35cm,
        ymajorgrids, tick align=inside,
        enlarge y limits={value=.1,upper},
        ymin=0, ymax=0.9,
        set layers,
        ytick={0.2, 0.4, 0.6, 0.8},
        axis x line*=bottom,
        axis y line*=left,
        y axis line style={opacity=0},
        tickwidth=0pt,
        enlarge x limits=true,
        extra y tick style={
            grid style={
                solid,
                line width = 2.5pt,
                /pgfplots/on layer=axis background,
            },
        },
        tick style={
            grid style={
                dashed,
                /pgfplots/on layer=axis background,
            },
        },
        legend style={
            at={(0.5,-0.30)},
            anchor=north,
            legend columns=-1,
            /tikz/every even column/.append style={column sep=0.325cm},
            font=\footnotesize
        },
        ylabel={Kappa ($\uparrow$)},
        ylabel style={font=\footnotesize},
        symbolic x coords={
           PANDA,
           Karolinska,
           Radboud, 
           Radboud $\rightarrow$\\ Karolinska,
           Radboud-U, 
           Radboud-U $\rightarrow$\\ Karolinska-U,
           Radboud-L, 
           Radboud-L $\rightarrow$\\ Radboud-R},
       xtick=data,
       x tick label style={text width = 2.25cm, align=center, font=\footnotesize},
    ]
    \addplot [fill=skyblue2, draw=skyblue2, line width=0mm, error bars/.cd, y dir=both, y explicit, error bar style={thick}, error bar style=black] coordinates { 
      (PANDA, 0.867677) +- (0, 0.0135075)
      (Karolinska, 0.863586) +- (0, 0.0270691)
      (Radboud, 0.838156) +- (0, 0.0233552)
      (Radboud $\rightarrow$\\ Karolinska, 0.61698) +- (0, 0.0590592)
      (Radboud-U, 0.829841) +- (0, 0.024966)
      (Radboud-U $\rightarrow$\\ Karolinska-U, 0.66363) +- (0, 0.0559497)
      (Radboud-L, 0.832021) +- (0, 0.0337764)
      (Radboud-L $\rightarrow$\\ Radboud-R, 0.720456) +- (0, 0.0152852)};
    \addplot [fill=caribbeangreen2, draw=caribbeangreen2, error bars/.cd, y dir=both, y explicit, error bar style={thick}, error bar style=black] coordinates {
      (PANDA, 0.808804) +- (0, 0.0201696)
      (Karolinska, 0.747719) +- (0, 0.031465)
      (Radboud, 0.791368) +- (0, 0.0292407)
      (Radboud $\rightarrow$\\ Karolinska, 0.340052) +- (0, 0.0686435)
      (Radboud-U, 0.782769) +- (0, 0.0274152)
      (Radboud-U $\rightarrow$\\ Karolinska-U, 0.349537) +- (0, 0.0430824)
      (Radboud-L, 0.784613) +- (0, 0.0266161)
      (Radboud-L $\rightarrow$\\ Radboud-R, 0.640242) +- (0, 0.00984075)};
    \addplot [fill=darkpastelpurple2, draw=darkpastelpurple2, error bars/.cd, y dir=both, y explicit, error bar style={thick}, error bar style=black] coordinates {
      (PANDA, 0.674131) +- (0, 0.0188681)
      (Karolinska, 0.555084) +- (0, 0.0419075)
      (Radboud, 0.672634) +- (0, 0.0195404)
      (Radboud $\rightarrow$\\ Karolinska, 0.025) +- (0, 0.0644499)
      (Radboud-U, 0.644478) +- (0, 0.0315855)
      (Radboud-U $\rightarrow$\\ Karolinska-U, 0.215219) +- (0, 0.0373707)
      (Radboud-L, 0.593542) +- (0, 0.0519831) 
      (Radboud-L $\rightarrow$\\ Radboud-R, 0.35055) +- (0, 0.00817425)};
    \legend{
            \textit{ABMIL},
            \textit{Mean Feature},
            \textit{kNN}
            }
  \end{axis}
  \end{tikzpicture}\vspace{-1.0mm}
    \caption{\textbf{Comparison of downstream ISUP grade classification models for additional feature extractors.} Performance of the three ISUP grade models \textit{ABMIL}, \textit{Mean Feature}, and \textit{kNN} when using different feature extractors: Phikon-v2 \textbf{(top row)}, UNI2-h, H-optimus-0, H0-mini, Prov-GigaPath, and Virchow \textbf{(bottom row)}. All results are reported as mean $\pm$ standard deviation over $10$ cross-validation folds. This is the same performance comparison as in Figure~\ref{fig:main_results_attmil_mean-pool_knn_kappa}, but for the remaining six feature extractors.}
    \label{fig:main_results_attmil_mean-pool_knn_kappa_remaining6}
\end{figure*}

\clearpage
\begin{figure*}[t]
    \centering
    \begin{tikzpicture}
  \centering
  \begin{axis}[
        ybar=2.5pt, axis on top,
        title={},
        height=4.925cm, width=23.0cm, 
        bar width=0.26cm, 
        ymajorgrids, tick align=inside,
        enlarge y limits={value=.1,upper},
        ymin=0, ymax=2.4,
        set layers,
        ytick={0.5, 1.0, 1.5, 2.0},
        axis x line*=bottom,
        axis y line*=left,
        y axis line style={opacity=0},
        tickwidth=0pt,
        enlarge x limits=0.20, 
        extra y tick style={
            grid style={
                solid,
                line width = 2.5pt,
                /pgfplots/on layer=axis background,
            },
        },
        tick style={
            grid style={
                dashed,
                /pgfplots/on layer=axis background,
            },
        },
        legend style={
            at={(0.5,-0.2)},
            anchor=north,
            legend columns=-1,
            /tikz/every even column/.append style={column sep=0.325cm},
            font=\footnotesize
        },
        ylabel={MAE ($\downarrow$)},
        ylabel style={font=\footnotesize},
        symbolic x coords={
           PANDA,
           Karolinska,
           Radboud, 
           Radboud $\rightarrow$ Karolinska},
       xtick=data,
       xticklabel=\empty,
       x tick label style={text width = 2.25cm, align=center, font=\footnotesize},
    ]
    \addplot [fill=OrangeRed2, draw=OrangeRed2, error bars/.cd, y dir=both, y explicit, error bar style={thick}, error bar style=black] coordinates {
      (PANDA, 0.621935) +- (0, 0.0233373)
      (Karolinska, 0.565257) +- (0, 0.037827)
      (Radboud, 0.7762) +- (0, 0.0422512)
      (Radboud $\rightarrow$ Karolinska, 1.55543) +- (0, 0.10277)};

    \addplot [fill=darkgray, draw=darkgray, error bars/.cd, y dir=both, y explicit, error bar style={thick}, error bar style=black] coordinates {
      (PANDA, 0.370307) +- (0, 0.0177045)
      (Karolinska, 0.307721) +- (0, 0.01665)
      (Radboud, 0.482) +- (0, 0.0242157)
      (Radboud $\rightarrow$ Karolinska, 1.45491) +- (0, 0.194843)};
    
    \addplot [fill=flamingopink, draw=flamingopink, line width=0mm, error bars/.cd, y dir=both, y explicit, error bar style={thick}, error bar style=black] coordinates { 
      (PANDA, 0.346935) +- (0, 0.0350706)
      (Karolinska, 0.299081) +- (0, 0.0178043)
      (Radboud, 0.4238) +- (0, 0.0450107)
      (Radboud $\rightarrow$ Karolinska, 1.52554) +- (0, 0.369422)};

    \addplot [fill=red-violet, draw=red-violet, line width=0mm, error bars/.cd, y dir=both, y explicit, error bar style={thick}, error bar style=black] coordinates { 
      (PANDA, 0.354693) +- (0, 0.0269327)
      (Karolinska, 0.298529) +- (0, 0.0272555)
      (Radboud, 0.439) +- (0, 0.0293837)
      (Radboud $\rightarrow$ Karolinska, 1.20905) +- (0, 0.220848)};

    \addplot [fill=caribbeangreen2, draw=caribbeangreen2, error bars/.cd, y dir=both, y explicit, error bar style={thick}, error bar style=black] coordinates {
      (PANDA, 0.404981) +- (0, 0.0184436)
      (Karolinska, 0.326287) +- (0, 0.0255012)
      (Radboud, 0.5378) +- (0, 0.0416793)
      (Radboud $\rightarrow$ Karolinska, 2.26281) +- (0, 0.108598)};

    \addplot [fill=ForestGreen, draw=ForestGreen, error bars/.cd, y dir=both, y explicit, error bar style={thick}, error bar style=black] coordinates {
      (PANDA, 0.431513) +- (0, 0.019965)
      (Karolinska, 0.334007) +- (0, 0.0308369)
      (Radboud, 0.5574) +- (0, 0.0330702)
      (Radboud $\rightarrow$ Karolinska, 0.572617) +- (0, 0.0595431)};

    \addplot [fill=skyblue2, draw=skyblue2, error bars/.cd, y dir=both, y explicit, error bar style={thick}, error bar style=black] coordinates {
      (PANDA, 0.333238) +- (0, 0.0295276)
      (Karolinska, 0.292831) +- (0, 0.0362095)
      (Radboud, 0.3952) +- (0, 0.045845)
      (Radboud $\rightarrow$ Karolinska, 0.877604) +- (0, 0.151029)};

    \addplot [fill=Cerulean, draw=Cerulean, line width=0mm, error bars/.cd, y dir=both, y explicit, error bar style={thick}, error bar style=black] coordinates { 
      (PANDA, 0.339368) +- (0, 0.0302417)
      (Karolinska, 0.29761) +- (0, 0.023215)
      (Radboud, 0.3896) +- (0, 0.0614479)
      (Radboud $\rightarrow$ Karolinska, 0.94078) +- (0, 0.155075)};

    \addplot [fill=darkcerulean2, draw=darkcerulean2, line width=0mm, error bars/.cd, y dir=both, y explicit, error bar style={thick}, error bar style=black] coordinates { 
      (PANDA, 0.340613) +- (0, 0.0277057)
      (Karolinska, 0.298897) +- (0, 0.0278928)
      (Radboud, 0.4146) +- (0, 0.0324968)
      (Radboud $\rightarrow$ Karolinska, 0.997368) +- (0, 0.115618)};

    \addplot [fill=darkpastelpurple2, draw=darkpastelpurple2, error bars/.cd, y dir=both, y explicit, error bar style={thick}, error bar style=black] coordinates {
      (PANDA, 0.345881) +- (0, 0.0244823)
      (Karolinska, 0.309191) +- (0, 0.0349052)
      (Radboud, 0.4102) +- (0, 0.0397638)
      (Radboud $\rightarrow$ Karolinska, 0.957527) +- (0, 0.118647)};

    \addplot [fill=goldenpoppy, draw=goldenpoppy, error bars/.cd, y dir=both, y explicit, error bar style={thick}, error bar style=black] coordinates {
      (PANDA, 0.393966) +- (0, 0.0235252)
      (Karolinska, 0.323346) +- (0, 0.0282688)
      (Radboud, 0.467) +- (0, 0.039792)
      (Radboud $\rightarrow$ Karolinska, 0.780401) +- (0, 0.0871312)};

    \addplot [fill=goldenpoppy3, draw=goldenpoppy3, error bars/.cd, y dir=both, y explicit, error bar style={thick}, error bar style=black] coordinates {
      (PANDA, 0.378161) +- (0, 0.0220922)
      (Karolinska, 0.302022) +- (0, 0.0251247)
      (Radboud, 0.459) +- (0, 0.0350457)
      (Radboud $\rightarrow$ Karolinska, 0.892455) +- (0, 0.227418)};
  \end{axis}
  \end{tikzpicture}\vspace{-3.0mm}
    \begin{tikzpicture}
  \centering
  \begin{axis}[
        ybar=2.5pt, axis on top,
        title={},
        height=4.925cm, width=23.0cm, 
        bar width=0.26cm, 
        ymajorgrids, tick align=inside,
        enlarge y limits={value=.1,upper},
        ymin=0, ymax=2.4,
        set layers,
        ytick={0.5, 1.0, 1.5, 2.0},
        axis x line*=bottom,
        axis y line*=left,
        y axis line style={opacity=0},
        tickwidth=0pt,
        enlarge x limits=0.20, 
        extra y tick style={
            grid style={
                solid,
                line width = 2.5pt,
                /pgfplots/on layer=axis background,
            },
        },
        tick style={
            grid style={
                dashed,
                /pgfplots/on layer=axis background,
            },
        },
        legend style={
            at={(0.5,-0.2)},
            anchor=north,
            legend columns=-1,
            /tikz/every even column/.append style={column sep=0.325cm},
            font=\footnotesize
        },
        ylabel={MAE ($\downarrow$)},
        ylabel style={font=\footnotesize},
        symbolic x coords={
           PANDA,
           Karolinska,
           Radboud, 
           Radboud $\rightarrow$ Karolinska},
       xtick=data,
       xticklabel=\empty,
       x tick label style={text width = 2.25cm, align=center, font=\footnotesize},
    ]
    \addplot [fill=OrangeRed2, draw=OrangeRed2, error bars/.cd, y dir=both, y explicit, error bar style={thick}, error bar style=black] coordinates {
      (PANDA, 0.975096) +- (0, 0.052894)
      (Karolinska, 1.10018) +- (0, 0.0761271)
      (Radboud, 1.128) +- (0, 0.0341292)
      (Radboud $\rightarrow$ Karolinska, 1.50294) +- (0, 0.0886444)};

    \addplot [fill=darkgray, draw=darkgray, error bars/.cd, y dir=both, y explicit, error bar style={thick}, error bar style=black] coordinates {
      (PANDA, 0.567337) +- (0, 0.03535)
      (Karolinska, 0.560846) +- (0, 0.0414843)
      (Radboud, 0.6158) +- (0, 0.042374)
      (Radboud $\rightarrow$ Karolinska, 1.41492) +- (0, 0.166942)};
    
    \addplot [fill=flamingopink, draw=flamingopink, line width=0mm, error bars/.cd, y dir=both, y explicit, error bar style={thick}, error bar style=black] coordinates { 
      (PANDA, 0.434579) +- (0, 0.0323268)
      (Karolinska, 0.496875) +- (0, 0.0336758)
      (Radboud, 0.4228) +- (0, 0.0318773)
      (Radboud $\rightarrow$ Karolinska, 0.972488) +- (0, 0.124728)};

    \addplot [fill=red-violet, draw=red-violet, line width=0mm, error bars/.cd, y dir=both, y explicit, error bar style={thick}, error bar style=black] coordinates { 
      (PANDA, 0.46159) +- (0, 0.0237712)
      (Karolinska, 0.5125) +- (0, 0.0438395)
      (Radboud, 0.4706) +- (0, 0.0449893)
      (Radboud $\rightarrow$ Karolinska, 0.979536) +- (0, 0.0627807)};

    \addplot [fill=caribbeangreen2, draw=caribbeangreen2, error bars/.cd, y dir=both, y explicit, error bar style={thick}, error bar style=black] coordinates {
      (PANDA, 0.620019) +- (0, 0.0276654)
      (Karolinska, 0.623897) +- (0, 0.0323884)
      (Radboud, 0.7124) +- (0, 0.039717)
      (Radboud $\rightarrow$ Karolinska, 2.2115) +- (0, 0.145679)};

    \addplot [fill=ForestGreen, draw=ForestGreen, error bars/.cd, y dir=both, y explicit, error bar style={thick}, error bar style=black] coordinates {
      (PANDA, 0.655268) +- (0, 0.0332789)
      (Karolinska, 0.62886) +- (0, 0.0400081)
      (Radboud, 0.7578) +- (0, 0.0367418)
      (Radboud $\rightarrow$ Karolinska, 0.807177) +- (0, 0.0288316)};

    \addplot [fill=skyblue2, draw=skyblue2, error bars/.cd, y dir=both, y explicit, error bar style={thick}, error bar style=black] coordinates {
      (PANDA, 0.435153) +- (0, 0.0276223)
      (Karolinska, 0.478309) +- (0, 0.0302478)
      (Radboud, 0.4524) +- (0, 0.0317843)
      (Radboud $\rightarrow$ Karolinska, 0.973703) +- (0, 0.0515132)};

    \addplot [fill=Cerulean, draw=Cerulean, line width=0mm, error bars/.cd, y dir=both, y explicit, error bar style={thick}, error bar style=black] coordinates { 
      (PANDA, 0.444061) +- (0, 0.0274729)
      (Karolinska, 0.47739) +- (0, 0.0243322)
      (Radboud, 0.457) +- (0, 0.0342374)
      (Radboud $\rightarrow$ Karolinska, 0.767777) +- (0, 0.0247762)};

    \addplot [fill=darkcerulean2, draw=darkcerulean2, line width=0mm, error bars/.cd, y dir=both, y explicit, error bar style={thick}, error bar style=black] coordinates { 
      (PANDA, 0.493966) +- (0, 0.0351653)
      (Karolinska, 0.523713) +- (0, 0.03708)
      (Radboud, 0.5228) +- (0, 0.0428504)
      (Radboud $\rightarrow$ Karolinska, 0.843614) +- (0, 0.0270459)};

    \addplot [fill=darkpastelpurple2, draw=darkpastelpurple2, error bars/.cd, y dir=both, y explicit, error bar style={thick}, error bar style=black] coordinates {
      (PANDA, 0.459866) +- (0, 0.0294934)
      (Karolinska, 0.537132) +- (0, 0.0380013)
      (Radboud, 0.4484) +- (0, 0.0324074)
      (Radboud $\rightarrow$ Karolinska, 1.05841) +- (0, 0.137268)};

    \addplot [fill=goldenpoppy, draw=goldenpoppy, error bars/.cd, y dir=both, y explicit, error bar style={thick}, error bar style=black] coordinates {
      (PANDA, 0.543295) +- (0, 0.0367517)
      (Karolinska, 0.559191) +- (0, 0.0422058)
      (Radboud, 0.5892) +- (0, 0.0536522)
      (Radboud $\rightarrow$ Karolinska, 1.15655) +- (0, 0.103995)};

    \addplot [fill=goldenpoppy3, draw=goldenpoppy3, error bars/.cd, y dir=both, y explicit, error bar style={thick}, error bar style=black] coordinates {
      (PANDA, 0.487452) +- (0, 0.0255819)
      (Karolinska, 0.496507) +- (0, 0.0275423)
      (Radboud, 0.5376) +- (0, 0.0371139)
      (Radboud $\rightarrow$ Karolinska, 1.66229) +- (0, 0.24416)};
  \end{axis}
  \end{tikzpicture}\vspace{-3.5mm}
    \begin{tikzpicture}
  \centering
  \begin{axis}[
        ybar=2.5pt, axis on top,
        title={},
        height=4.925cm, width=23.0cm, 
        bar width=0.26cm, 
        ymajorgrids, tick align=inside,
        enlarge y limits={value=.1,upper},
        ymin=0, ymax=2.4,
        set layers,
        ytick={0.5, 1.0, 1.5, 2.0},
        axis x line*=bottom,
        axis y line*=left,
        y axis line style={opacity=0},
        tickwidth=0pt,
        enlarge x limits=0.20, 
        extra y tick style={
            grid style={
                solid,
                line width = 2.5pt,
                /pgfplots/on layer=axis background,
            },
        },
        tick style={
            grid style={
                dashed,
                /pgfplots/on layer=axis background,
            },
        },
        legend style={
            at={(0.5,-0.225)},
            anchor=north,
            legend columns=-1,
            /tikz/every even column/.append style={column sep=0.175cm},
            font=\footnotesize
        },
        ylabel={MAE ($\downarrow$)},
        ylabel style={font=\footnotesize},
        symbolic x coords={
           PANDA,
           Karolinska,
           Radboud, 
           Radboud $\rightarrow$ Karolinska},
       xtick=data,
       x tick label style={text width = 5.0cm, align=center, font={\bfseries}},
    ]
    \addplot [fill=OrangeRed2, draw=OrangeRed2, error bars/.cd, y dir=both, y explicit, error bar style={thick}, error bar style=black] coordinates {
      (PANDA, 0.864464) +- (0, 0.0243048)
      (Karolinska, 0.845037) +- (0, 0.0344692)
      (Radboud, 0.8698) +- (0, 0.0669235)
      (Radboud $\rightarrow$ Karolinska, 1.97403) +- (0, 0.10194)};

    \addplot [fill=darkgray, draw=darkgray, error bars/.cd, y dir=both, y explicit, error bar style={thick}, error bar style=black] coordinates {
      (PANDA, 0.736686) +- (0, 0.0257677)
      (Karolinska, 0.751471) +- (0, 0.0254554)
      (Radboud, 0.6716) +- (0, 0.0350063)
      (Radboud $\rightarrow$ Karolinska, 1.18147) +- (0, 0.131297)};
    
    \addplot [fill=flamingopink, draw=flamingopink, line width=0mm, error bars/.cd, y dir=both, y explicit, error bar style={thick}, error bar style=black] coordinates { 
      (PANDA, 0.67069) +- (0, 0.0272098)
      (Karolinska, 0.730331) +- (0, 0.0267405)
      (Radboud, 0.6302) +- (0, 0.0519111)
      (Radboud $\rightarrow$ Karolinska, 1.28081) +- (0, 0.0565518)};

    \addplot [fill=red-violet, draw=red-violet, line width=0mm, error bars/.cd, y dir=both, y explicit, error bar style={thick}, error bar style=black] coordinates { 
      (PANDA, 0.685249) +- (0, 0.0267624)
      (Karolinska, 0.750551) +- (0, 0.0350043)
      (Radboud, 0.6054) +- (0, 0.0601269)
      (Radboud $\rightarrow$ Karolinska, 1.58796) +- (0, 0.0747343)};

    \addplot [fill=caribbeangreen2, draw=caribbeangreen2, error bars/.cd, y dir=both, y explicit, error bar style={thick}, error bar style=black] coordinates {
      (PANDA, 0.755077) +- (0, 0.026295)
      (Karolinska, 0.727941) +- (0, 0.0260615)
      (Radboud, 0.7634) +- (0, 0.0659639)
      (Radboud $\rightarrow$ Karolinska, 1.46491) +- (0, 0.0411191)};

    \addplot [fill=ForestGreen, draw=ForestGreen, error bars/.cd, y dir=both, y explicit, error bar style={thick}, error bar style=black] coordinates {
      (PANDA, 0.730843) +- (0, 0.0263401)
      (Karolinska, 0.698346) +- (0, 0.024407)
      (Radboud, 0.7658) +- (0, 0.0728475)
      (Radboud $\rightarrow$ Karolinska, 1.25585) +- (0, 0.0251673)};

    \addplot [fill=skyblue2, draw=skyblue2, error bars/.cd, y dir=both, y explicit, error bar style={thick}, error bar style=black] coordinates {
      (PANDA, 0.682759) +- (0, 0.0208926)
      (Karolinska, 0.726103) +- (0, 0.0359562)
      (Radboud, 0.6148) +- (0, 0.0685519)
      (Radboud $\rightarrow$ Karolinska, 1.14021) +- (0, 0.0152437)};

    \addplot [fill=Cerulean, draw=Cerulean, line width=0mm, error bars/.cd, y dir=both, y explicit, error bar style={thick}, error bar style=black] coordinates { 
      (PANDA, 0.668487) +- (0, 0.027467)
      (Karolinska, 0.707904) +- (0, 0.0320344)
      (Radboud, 0.6122) +- (0, 0.0632737)
      (Radboud $\rightarrow$ Karolinska, 1.08831) +- (0, 0.0276389)};

    \addplot [fill=darkcerulean2, draw=darkcerulean2, line width=0mm, error bars/.cd, y dir=both, y explicit, error bar style={thick}, error bar style=black] coordinates { 
      (PANDA, 0.73908) +- (0, 0.0302057)
      (Karolinska, 0.780331) +- (0, 0.0361553)
      (Radboud, 0.7012) +- (0, 0.0686801)
      (Radboud $\rightarrow$ Karolinska, 1.09262) +- (0, 0.0126168)};

    \addplot [fill=darkpastelpurple2, draw=darkpastelpurple2, error bars/.cd, y dir=both, y explicit, error bar style={thick}, error bar style=black] coordinates {
      (PANDA, 0.680077) +- (0, 0.0367422)
      (Karolinska, 0.742647) +- (0, 0.0282275)
      (Radboud, 0.6262) +- (0, 0.0420328)
      (Radboud $\rightarrow$ Karolinska, 1.3429) +- (0, 0.0864625)};

    \addplot [fill=goldenpoppy, draw=goldenpoppy, error bars/.cd, y dir=both, y explicit, error bar style={thick}, error bar style=black] coordinates {
      (PANDA, 0.773372) +- (0, 0.0314381)
      (Karolinska, 0.765257) +- (0, 0.0312073)
      (Radboud, 0.7968) +- (0, 0.0416)
      (Radboud $\rightarrow$ Karolinska, 1.96327) +- (0, 0.0948864)};

    \addplot [fill=goldenpoppy3, draw=goldenpoppy3, error bars/.cd, y dir=both, y explicit, error bar style={thick}, error bar style=black] coordinates {
      (PANDA, 0.742625) +- (0, 0.0201607)
      (Karolinska, 0.723162) +- (0, 0.0265012)
      (Radboud, 0.744) +- (0, 0.0564978)
      (Radboud $\rightarrow$ Karolinska, 1.11205) +- (0, 0.0130528)};
    \legend{
            Resnet-IN,
            Phikon-v2,
            UNI,
            UNI2-h,
            CONCH,
            CONCHv1.5,
            H-optimus-0,
            H-optimus-1,
            H0-mini,
            Prov-GigaPath,
            Virchow,
            Virchow2
            }
  \end{axis}
  \end{tikzpicture}\vspace{-1.0mm}
    \caption{\textbf{Main model comparison across PANDA and its primary subsets (MAE).} Performance of all twelve evaluated models across the full PANDA dataset and its two site-specific subsets (Karolinska, Radboud), when used as frozen patch-level feature extractors in the \textit{ABMIL} \textbf{(top)}, \textit{Mean Feature} \textbf{(middle)}, and \textit{kNN} \textbf{(bottom)} ISUP grade classification models. All results are reported as mean $\pm$ standard deviation over $10$ cross-validation folds. This figure presents the same experiments as Figure~\ref{fig:main_results3a_12fms_kappa}, but using mean absolute error (MAE, lower is better) instead of quadratically weighted Cohen's kappa.}
    \label{fig:main_results3a_12fms_mae}
\end{figure*}
\begin{figure*}[t]
    \centering
    \begin{tikzpicture}
  \centering
  \begin{axis}[
        ybar=2.5pt, axis on top,
        title={},
        height=4.925cm, width=23.0cm, 
        bar width=0.26cm, 
        ymajorgrids, tick align=inside,
        enlarge y limits={value=.1,upper},
        ymin=0, ymax=2.4,
        set layers,
        ytick={0.5, 1.0, 1.5, 2.0},
        axis x line*=bottom,
        axis y line*=left,
        y axis line style={opacity=0},
        tickwidth=0pt,
        enlarge x limits=0.20, 
        extra y tick style={
            grid style={
                solid,
                line width = 2.5pt,
                /pgfplots/on layer=axis background,
            },
        },
        tick style={
            grid style={
                dashed,
                /pgfplots/on layer=axis background,
            },
        },
        legend style={
            at={(0.5,-0.2)},
            anchor=north,
            legend columns=-1,
            /tikz/every even column/.append style={column sep=0.325cm},
            font=\footnotesize
        },
        ylabel={MAE ($\downarrow$)},
        ylabel style={font=\footnotesize},
        symbolic x coords={
           Radboud-U, 
           Radboud-U $\rightarrow$ Karolinska-U,
           Radboud-L, 
           Radboud-L $\rightarrow$ Radboud-R},
       xtick=data,
       xticklabel=\empty,
       x tick label style={text width = 2.25cm, align=center, font=\footnotesize},
    ]
    \addplot [fill=OrangeRed2, draw=OrangeRed2, error bars/.cd, y dir=both, y explicit, error bar style={thick}, error bar style=black] coordinates {
      (Radboud-U, 0.780348) +- (0, 0.0283855)
      (Radboud-U $\rightarrow$ Karolinska-U, 1.29024) +- (0, 0.0282161)
      (Radboud-L, 0.697189) +- (0, 0.0465656)
      (Radboud-L $\rightarrow$ Radboud-R, 1.08342) +- (0, 0.0449268)};

    \addplot [fill=darkgray, draw=darkgray, error bars/.cd, y dir=both, y explicit, error bar style={thick}, error bar style=black] coordinates {
      (Radboud-U, 0.492786) +- (0, 0.0522204)
      (Radboud-U $\rightarrow$ Karolinska-U, 1.05246) +- (0, 0.074783)
      (Radboud-L, 0.46988) +- (0, 0.046001)
      (Radboud-L $\rightarrow$ Radboud-R, 0.701869) +- (0, 0.0294566)};
    
    \addplot [fill=flamingopink, draw=flamingopink, line width=0mm, error bars/.cd, y dir=both, y explicit, error bar style={thick}, error bar style=black] coordinates { 
      (Radboud-U, 0.447264) +- (0, 0.0397512)
      (Radboud-U $\rightarrow$ Karolinska-U, 1.15465) +- (0, 0.169618)
      (Radboud-L, 0.455422) +- (0, 0.0500705)
      (Radboud-L $\rightarrow$ Radboud-R, 0.663022) +- (0, 0.0320972)};

    \addplot [fill=red-violet, draw=red-violet, line width=0mm, error bars/.cd, y dir=both, y explicit, error bar style={thick}, error bar style=black] coordinates { 
      (Radboud-U, 0.443035) +- (0, 0.0268484)
      (Radboud-U $\rightarrow$ Karolinska-U, 1.04548) +- (0, 0.0756997)
      (Radboud-L, 0.428514) +- (0, 0.03389)
      (Radboud-L $\rightarrow$ Radboud-R, 0.679046) +- (0, 0.0406002)};

    \addplot [fill=caribbeangreen2, draw=caribbeangreen2, error bars/.cd, y dir=both, y explicit, error bar style={thick}, error bar style=black] coordinates {
      (Radboud-U, 0.546269) +- (0, 0.0449994)
      (Radboud-U $\rightarrow$ Karolinska-U, 1.4664) +- (0, 0.112544)
      (Radboud-L, 0.49759) +- (0, 0.0626158)
      (Radboud-L $\rightarrow$ Radboud-R, 0.726441) +- (0, 0.0344725)};

    \addplot [fill=ForestGreen, draw=ForestGreen, error bars/.cd, y dir=both, y explicit, error bar style={thick}, error bar style=black] coordinates {
      (Radboud-U, 0.565423) +- (0, 0.0445135)
      (Radboud-U $\rightarrow$ Karolinska-U, 0.810757) +- (0, 0.0351173)
      (Radboud-L, 0.500803) +- (0, 0.0484947)
      (Radboud-L $\rightarrow$ Radboud-R, 0.761511) +- (0, 0.0296167)};

    \addplot [fill=skyblue2, draw=skyblue2, error bars/.cd, y dir=both, y explicit, error bar style={thick}, error bar style=black] coordinates {
      (Radboud-U, 0.437313) +- (0, 0.05033)
      (Radboud-U $\rightarrow$ Karolinska-U, 0.875697) +- (0, 0.0546145)
      (Radboud-L, 0.423293) +- (0, 0.037554)
      (Radboud-L $\rightarrow$ Radboud-R, 0.667356) +- (0, 0.0480552)};

    \addplot [fill=Cerulean, draw=Cerulean, line width=0mm, error bars/.cd, y dir=both, y explicit, error bar style={thick}, error bar style=black] coordinates { 
      (Radboud-U, 0.420398) +- (0, 0.027476)
      (Radboud-U $\rightarrow$ Karolinska-U, 0.886587) +- (0, 0.0894473)
      (Radboud-L, 0.415261) +- (0, 0.0570281)
      (Radboud-L $\rightarrow$ Radboud-R, 0.653837) +- (0, 0.0421783)};

    \addplot [fill=darkcerulean2, draw=darkcerulean2, line width=0mm, error bars/.cd, y dir=both, y explicit, error bar style={thick}, error bar style=black] coordinates { 
      (Radboud-U, 0.426617) +- (0, 0.0438439)
      (Radboud-U $\rightarrow$ Karolinska-U, 0.888977) +- (0, 0.026325)
      (Radboud-L, 0.451406) +- (0, 0.0710333)
      (Radboud-L $\rightarrow$ Radboud-R, 0.678887) +- (0, 0.039581)};

    \addplot [fill=darkpastelpurple2, draw=darkpastelpurple2, error bars/.cd, y dir=both, y explicit, error bar style={thick}, error bar style=black] coordinates {
      (Radboud-U, 0.460448) +- (0, 0.0322511)
      (Radboud-U $\rightarrow$ Karolinska-U, 0.843094) +- (0, 0.080393)
      (Radboud-L, 0.436546) +- (0, 0.0495801)
      (Radboud-L $\rightarrow$ Radboud-R, 0.677416) +- (0, 0.0264209)};

    \addplot [fill=goldenpoppy, draw=goldenpoppy, error bars/.cd, y dir=both, y explicit, error bar style={thick}, error bar style=black] coordinates {
      (Radboud-U, 0.502488) +- (0, 0.0466045)
      (Radboud-U $\rightarrow$ Karolinska-U, 0.864608) +- (0, 0.0806058)
      (Radboud-L, 0.452209) +- (0, 0.0672544)
      (Radboud-L $\rightarrow$ Radboud-R, 0.691372) +- (0, 0.0433038)};

    \addplot [fill=goldenpoppy3, draw=goldenpoppy3, error bars/.cd, y dir=both, y explicit, error bar style={thick}, error bar style=black] coordinates {
      (Radboud-U, 0.490796) +- (0, 0.0538722)
      (Radboud-U $\rightarrow$ Karolinska-U, 0.831939) +- (0, 0.0712366)
      (Radboud-L, 0.483133) +- (0, 0.0371155)
      (Radboud-L $\rightarrow$ Radboud-R, 0.699205) +- (0, 0.039343)};
  \end{axis}
  \end{tikzpicture}\vspace{-3.0mm}
    \begin{tikzpicture}
  \centering
  \begin{axis}[
        ybar=2.5pt, axis on top,
        title={},
        height=4.925cm, width=23.0cm, 
        bar width=0.26cm, 
        ymajorgrids, tick align=inside,
        enlarge y limits={value=.1,upper},
        ymin=0, ymax=2.4,
        set layers,
        ytick={0.5, 1.0, 1.5, 2.0},
        axis x line*=bottom,
        axis y line*=left,
        y axis line style={opacity=0},
        tickwidth=0pt,
        enlarge x limits=0.20, 
        extra y tick style={
            grid style={
                solid,
                line width = 2.5pt,
                /pgfplots/on layer=axis background,
            },
        },
        tick style={
            grid style={
                dashed,
                /pgfplots/on layer=axis background,
            },
        },
        legend style={
            at={(0.5,-0.2)},
            anchor=north,
            legend columns=-1,
            /tikz/every even column/.append style={column sep=0.325cm},
            font=\footnotesize
        },
        ylabel={MAE ($\downarrow$)},
        ylabel style={font=\footnotesize},
        symbolic x coords={
           Radboud-U, 
           Radboud-U $\rightarrow$ Karolinska-U,
           Radboud-L, 
           Radboud-L $\rightarrow$ Radboud-R},
       xtick=data,
       xticklabel=\empty,
       x tick label style={text width = 2.25cm, align=center, font=\footnotesize},
    ]
    \addplot [fill=OrangeRed2, draw=OrangeRed2, error bars/.cd, y dir=both, y explicit, error bar style={thick}, error bar style=black] coordinates {
      (Radboud-U, 1.15572) +- (0, 0.0647359)
      (Radboud-U $\rightarrow$ Karolinska-U, 1.33367) +- (0, 0.0213751)
      (Radboud-L, 1.12129) +- (0, 0.0777459)
      (Radboud-L $\rightarrow$ Radboud-R, 1.39189) +- (0, 0.0245475)};

    \addplot [fill=darkgray, draw=darkgray, error bars/.cd, y dir=both, y explicit, error bar style={thick}, error bar style=black] coordinates {
      (Radboud-U, 0.649751) +- (0, 0.0449664)
      (Radboud-U $\rightarrow$ Karolinska-U, 1.40757) +- (0, 0.0381997)
      (Radboud-L, 0.59759) +- (0, 0.0389289)
      (Radboud-L $\rightarrow$ Radboud-R, 1) +- (0.867475, 0.0229779)};
    
    \addplot [fill=flamingopink, draw=flamingopink, line width=0mm, error bars/.cd, y dir=both, y explicit, error bar style={thick}, error bar style=black] coordinates { 
      (Radboud-U, 0.460697) +- (0, 0.0503791)
      (Radboud-U $\rightarrow$ Karolinska-U, 1.20584) +- (0, 0.0465443)
      (Radboud-L, 0.476305) +- (0, 0.0461131)
      (Radboud-L $\rightarrow$ Radboud-R, 0.720596) +- (0, 0.0473235)};

    \addplot [fill=red-violet, draw=red-violet, line width=0mm, error bars/.cd, y dir=both, y explicit, error bar style={thick}, error bar style=black] coordinates { 
      (Radboud-U, 0.498756) +- (0, 0.0584365)
      (Radboud-U $\rightarrow$ Karolinska-U, 1.35598) +- (0, 0.0348709)
      (Radboud-L, 0.496386) +- (0, 0.0348357)
      (Radboud-L $\rightarrow$ Radboud-R, 0.744771) +- (0, 0.0075598)};

    \addplot [fill=caribbeangreen2, draw=caribbeangreen2, error bars/.cd, y dir=both, y explicit, error bar style={thick}, error bar style=black] coordinates {
      (Radboud-U, 0.717662) +- (0, 0.0550673)
      (Radboud-U $\rightarrow$ Karolinska-U, 1.66972) +- (0, 0.105759)
      (Radboud-L, 0.660643) +- (0, 0.0808666)
      (Radboud-L $\rightarrow$ Radboud-R, 0.986561) +- (0, 0.0460075)};

    \addplot [fill=ForestGreen, draw=ForestGreen, error bars/.cd, y dir=both, y explicit, error bar style={thick}, error bar style=black] coordinates {
      (Radboud-U, 0.7801) +- (0, 0.0521701)
      (Radboud-U $\rightarrow$ Karolinska-U, 1.1077) +- (0, 0.023015)
      (Radboud-L, 0.666667) +- (0, 0.0525782)
      (Radboud-L $\rightarrow$ Radboud-R, 0.998608) +- (0, 0.0518634)};

    \addplot [fill=skyblue2, draw=skyblue2, error bars/.cd, y dir=both, y explicit, error bar style={thick}, error bar style=black] coordinates {
      (Radboud-U, 0.477861) +- (0, 0.0565771)
      (Radboud-U $\rightarrow$ Karolinska-U, 1.28871) +- (0, 0.0446399)
      (Radboud-L, 0.473092) +- (0, 0.0506024)
      (Radboud-L $\rightarrow$ Radboud-R, 0.722227) +- (0, 0.0116131)};

    \addplot [fill=Cerulean, draw=Cerulean, line width=0mm, error bars/.cd, y dir=both, y explicit, error bar style={thick}, error bar style=black] coordinates { 
      (Radboud-U, 0.474378) +- (0, 0.0459095)
      (Radboud-U $\rightarrow$ Karolinska-U, 1.21228) +- (0, 0.0440255)
      (Radboud-L, 0.467068) +- (0, 0.039022)
      (Radboud-L $\rightarrow$ Radboud-R, 0.713956) +- (0, 0.0135798)};

    \addplot [fill=darkcerulean2, draw=darkcerulean2, line width=0mm, error bars/.cd, y dir=both, y explicit, error bar style={thick}, error bar style=black] coordinates { 
      (Radboud-U, 0.558706) +- (0, 0.0593084)
      (Radboud-U $\rightarrow$ Karolinska-U, 1.38426) +- (0, 0.0454923)
      (Radboud-L, 0.514859) +- (0, 0.0523877)
      (Radboud-L $\rightarrow$ Radboud-R, 0.790577) +- (0, 0.0264311)};

    \addplot [fill=darkpastelpurple2, draw=darkpastelpurple2, error bars/.cd, y dir=both, y explicit, error bar style={thick}, error bar style=black] coordinates {
      (Radboud-U, 0.478856) +- (0, 0.0659486)
      (Radboud-U $\rightarrow$ Karolinska-U, 1.2674) +- (0, 0.0421948)
      (Radboud-L, 0.482731) +- (0, 0.0405523)
      (Radboud-L $\rightarrow$ Radboud-R, 0.723062) +- (0, 0.0167067)};

    \addplot [fill=goldenpoppy, draw=goldenpoppy, error bars/.cd, y dir=both, y explicit, error bar style={thick}, error bar style=black] coordinates {
      (Radboud-U, 0.606219) +- (0, 0.0389373)
      (Radboud-U $\rightarrow$ Karolinska-U, 1.34475) +- (0, 0.0302089)
      (Radboud-L, 0.573494) +- (0, 0.03514)
      (Radboud-L $\rightarrow$ Radboud-R, 0.845328) +- (0, 0.0183696)};

    \addplot [fill=goldenpoppy3, draw=goldenpoppy3, error bars/.cd, y dir=both, y explicit, error bar style={thick}, error bar style=black] coordinates {
      (Radboud-U, 0.537811) +- (0, 0.059896)
      (Radboud-U $\rightarrow$ Karolinska-U, 1.34303) +- (0, 0.079169)
      (Radboud-L, 0.525703) +- (0, 0.0309237)
      (Radboud-L $\rightarrow$ Radboud-R, 0.794394) +- (0, 0.0316998)};
  \end{axis}
  \end{tikzpicture}\vspace{-3.5mm}
    \begin{tikzpicture}
  \centering
  \begin{axis}[
        ybar=2.5pt, axis on top,
        title={},
        height=4.925cm, width=23.0cm, 
        bar width=0.26cm, 
        ymajorgrids, tick align=inside,
        enlarge y limits={value=.1,upper},
        ymin=0, ymax=2.4,
        set layers,
        ytick={0.5, 1.0, 1.5, 2.0},
        axis x line*=bottom,
        axis y line*=left,
        y axis line style={opacity=0},
        tickwidth=0pt,
        enlarge x limits=0.20, 
        extra y tick style={
            grid style={
                solid,
                line width = 2.5pt,
                /pgfplots/on layer=axis background,
            },
        },
        tick style={
            grid style={
                dashed,
                /pgfplots/on layer=axis background,
            },
        },
        legend style={
            at={(0.5,-0.225)},
            anchor=north,
            legend columns=-1,
            /tikz/every even column/.append style={column sep=0.175cm},
            font=\footnotesize
        },
        ylabel={MAE ($\downarrow$)},
        ylabel style={font=\footnotesize},
        symbolic x coords={
           Radboud-U, 
           Radboud-U $\rightarrow$ Karolinska-U,
           Radboud-L, 
           Radboud-L $\rightarrow$ Radboud-R},
       xtick=data,
       x tick label style={text width = 5.0cm, align=center, font={\bfseries}},
    ]
    \addplot [fill=OrangeRed2, draw=OrangeRed2, error bars/.cd, y dir=both, y explicit, error bar style={thick}, error bar style=black] coordinates {
      (Radboud-U, 0.908209) +- (0, 0.083409)
      (Radboud-U $\rightarrow$ Karolinska-U, 1.70093) +- (0, 0.0531301)
      (Radboud-L, 0.869076) +- (0, 0.0540366)
      (Radboud-L $\rightarrow$ Radboud-R, 1.62286) +- (0, 0.017535)};

    \addplot [fill=darkgray, draw=darkgray, error bars/.cd, y dir=both, y explicit, error bar style={thick}, error bar style=black] coordinates {
      (Radboud-U, 0.760697) +- (0, 0.0460676)
      (Radboud-U $\rightarrow$ Karolinska-U, 1.51866) +- (0, 0.0172399)
      (Radboud-L, 0.751406) +- (0, 0.0655357)
      (Radboud-L $\rightarrow$ Radboud-R, 1.42887) +- (0, 0.0155357)};
    
    \addplot [fill=flamingopink, draw=flamingopink, line width=0mm, error bars/.cd, y dir=both, y explicit, error bar style={thick}, error bar style=black] coordinates { 
      (Radboud-U, 0.685572) +- (0, 0.0625977)
      (Radboud-U $\rightarrow$ Karolinska-U, 1.41195) +- (0, 0.0471749)
      (Radboud-L, 0.662249) +- (0, 0.0390798)
      (Radboud-L $\rightarrow$ Radboud-R, 1.31765) +- (0, 0.0103494)};

    \addplot [fill=red-violet, draw=red-violet, line width=0mm, error bars/.cd, y dir=both, y explicit, error bar style={thick}, error bar style=black] coordinates { 
      (Radboud-U, 0.698259) +- (0, 0.0604846)
      (Radboud-U $\rightarrow$ Karolinska-U, 1.43725) +- (0, 0.0294423)
      (Radboud-L, 0.65743) +- (0, 0.0304295)
      (Radboud-L $\rightarrow$ Radboud-R, 1.27062) +- (0, 0.015859)};

    \addplot [fill=caribbeangreen2, draw=caribbeangreen2, error bars/.cd, y dir=both, y explicit, error bar style={thick}, error bar style=black] coordinates {
      (Radboud-U, 0.805224) +- (0, 0.0454489)
      (Radboud-U $\rightarrow$ Karolinska-U, 1.38181) +- (0, 0.0280272)
      (Radboud-L, 0.718072) +- (0, 0.0467246)
      (Radboud-L $\rightarrow$ Radboud-R, 1.38143) +- (0, 0.0145823)};

    \addplot [fill=ForestGreen, draw=ForestGreen, error bars/.cd, y dir=both, y explicit, error bar style={thick}, error bar style=black] coordinates {
      (Radboud-U, 0.814179) +- (0, 0.0689019)
      (Radboud-U $\rightarrow$ Karolinska-U, 1.33108) +- (0, 0.0241707)
      (Radboud-L, 0.720884) +- (0, 0.0462894)
      (Radboud-L $\rightarrow$ Radboud-R, 1.35316) +- (0, 0.0106287)};

    \addplot [fill=skyblue2, draw=skyblue2, error bars/.cd, y dir=both, y explicit, error bar style={thick}, error bar style=black] coordinates {
      (Radboud-U, 0.692786) +- (0, 0.0660329)
      (Radboud-U $\rightarrow$ Karolinska-U, 1.55797) +- (0, 0.0139222)
      (Radboud-L, 0.657831) +- (0, 0.023196)
      (Radboud-L $\rightarrow$ Radboud-R, 1.29678) +- (0, 0.0174482)};

    \addplot [fill=Cerulean, draw=Cerulean, line width=0mm, error bars/.cd, y dir=both, y explicit, error bar style={thick}, error bar style=black] coordinates { 
      (Radboud-U, 0.68806) +- (0, 0.0685701)
      (Radboud-U $\rightarrow$ Karolinska-U, 1.567) +- (0, 0.0202723)
      (Radboud-L, 0.603614) +- (0, 0.0346431)
      (Radboud-L $\rightarrow$ Radboud-R, 1.21233) +- (0, 0.0146719)};

    \addplot [fill=darkcerulean2, draw=darkcerulean2, line width=0mm, error bars/.cd, y dir=both, y explicit, error bar style={thick}, error bar style=black] coordinates { 
      (Radboud-U, 0.760448) +- (0, 0.0526171)
      (Radboud-U $\rightarrow$ Karolinska-U, 1.58745) +- (0, 0.0285942)
      (Radboud-L, 0.704819) +- (0, 0.0355711)
      (Radboud-L $\rightarrow$ Radboud-R, 1.38091) +- (0, 0.0100164)};

    \addplot [fill=darkpastelpurple2, draw=darkpastelpurple2, error bars/.cd, y dir=both, y explicit, error bar style={thick}, error bar style=black] coordinates {
      (Radboud-U, 0.685572) +- (0, 0.0546721)
      (Radboud-U $\rightarrow$ Karolinska-U, 1.43672) +- (0, 0.0216737)
      (Radboud-L, 0.681928) +- (0, 0.0541439)
      (Radboud-L $\rightarrow$ Radboud-R, 1.35618) +- (0, 0.0118524)};

    \addplot [fill=goldenpoppy, draw=goldenpoppy, error bars/.cd, y dir=both, y explicit, error bar style={thick}, error bar style=black] coordinates {
      (Radboud-U, 0.824627) +- (0, 0.0553252)
      (Radboud-U $\rightarrow$ Karolinska-U, 1.89608) +- (0, 0.0477629)
      (Radboud-L, 0.76988) +- (0, 0.0697698)
      (Radboud-L $\rightarrow$ Radboud-R, 1.44445) +- (0, 0.0176621)};

    \addplot [fill=goldenpoppy3, draw=goldenpoppy3, error bars/.cd, y dir=both, y explicit, error bar style={thick}, error bar style=black] coordinates {
      (Radboud-U, 0.784826) +- (0, 0.0608335)
      (Radboud-U $\rightarrow$ Karolinska-U, 1.51441) +- (0, 0.0268888)
      (Radboud-L, 0.71004) +- (0, 0.0366233)
      (Radboud-L $\rightarrow$ Radboud-R, 1.3695) +- (0, 0.0113666)};
    \legend{
            Resnet-IN,
            Phikon-v2,
            UNI,
            UNI2-h,
            CONCH,
            CONCHv1.5,
            H-optimus-0,
            H-optimus-1,
            H0-mini,
            Prov-GigaPath,
            Virchow,
            Virchow2
            }
  \end{axis}
  \end{tikzpicture}\vspace{-1.0mm}
    \caption{\textbf{Model performance under controlled label-distribution and cross-site shifts (MAE).} The same performance comparison as in Figure~\ref{fig:main_results3a_12fms_mae}, evaluated across four additional PANDA subsets designed to isolate the effects of label-distribution and domain shift. All results are reported as mean $\pm$ standard deviation over $10$ cross-validation folds. This figure presents the same experiments as Figure~\ref{fig:main_results3b_12fms_kappa}, but using mean absolute error (MAE, lower is better) instead of quadratically weighted Cohen's kappa.}
    \label{fig:main_results3b_12fms_mae}
\end{figure*}

\begin{figure*}[t]
    \centering
    \begin{tikzpicture}
  \centering
  \begin{axis}[
        ybar, axis on top,
        title={},
        height=4.525cm, width=18.0cm, 
        bar width=0.35cm,
        ymajorgrids, tick align=inside,
        enlarge y limits={value=.1,upper},
        ymin=0, ymax=2.4,
        set layers,
        ytick={0.5, 1.0, 1.5, 2.0},
        axis x line*=bottom,
        axis y line*=left,
        y axis line style={opacity=0},
        tickwidth=0pt,
        enlarge x limits=true,
        extra y tick style={
            grid style={
                solid,
                line width = 2.5pt,
                /pgfplots/on layer=axis background,
            },
        },
        tick style={
            grid style={
                dashed,
                /pgfplots/on layer=axis background,
            },
        },
        legend style={
            at={(0.5,-0.30)},
            anchor=north,
            legend columns=-1,
            /tikz/every even column/.append style={column sep=0.325cm},
            font=\footnotesize
        },
        ylabel={MAE ($\downarrow$)},
        ylabel style={font=\footnotesize},
        symbolic x coords={
           PANDA,
           Karolinska,
           Radboud, 
           Radboud $\rightarrow$\\ Karolinska,
           Radboud-U, 
           Radboud-U $\rightarrow$\\ Karolinska-U,
           Radboud-L, 
           Radboud-L $\rightarrow$\\ Radboud-R},
       xtick=data,
       xticklabel=\empty,
       x tick label style={text width = 2.25cm, align=center, font=\footnotesize},
    ]
    \addplot [fill=skyblue2, draw=skyblue2, line width=0mm, error bars/.cd, y dir=both, y explicit, error bar style={thick}, error bar style=black] coordinates { 
      (PANDA, 0.621935) +- (0, 0.0233373)
      (Karolinska, 0.565257) +- (0, 0.037827)
      (Radboud, 0.7762) +- (0, 0.0422512)
      (Radboud $\rightarrow$\\ Karolinska, 1.55543) +- (0, 0.10277)
      (Radboud-U, 0.780348) +- (0, 0.0283855)
      (Radboud-U $\rightarrow$\\ Karolinska-U, 1.29024) +- (0, 0.0282161)
      (Radboud-L, 0.697189) +- (0, 0.0465656)
      (Radboud-L $\rightarrow$\\ Radboud-R, 1.08342) +- (0, 0.0449268)};
    \addplot [fill=caribbeangreen2, draw=caribbeangreen2, error bars/.cd, y dir=both, y explicit, error bar style={thick}, error bar style=black] coordinates {
      (PANDA, 0.975096) +- (0, 0.052894)
      (Karolinska, 1.10018) +- (0, 0.0761271)
      (Radboud, 1.128) +- (0, 0.0341292)
      (Radboud $\rightarrow$\\ Karolinska, 1.50294) +- (0, 0.0886444)
      (Radboud-U, 1.15572) +- (0, 0.0647359)
      (Radboud-U $\rightarrow$\\ Karolinska-U, 1.33367) +- (0, 0.0213751)
      (Radboud-L, 1.12129) +- (0, 0.0777459)
      (Radboud-L $\rightarrow$\\ Radboud-R, 1.39189) +- (0, 0.0245475)};
    \addplot [fill=darkpastelpurple2, draw=darkpastelpurple2, error bars/.cd, y dir=both, y explicit, error bar style={thick}, error bar style=black] coordinates {
      (PANDA, 0.864464) +- (0, 0.0243048)
      (Karolinska, 0.845037) +- (0, 0.0344692)
      (Radboud, 0.8698) +- (0, 0.0669235)
      (Radboud $\rightarrow$\\ Karolinska, 1.97403) +- (0, 0.10194)
      (Radboud-U, 0.908209) +- (0, 0.083409)
      (Radboud-U $\rightarrow$\\ Karolinska-U, 1.70093) +- (0, 0.0531301)
      (Radboud-L, 0.869076) +- (0, 0.0540366)
      (Radboud-L $\rightarrow$\\ Radboud-R, 1.62286) +- (0, 0.017535)};
  \end{axis}
  \end{tikzpicture}\vspace{-2.5mm}
    \begin{tikzpicture}
  \centering
  \begin{axis}[
        ybar, axis on top,
        title={},
        height=4.525cm, width=18.0cm, 
        bar width=0.35cm,
        ymajorgrids, tick align=inside,
        enlarge y limits={value=.1,upper},
        ymin=0, ymax=2.4,
        set layers,
        ytick={0.5, 1.0, 1.5, 2.0},
        axis x line*=bottom,
        axis y line*=left,
        y axis line style={opacity=0},
        tickwidth=0pt,
        enlarge x limits=true,
        extra y tick style={
            grid style={
                solid,
                line width = 2.5pt,
                /pgfplots/on layer=axis background,
            },
        },
        tick style={
            grid style={
                dashed,
                /pgfplots/on layer=axis background,
            },
        },
        legend style={
            at={(0.5,-0.2)},
            anchor=north,
            legend columns=-1,
            /tikz/every even column/.append style={column sep=0.325cm},
            font=\footnotesize
        },
        ylabel={MAE ($\downarrow$)},
        ylabel style={font=\footnotesize},
        symbolic x coords={
           PANDA,
           Karolinska,
           Radboud, 
           Radboud $\rightarrow$\\ Karolinska,
           Radboud-U, 
           Radboud-U $\rightarrow$\\ Karolinska-U,
           Radboud-L, 
           Radboud-L $\rightarrow$\\ Radboud-R},
       xtick=data,
       xticklabel=\empty,
       x tick label style={text width = 2.25cm, align=center, font=\footnotesize},
    ]
    \addplot [fill=skyblue2, draw=skyblue2, line width=0mm, error bars/.cd, y dir=both, y explicit, error bar style={thick}, error bar style=black] coordinates { 
      (PANDA, 0.346935) +- (0, 0.0350706)
      (Karolinska, 0.299081) +- (0, 0.0178043)
      (Radboud, 0.4238) +- (0, 0.0450107)
      (Radboud $\rightarrow$\\ Karolinska, 1.52554) +- (0, 0.369422)
      (Radboud-U, 0.447264) +- (0, 0.0397512)
      (Radboud-U $\rightarrow$\\ Karolinska-U, 1.15465) +- (0, 0.169618)
      (Radboud-L, 0.455422) +- (0, 0.0500705)
      (Radboud-L $\rightarrow$\\ Radboud-R, 0.663022) +- (0, 0.0320972)};
    \addplot [fill=caribbeangreen2, draw=caribbeangreen2, error bars/.cd, y dir=both, y explicit, error bar style={thick}, error bar style=black] coordinates {
      (PANDA, 0.434579) +- (0, 0.0323268)
      (Karolinska, 0.496875) +- (0, 0.0336758)
      (Radboud, 0.4228) +- (0, 0.0318773)
      (Radboud $\rightarrow$\\ Karolinska, 0.972488) +- (0, 0.124728)
      (Radboud-U, 0.460697) +- (0, 0.0503791)
      (Radboud-U $\rightarrow$\\ Karolinska-U, 1.20584) +- (0, 0.0465443)
      (Radboud-L, 0.476305) +- (0, 0.0461131)
      (Radboud-L $\rightarrow$\\ Radboud-R, 0.720596) +- (0, 0.0473235)};
    \addplot [fill=darkpastelpurple2, draw=darkpastelpurple2, error bars/.cd, y dir=both, y explicit, error bar style={thick}, error bar style=black] coordinates {
      (PANDA, 0.67069) +- (0, 0.0272098)
      (Karolinska, 0.730331) +- (0, 0.0267405)
      (Radboud, 0.6302) +- (0, 0.0519111)
      (Radboud $\rightarrow$\\ Karolinska, 1.28081) +- (0, 0.0565518)
      (Radboud-U, 0.685572) +- (0, 0.0625977)
      (Radboud-U $\rightarrow$\\ Karolinska-U, 1.41195) +- (0, 0.0471749)
      (Radboud-L, 0.662249) +- (0, 0.0390798)
      (Radboud-L $\rightarrow$\\ Radboud-R, 1.31765) +- (0, 0.0103494)};
  \end{axis}
  \end{tikzpicture}\vspace{-2.5mm}
    \begin{tikzpicture}
  \centering
  \begin{axis}[
        ybar, axis on top,
        title={},
        height=4.525cm, width=18.0cm, 
        bar width=0.35cm,
        ymajorgrids, tick align=inside,
        enlarge y limits={value=.1,upper},
        ymin=0, ymax=2.4,
        set layers,
        ytick={0.5, 1.0, 1.5, 2.0},
        axis x line*=bottom,
        axis y line*=left,
        y axis line style={opacity=0},
        tickwidth=0pt,
        enlarge x limits=true,
        extra y tick style={
            grid style={
                solid,
                line width = 2.5pt,
                /pgfplots/on layer=axis background,
            },
        },
        tick style={
            grid style={
                dashed,
                /pgfplots/on layer=axis background,
            },
        },
        legend style={
            at={(0.5,-0.2)},
            anchor=north,
            legend columns=-1,
            /tikz/every even column/.append style={column sep=0.325cm},
            font=\footnotesize
        },
        ylabel={MAE ($\downarrow$)},
        ylabel style={font=\footnotesize},
        symbolic x coords={
           PANDA,
           Karolinska,
           Radboud, 
           Radboud $\rightarrow$\\ Karolinska,
           Radboud-U, 
           Radboud-U $\rightarrow$\\ Karolinska-U,
           Radboud-L, 
           Radboud-L $\rightarrow$\\ Radboud-R},
       xtick=data,
       xticklabel=\empty,
       x tick label style={text width = 2.25cm, align=center, font=\footnotesize},
    ]
    \addplot [fill=skyblue2, draw=skyblue2, line width=0mm, error bars/.cd, y dir=both, y explicit, error bar style={thick}, error bar style=black] coordinates { 
      (PANDA, 0.404981) +- (0, 0.0184436)
      (Karolinska, 0.326287) +- (0, 0.0255012)
      (Radboud, 0.5378) +- (0, 0.0416793)
      (Radboud $\rightarrow$\\ Karolinska, 2.26281) +- (0, 0.108598)
      (Radboud-U, 0.546269) +- (0, 0.0449994)
      (Radboud-U $\rightarrow$\\ Karolinska-U, 1.4664) +- (0, 0.112544)
      (Radboud-L, 0.49759) +- (0, 0.0626158)
      (Radboud-L $\rightarrow$\\ Radboud-R, 0.726441) +- (0, 0.0344725)};
    \addplot [fill=caribbeangreen2, draw=caribbeangreen2, error bars/.cd, y dir=both, y explicit, error bar style={thick}, error bar style=black] coordinates {
      (PANDA, 0.620019) +- (0, 0.0276654)
      (Karolinska, 0.623897) +- (0, 0.0323884)
      (Radboud, 0.7124) +- (0, 0.039717)
      (Radboud $\rightarrow$\\ Karolinska, 2.2115) +- (0, 0.145679)
      (Radboud-U, 0.717662) +- (0, 0.0550673)
      (Radboud-U $\rightarrow$\\ Karolinska-U, 1.66972) +- (0, 0.105759)
      (Radboud-L, 0.660643) +- (0, 0.0808666)
      (Radboud-L $\rightarrow$\\ Radboud-R, 0.986561) +- (0, 0.0460075)};
    \addplot [fill=darkpastelpurple2, draw=darkpastelpurple2, error bars/.cd, y dir=both, y explicit, error bar style={thick}, error bar style=black] coordinates {
      (PANDA, 0.755077) +- (0, 0.026295)
      (Karolinska, 0.727941) +- (0, 0.0260615)
      (Radboud, 0.7634) +- (0, 0.0659639)
      (Radboud $\rightarrow$\\ Karolinska, 1.46491) +- (0, 0.0411191)
      (Radboud-U, 0.805224) +- (0, 0.0454489)
      (Radboud-U $\rightarrow$\\ Karolinska-U, 1.38181) +- (0, 0.0280272)
      (Radboud-L, 0.718072) +- (0, 0.0467246)
      (Radboud-L $\rightarrow$\\ Radboud-R, 1.38143) +- (0, 0.0145823)};
  \end{axis}
  \end{tikzpicture}\vspace{-2.5mm}
    \begin{tikzpicture}
  \centering
  \begin{axis}[
        ybar, axis on top,
        title={},
        height=4.525cm, width=18.0cm, 
        bar width=0.35cm,
        ymajorgrids, tick align=inside,
        enlarge y limits={value=.1,upper},
        ymin=0, ymax=2.4,
        set layers,
        ytick={0.5, 1.0, 1.5, 2.0},
        axis x line*=bottom,
        axis y line*=left,
        y axis line style={opacity=0},
        tickwidth=0pt,
        enlarge x limits=true,
        extra y tick style={
            grid style={
                solid,
                line width = 2.5pt,
                /pgfplots/on layer=axis background,
            },
        },
        tick style={
            grid style={
                dashed,
                /pgfplots/on layer=axis background,
            },
        },
        legend style={
            at={(0.5,-0.2)},
            anchor=north,
            legend columns=-1,
            /tikz/every even column/.append style={column sep=0.325cm},
            font=\footnotesize
        },
        ylabel={MAE ($\downarrow$)},
        ylabel style={font=\footnotesize},
        symbolic x coords={
           PANDA,
           Karolinska,
           Radboud, 
           Radboud $\rightarrow$\\ Karolinska,
           Radboud-U, 
           Radboud-U $\rightarrow$\\ Karolinska-U,
           Radboud-L, 
           Radboud-L $\rightarrow$\\ Radboud-R},
       xtick=data,
       xticklabel=\empty,
       x tick label style={text width = 2.25cm, align=center, font=\footnotesize},
    ]
    \addplot [fill=skyblue2, draw=skyblue2, line width=0mm, error bars/.cd, y dir=both, y explicit, error bar style={thick}, error bar style=black] coordinates { 
      (PANDA, 0.431513) +- (0, 0.019965)
      (Karolinska, 0.334007) +- (0, 0.0308369)
      (Radboud, 0.5574) +- (0, 0.0330702)
      (Radboud $\rightarrow$\\ Karolinska, 0.572617) +- (0, 0.0595431)
      (Radboud-U, 0.565423) +- (0, 0.0445135)
      (Radboud-U $\rightarrow$\\ Karolinska-U, 0.810757) +- (0, 0.0351173)
      (Radboud-L, 0.500803) +- (0, 0.0484947)
      (Radboud-L $\rightarrow$\\ Radboud-R, 0.761511) +- (0, 0.0296167)};
    \addplot [fill=caribbeangreen2, draw=caribbeangreen2, error bars/.cd, y dir=both, y explicit, error bar style={thick}, error bar style=black] coordinates {
      (PANDA, 0.655268) +- (0, 0.0332789)
      (Karolinska, 0.62886) +- (0, 0.0400081)
      (Radboud, 0.7578) +- (0, 0.0367418)
      (Radboud $\rightarrow$\\ Karolinska, 0.807177) +- (0, 0.0288316)
      (Radboud-U, 0.7801) +- (0, 0.0521701)
      (Radboud-U $\rightarrow$\\ Karolinska-U, 1.1077) +- (0, 0.023015)
      (Radboud-L, 0.666667) +- (0, 0.0525782)
      (Radboud-L $\rightarrow$\\ Radboud-R, 0.998608) +- (0, 0.0518634)};
    \addplot [fill=darkpastelpurple2, draw=darkpastelpurple2, error bars/.cd, y dir=both, y explicit, error bar style={thick}, error bar style=black] coordinates {
      (PANDA, 0.730843) +- (0, 0.0263401)
      (Karolinska, 0.698346) +- (0, 0.024407)
      (Radboud, 0.7658) +- (0, 0.0728475)
      (Radboud $\rightarrow$\\ Karolinska, 1.25585) +- (0, 0.0251673)
      (Radboud-U, 0.814179) +- (0, 0.0689019)
      (Radboud-U $\rightarrow$\\ Karolinska-U, 1.33108) +- (0, 0.0241707)
      (Radboud-L, 0.720884) +- (0, 0.0462894)
      (Radboud-L $\rightarrow$\\ Radboud-R, 1.35316) +- (0, 0.0106287)};
  \end{axis}
  \end{tikzpicture}\vspace{-2.5mm}
    \begin{tikzpicture}
  \centering
  \begin{axis}[
        ybar, axis on top,
        title={},
        height=4.525cm, width=18.0cm, 
        bar width=0.35cm,
        ymajorgrids, tick align=inside,
        enlarge y limits={value=.1,upper},
        ymin=0, ymax=2.4,
        set layers,
        ytick={0.5, 1.0, 1.5, 2.0},
        axis x line*=bottom,
        axis y line*=left,
        y axis line style={opacity=0},
        tickwidth=0pt,
        enlarge x limits=true,
        extra y tick style={
            grid style={
                solid,
                line width = 2.5pt,
                /pgfplots/on layer=axis background,
            },
        },
        tick style={
            grid style={
                dashed,
                /pgfplots/on layer=axis background,
            },
        },
        legend style={
            at={(0.5,-0.2)},
            anchor=north,
            legend columns=-1,
            /tikz/every even column/.append style={column sep=0.325cm},
            font=\footnotesize
        },
        ylabel={MAE ($\downarrow$)},
        ylabel style={font=\footnotesize},
        symbolic x coords={
           PANDA,
           Karolinska,
           Radboud, 
           Radboud $\rightarrow$\\ Karolinska,
           Radboud-U, 
           Radboud-U $\rightarrow$\\ Karolinska-U,
           Radboud-L, 
           Radboud-L $\rightarrow$\\ Radboud-R},
       xtick=data,
       xticklabel=\empty,
       x tick label style={text width = 2.25cm, align=center, font=\footnotesize},
    ]
    \addplot [fill=skyblue2, draw=skyblue2, line width=0mm, error bars/.cd, y dir=both, y explicit, error bar style={thick}, error bar style=black] coordinates { 
      (PANDA, 0.339368) +- (0, 0.0302417)
      (Karolinska, 0.29761) +- (0, 0.023215)
      (Radboud, 0.3896) +- (0, 0.0614479)
      (Radboud $\rightarrow$\\ Karolinska, 0.94078) +- (0, 0.155075)
      (Radboud-U, 0.420398) +- (0, 0.027476)
      (Radboud-U $\rightarrow$\\ Karolinska-U, 0.886587) +- (0, 0.0894473)
      (Radboud-L, 0.415261) +- (0, 0.0570281)
      (Radboud-L $\rightarrow$\\ Radboud-R, 0.653837) +- (0, 0.0421783)};
    \addplot [fill=caribbeangreen2, draw=caribbeangreen2, error bars/.cd, y dir=both, y explicit, error bar style={thick}, error bar style=black] coordinates {
      (PANDA, 0.444061) +- (0, 0.0274729)
      (Karolinska, 0.47739) +- (0, 0.0243322)
      (Radboud, 0.457) +- (0, 0.0342374)
      (Radboud $\rightarrow$\\ Karolinska, 0.767777) +- (0, 0.0247762)
      (Radboud-U, 0.474378) +- (0, 0.0459095)
      (Radboud-U $\rightarrow$\\ Karolinska-U, 1.21228) +- (0, 0.0440255)
      (Radboud-L, 0.467068) +- (0, 0.039022)
      (Radboud-L $\rightarrow$\\ Radboud-R, 0.713956) +- (0, 0.0135798)};
    \addplot [fill=darkpastelpurple2, draw=darkpastelpurple2, error bars/.cd, y dir=both, y explicit, error bar style={thick}, error bar style=black] coordinates {
      (PANDA, 0.668487) +- (0, 0.027467)
      (Karolinska, 0.707904) +- (0, 0.0320344)
      (Radboud, 0.6122) +- (0, 0.0632737)
      (Radboud $\rightarrow$\\ Karolinska, 1.08831) +- (0, 0.0276389)
      (Radboud-U, 0.68806) +- (0, 0.0685701)
      (Radboud-U $\rightarrow$\\ Karolinska-U, 1.567) +- (0, 0.0202723)
      (Radboud-L, 0.603614) +- (0, 0.0346431)
      (Radboud-L $\rightarrow$\\ Radboud-R, 1.21233) +- (0, 0.0146719)};
  \end{axis}
  \end{tikzpicture}\vspace{-2.5mm}
    \begin{tikzpicture}
  \centering
  \begin{axis}[
        ybar, axis on top,
        title={},
        height=4.525cm, width=18.0cm, 
        bar width=0.35cm,
        ymajorgrids, tick align=inside,
        enlarge y limits={value=.1,upper},
        ymin=0, ymax=2.4,
        set layers,
        ytick={0.5, 1.0, 1.5, 2.0},
        axis x line*=bottom,
        axis y line*=left,
        y axis line style={opacity=0},
        tickwidth=0pt,
        enlarge x limits=true,
        extra y tick style={
            grid style={
                solid,
                line width = 2.5pt,
                /pgfplots/on layer=axis background,
            },
        },
        tick style={
            grid style={
                dashed,
                /pgfplots/on layer=axis background,
            },
        },
        legend style={
            at={(0.5,-0.30)},
            anchor=north,
            legend columns=-1,
            /tikz/every even column/.append style={column sep=0.325cm},
            font=\footnotesize
        },
        ylabel={MAE ($\downarrow$)},
        ylabel style={font=\footnotesize},
        symbolic x coords={
           PANDA,
           Karolinska,
           Radboud, 
           Radboud $\rightarrow$\\ Karolinska,
           Radboud-U, 
           Radboud-U $\rightarrow$\\ Karolinska-U,
           Radboud-L, 
           Radboud-L $\rightarrow$\\ Radboud-R},
       xtick=data,
       x tick label style={text width = 2.25cm, align=center, font=\footnotesize},
    ]
    \addplot [fill=skyblue2, draw=skyblue2, line width=0mm, error bars/.cd, y dir=both, y explicit, error bar style={thick}, error bar style=black] coordinates { 
      (PANDA, 0.378161) +- (0, 0.0220922)
      (Karolinska, 0.302022) +- (0, 0.0251247)
      (Radboud, 0.459) +- (0, 0.0350457)
      (Radboud $\rightarrow$\\ Karolinska, 0.892455) +- (0, 0.227418)
      (Radboud-U, 0.490796) +- (0, 0.0538722)
      (Radboud-U $\rightarrow$\\ Karolinska-U, 0.831939) +- (0, 0.0712366)
      (Radboud-L, 0.483133) +- (0, 0.0371155)
      (Radboud-L $\rightarrow$\\ Radboud-R, 0.699205) +- (0, 0.039343)};
    \addplot [fill=caribbeangreen2, draw=caribbeangreen2, error bars/.cd, y dir=both, y explicit, error bar style={thick}, error bar style=black] coordinates {
      (PANDA, 0.487452) +- (0, 0.0255819)
      (Karolinska, 0.496507) +- (0, 0.0275423)
      (Radboud, 0.5376) +- (0, 0.0371139)
      (Radboud $\rightarrow$\\ Karolinska, 1.66229) +- (0, 0.24416)
      (Radboud-U, 0.537811) +- (0, 0.059896)
      (Radboud-U $\rightarrow$\\ Karolinska-U, 1.34303) +- (0, 0.079169)
      (Radboud-L, 0.525703) +- (0, 0.0309237)
      (Radboud-L $\rightarrow$\\ Radboud-R, 0.794394) +- (0, 0.0316998)};
    \addplot [fill=darkpastelpurple2, draw=darkpastelpurple2, error bars/.cd, y dir=both, y explicit, error bar style={thick}, error bar style=black] coordinates {
      (PANDA, 0.742625) +- (0, 0.0201607)
      (Karolinska, 0.723162) +- (0, 0.0265012)
      (Radboud, 0.744) +- (0, 0.0564978)
      (Radboud $\rightarrow$\\ Karolinska, 1.11205) +- (0, 0.0130528)
      (Radboud-U, 0.784826) +- (0, 0.0608335)
      (Radboud-U $\rightarrow$\\ Karolinska-U, 1.51441) +- (0, 0.0268888)
      (Radboud-L, 0.71004) +- (0, 0.0366233)
      (Radboud-L $\rightarrow$\\ Radboud-R, 1.3695) +- (0, 0.0113666)};
    \legend{
            \textit{ABMIL},
            \textit{Mean Feature},
            \textit{kNN}
            }
  \end{axis}
  \end{tikzpicture}\vspace{-1.0mm}
    \caption{\textbf{Comparison of downstream ISUP grade classification models (MAE).} Performance of the three ISUP grade models \textit{ABMIL}, \textit{Mean Feature}, and \textit{kNN} when using different feature extractors: Resnet-IN \textbf{(top row)}, UNI, CONCH, CONCHv1.5, H-optimus-1, and Virchow2 \textbf{(bottom row)}. All results are reported as mean $\pm$ standard deviation over $10$ cross-validation folds. This figure presents the same comparison as Figure~\ref{fig:main_results_attmil_mean-pool_knn_kappa}, but using mean absolute error (MAE, lower is better) instead of quadratically weighted Cohen's kappa.}
    \label{fig:main_results_attmil_mean-pool_knn_mae}
\end{figure*}
\begin{figure*}[t]
    \centering
    \begin{tikzpicture}
  \centering
  \begin{axis}[
        ybar, axis on top,
        title={},
        height=4.525cm, width=18.0cm, 
        bar width=0.35cm,
        ymajorgrids, tick align=inside,
        enlarge y limits={value=.1,upper},
        ymin=0, ymax=2.4,
        set layers,
        ytick={0.5, 1.0, 1.5, 2.0},
        axis x line*=bottom,
        axis y line*=left,
        y axis line style={opacity=0},
        tickwidth=0pt,
        enlarge x limits=true,
        extra y tick style={
            grid style={
                solid,
                line width = 2.5pt,
                /pgfplots/on layer=axis background,
            },
        },
        tick style={
            grid style={
                dashed,
                /pgfplots/on layer=axis background,
            },
        },
        legend style={
            at={(0.5,-0.2)},
            anchor=north,
            legend columns=-1,
            /tikz/every even column/.append style={column sep=0.325cm},
            font=\footnotesize
        },
        ylabel={MAE ($\downarrow$)},
        ylabel style={font=\footnotesize},
        symbolic x coords={
           PANDA,
           Karolinska,
           Radboud, 
           Radboud $\rightarrow$\\ Karolinska,
           Radboud-U, 
           Radboud-U $\rightarrow$\\ Karolinska-U,
           Radboud-L, 
           Radboud-L $\rightarrow$\\ Radboud-R},
       xtick=data,
       xticklabel=\empty,
       x tick label style={text width = 2.25cm, align=center, font=\footnotesize},
    ]
    \addplot [fill=skyblue2, draw=skyblue2, line width=0mm, error bars/.cd, y dir=both, y explicit, error bar style={thick}, error bar style=black] coordinates { 
      (PANDA, 0.370307) +- (0, 0.0177045)
      (Karolinska, 0.307721) +- (0, 0.01665)
      (Radboud, 0.482) +- (0, 0.0242157)
      (Radboud $\rightarrow$\\ Karolinska, 1.45491) +- (0, 0.194843)
      (Radboud-U, 0.492786) +- (0, 0.0522204)
      (Radboud-U $\rightarrow$\\ Karolinska-U, 1.05246) +- (0, 0.074783)
      (Radboud-L, 0.46988) +- (0, 0.046001)
      (Radboud-L $\rightarrow$\\ Radboud-R, 0.701869) +- (0, 0.0294566)};
    \addplot [fill=caribbeangreen2, draw=caribbeangreen2, error bars/.cd, y dir=both, y explicit, error bar style={thick}, error bar style=black] coordinates {
      (PANDA, 0.567337) +- (0, 0.03535)
      (Karolinska, 0.560846) +- (0, 0.0414843)
      (Radboud, 0.6158) +- (0, 0.042374)
      (Radboud $\rightarrow$\\ Karolinska, 1.41492) +- (0, 0.166942)
      (Radboud-U, 0.649751) +- (0, 0.0449664)
      (Radboud-U $\rightarrow$\\ Karolinska-U, 1.40757) +- (0, 0.0381997)
      (Radboud-L, 0.59759) +- (0, 0.0389289)
      (Radboud-L $\rightarrow$\\ Radboud-R, 1) +- (0.867475, 0.0229779)};
    \addplot [fill=darkpastelpurple2, draw=darkpastelpurple2, error bars/.cd, y dir=both, y explicit, error bar style={thick}, error bar style=black] coordinates {
      (PANDA, 0.736686) +- (0, 0.0257677)
      (Karolinska, 0.751471) +- (0, 0.0254554)
      (Radboud, 0.6716) +- (0, 0.0350063)
      (Radboud $\rightarrow$\\ Karolinska, 1.18147) +- (0, 0.131297)
      (Radboud-U, 0.760697) +- (0, 0.0460676)
      (Radboud-U $\rightarrow$\\ Karolinska-U, 1.51866) +- (0, 0.0172399)
      (Radboud-L, 0.751406) +- (0, 0.0655357)
      (Radboud-L $\rightarrow$\\ Radboud-R, 1.42887) +- (0, 0.0155357)};
  \end{axis}
  \end{tikzpicture}\vspace{-2.5mm}
    \begin{tikzpicture}
  \centering
  \begin{axis}[
        ybar, axis on top,
        title={},
        height=4.525cm, width=18.0cm, 
        bar width=0.35cm,
        ymajorgrids, tick align=inside,
        enlarge y limits={value=.1,upper},
        ymin=0, ymax=2.4,
        set layers,
        ytick={0.5, 1.0, 1.5, 2.0},
        axis x line*=bottom,
        axis y line*=left,
        y axis line style={opacity=0},
        tickwidth=0pt,
        enlarge x limits=true,
        extra y tick style={
            grid style={
                solid,
                line width = 2.5pt,
                /pgfplots/on layer=axis background,
            },
        },
        tick style={
            grid style={
                dashed,
                /pgfplots/on layer=axis background,
            },
        },
        legend style={
            at={(0.5,-0.2)},
            anchor=north,
            legend columns=-1,
            /tikz/every even column/.append style={column sep=0.325cm},
            font=\footnotesize
        },
        ylabel={MAE ($\downarrow$)},
        ylabel style={font=\footnotesize},
        symbolic x coords={
           PANDA,
           Karolinska,
           Radboud, 
           Radboud $\rightarrow$\\ Karolinska,
           Radboud-U, 
           Radboud-U $\rightarrow$\\ Karolinska-U,
           Radboud-L, 
           Radboud-L $\rightarrow$\\ Radboud-R},
       xtick=data,
       xticklabel=\empty,
       x tick label style={text width = 2.25cm, align=center, font=\footnotesize},
    ]
    \addplot [fill=skyblue2, draw=skyblue2, line width=0mm, error bars/.cd, y dir=both, y explicit, error bar style={thick}, error bar style=black] coordinates { 
      (PANDA, 0.354693) +- (0, 0.0269327)
      (Karolinska, 0.298529) +- (0, 0.0272555)
      (Radboud, 0.439) +- (0, 0.0293837)
      (Radboud $\rightarrow$\\ Karolinska, 1.20905) +- (0, 0.220848)
      (Radboud-U, 0.443035) +- (0, 0.0268484)
      (Radboud-U $\rightarrow$\\ Karolinska-U, 1.04548) +- (0, 0.0756997)
      (Radboud-L, 0.428514) +- (0, 0.03389)
      (Radboud-L $\rightarrow$\\ Radboud-R, 0.679046) +- (0, 0.0406002)};
    \addplot [fill=caribbeangreen2, draw=caribbeangreen2, error bars/.cd, y dir=both, y explicit, error bar style={thick}, error bar style=black] coordinates {
      (PANDA, 0.46159) +- (0, 0.0237712)
      (Karolinska, 0.5125) +- (0, 0.0438395)
      (Radboud, 0.4706) +- (0, 0.0449893)
      (Radboud $\rightarrow$\\ Karolinska, 0.979536) +- (0, 0.0627807)
      (Radboud-U, 0.498756) +- (0, 0.0584365)
      (Radboud-U $\rightarrow$\\ Karolinska-U, 1.35598) +- (0, 0.0348709)
      (Radboud-L, 0.496386) +- (0, 0.0348357)
      (Radboud-L $\rightarrow$\\ Radboud-R, 0.744771) +- (0, 0.0075598)};
    \addplot [fill=darkpastelpurple2, draw=darkpastelpurple2, error bars/.cd, y dir=both, y explicit, error bar style={thick}, error bar style=black] coordinates {
      (PANDA, 0.685249) +- (0, 0.0267624)
      (Karolinska, 0.750551) +- (0, 0.0350043)
      (Radboud, 0.6054) +- (0, 0.0601269)
      (Radboud $\rightarrow$\\ Karolinska, 1.58796) +- (0, 0.0747343)
      (Radboud-U, 0.698259) +- (0, 0.0604846)
      (Radboud-U $\rightarrow$\\ Karolinska-U, 1.43725) +- (0, 0.0294423)
      (Radboud-L, 0.65743) +- (0, 0.0304295)
      (Radboud-L $\rightarrow$\\ Radboud-R, 1.27062) +- (0, 0.015859)};
  \end{axis}
  \end{tikzpicture}\vspace{-2.5mm}
    \begin{tikzpicture}
  \centering
  \begin{axis}[
        ybar, axis on top,
        title={},
        height=4.525cm, width=18.0cm, 
        bar width=0.35cm,
        ymajorgrids, tick align=inside,
        enlarge y limits={value=.1,upper},
        ymin=0, ymax=2.4,
        set layers,
        ytick={0.5, 1.0, 1.5, 2.0},
        axis x line*=bottom,
        axis y line*=left,
        y axis line style={opacity=0},
        tickwidth=0pt,
        enlarge x limits=true,
        extra y tick style={
            grid style={
                solid,
                line width = 2.5pt,
                /pgfplots/on layer=axis background,
            },
        },
        tick style={
            grid style={
                dashed,
                /pgfplots/on layer=axis background,
            },
        },
        legend style={
            at={(0.5,-0.2)},
            anchor=north,
            legend columns=-1,
            /tikz/every even column/.append style={column sep=0.325cm},
            font=\footnotesize
        },
        ylabel={MAE ($\downarrow$)},
        ylabel style={font=\footnotesize},
        symbolic x coords={
           PANDA,
           Karolinska,
           Radboud, 
           Radboud $\rightarrow$\\ Karolinska,
           Radboud-U, 
           Radboud-U $\rightarrow$\\ Karolinska-U,
           Radboud-L, 
           Radboud-L $\rightarrow$\\ Radboud-R},
       xtick=data,
       xticklabel=\empty,
       x tick label style={text width = 2.25cm, align=center, font=\footnotesize},
    ]
    \addplot [fill=skyblue2, draw=skyblue2, line width=0mm, error bars/.cd, y dir=both, y explicit, error bar style={thick}, error bar style=black] coordinates { 
      (PANDA, 0.333238) +- (0, 0.0295276)
      (Karolinska, 0.292831) +- (0, 0.0362095)
      (Radboud, 0.3952) +- (0, 0.045845)
      (Radboud $\rightarrow$\\ Karolinska, 0.877604) +- (0, 0.151029)
      (Radboud-U, 0.437313) +- (0, 0.05033)
      (Radboud-U $\rightarrow$\\ Karolinska-U, 0.875697) +- (0, 0.0546145)
      (Radboud-L, 0.423293) +- (0, 0.037554)
      (Radboud-L $\rightarrow$\\ Radboud-R, 0.667356) +- (0, 0.0480552)};
    \addplot [fill=caribbeangreen2, draw=caribbeangreen2, error bars/.cd, y dir=both, y explicit, error bar style={thick}, error bar style=black] coordinates {
      (PANDA, 0.435153) +- (0, 0.0276223)
      (Karolinska, 0.478309) +- (0, 0.0302478)
      (Radboud, 0.4524) +- (0, 0.0317843)
      (Radboud $\rightarrow$\\ Karolinska, 0.973703) +- (0, 0.0515132)
      (Radboud-U, 0.477861) +- (0, 0.0565771)
      (Radboud-U $\rightarrow$\\ Karolinska-U, 1.28871) +- (0, 0.0446399)
      (Radboud-L, 0.473092) +- (0, 0.0506024)
      (Radboud-L $\rightarrow$\\ Radboud-R, 0.722227) +- (0, 0.0116131)};
    \addplot [fill=darkpastelpurple2, draw=darkpastelpurple2, error bars/.cd, y dir=both, y explicit, error bar style={thick}, error bar style=black] coordinates {
      (PANDA, 0.682759) +- (0, 0.0208926)
      (Karolinska, 0.726103) +- (0, 0.0359562)
      (Radboud, 0.6148) +- (0, 0.0685519)
      (Radboud $\rightarrow$\\ Karolinska, 1.14021) +- (0, 0.0152437)
      (Radboud-U, 0.692786) +- (0, 0.0660329)
      (Radboud-U $\rightarrow$\\ Karolinska-U, 1.55797) +- (0, 0.0139222)
      (Radboud-L, 0.657831) +- (0, 0.023196)
      (Radboud-L $\rightarrow$\\ Radboud-R, 1.29678) +- (0, 0.0174482)};
  \end{axis}
  \end{tikzpicture}\vspace{-2.5mm}
    \begin{tikzpicture}
  \centering
  \begin{axis}[
        ybar, axis on top,
        title={},
        height=4.525cm, width=18.0cm, 
        bar width=0.35cm,
        ymajorgrids, tick align=inside,
        enlarge y limits={value=.1,upper},
        ymin=0, ymax=2.4,
        set layers,
        ytick={0.5, 1.0, 1.5, 2.0},
        axis x line*=bottom,
        axis y line*=left,
        y axis line style={opacity=0},
        tickwidth=0pt,
        enlarge x limits=true,
        extra y tick style={
            grid style={
                solid,
                line width = 2.5pt,
                /pgfplots/on layer=axis background,
            },
        },
        tick style={
            grid style={
                dashed,
                /pgfplots/on layer=axis background,
            },
        },
        legend style={
            at={(0.5,-0.2)},
            anchor=north,
            legend columns=-1,
            /tikz/every even column/.append style={column sep=0.325cm},
            font=\footnotesize
        },
        ylabel={MAE ($\downarrow$)},
        ylabel style={font=\footnotesize},
        symbolic x coords={
           PANDA,
           Karolinska,
           Radboud, 
           Radboud $\rightarrow$\\ Karolinska,
           Radboud-U, 
           Radboud-U $\rightarrow$\\ Karolinska-U,
           Radboud-L, 
           Radboud-L $\rightarrow$\\ Radboud-R},
       xtick=data,
       xticklabel=\empty,
       x tick label style={text width = 2.25cm, align=center, font=\footnotesize},
    ]
    \addplot [fill=skyblue2, draw=skyblue2, line width=0mm, error bars/.cd, y dir=both, y explicit, error bar style={thick}, error bar style=black] coordinates { 
      (PANDA, 0.340613) +- (0, 0.0277057)
      (Karolinska, 0.298897) +- (0, 0.0278928)
      (Radboud, 0.4146) +- (0, 0.0324968)
      (Radboud $\rightarrow$\\ Karolinska, 0.997368) +- (0, 0.115618)
      (Radboud-U, 0.426617) +- (0, 0.0438439)
      (Radboud-U $\rightarrow$\\ Karolinska-U, 0.888977) +- (0, 0.026325)
      (Radboud-L, 0.451406) +- (0, 0.0710333)
      (Radboud-L $\rightarrow$\\ Radboud-R, 0.678887) +- (0, 0.039581)};
    \addplot [fill=caribbeangreen2, draw=caribbeangreen2, error bars/.cd, y dir=both, y explicit, error bar style={thick}, error bar style=black] coordinates {
      (PANDA, 0.493966) +- (0, 0.0351653)
      (Karolinska, 0.523713) +- (0, 0.03708)
      (Radboud, 0.5228) +- (0, 0.0428504)
      (Radboud $\rightarrow$\\ Karolinska, 0.843614) +- (0, 0.0270459)
      (Radboud-U, 0.558706) +- (0, 0.0593084)
      (Radboud-U $\rightarrow$\\ Karolinska-U, 1.38426) +- (0, 0.0454923)
      (Radboud-L, 0.514859) +- (0, 0.0523877)
      (Radboud-L $\rightarrow$\\ Radboud-R, 0.790577) +- (0, 0.0264311)};
    \addplot [fill=darkpastelpurple2, draw=darkpastelpurple2, error bars/.cd, y dir=both, y explicit, error bar style={thick}, error bar style=black] coordinates {
      (PANDA, 0.73908) +- (0, 0.0302057)
      (Karolinska, 0.780331) +- (0, 0.0361553)
      (Radboud, 0.7012) +- (0, 0.0686801)
      (Radboud $\rightarrow$\\ Karolinska, 1.09262) +- (0, 0.0126168)
      (Radboud-U, 0.760448) +- (0, 0.0526171)
      (Radboud-U $\rightarrow$\\ Karolinska-U, 1.58745) +- (0, 0.0285942)
      (Radboud-L, 0.704819) +- (0, 0.0355711)
      (Radboud-L $\rightarrow$\\ Radboud-R, 1.38091) +- (0, 0.0100164)};
  \end{axis}
  \end{tikzpicture}\vspace{-2.5mm}
    \begin{tikzpicture}
  \centering
  \begin{axis}[
        ybar, axis on top,
        title={},
        height=4.525cm, width=18.0cm, 
        bar width=0.35cm,
        ymajorgrids, tick align=inside,
        enlarge y limits={value=.1,upper},
        ymin=0, ymax=2.4,
        set layers,
        ytick={0.5, 1.0, 1.5, 2.0},
        axis x line*=bottom,
        axis y line*=left,
        y axis line style={opacity=0},
        tickwidth=0pt,
        enlarge x limits=true,
        extra y tick style={
            grid style={
                solid,
                line width = 2.5pt,
                /pgfplots/on layer=axis background,
            },
        },
        tick style={
            grid style={
                dashed,
                /pgfplots/on layer=axis background,
            },
        },
        legend style={
            at={(0.5,-0.2)},
            anchor=north,
            legend columns=-1,
            /tikz/every even column/.append style={column sep=0.325cm},
            font=\footnotesize
        },
        ylabel={MAE ($\downarrow$)},
        ylabel style={font=\footnotesize},
        symbolic x coords={
           PANDA,
           Karolinska,
           Radboud, 
           Radboud $\rightarrow$\\ Karolinska,
           Radboud-U, 
           Radboud-U $\rightarrow$\\ Karolinska-U,
           Radboud-L, 
           Radboud-L $\rightarrow$\\ Radboud-R},
       xtick=data,
       xticklabel=\empty,
       x tick label style={text width = 2.25cm, align=center, font=\footnotesize},
    ]
    \addplot [fill=skyblue2, draw=skyblue2, line width=0mm, error bars/.cd, y dir=both, y explicit, error bar style={thick}, error bar style=black] coordinates { 
      (PANDA, 0.345881) +- (0, 0.0244823)
      (Karolinska, 0.309191) +- (0, 0.0349052)
      (Radboud, 0.4102) +- (0, 0.0397638)
      (Radboud $\rightarrow$\\ Karolinska, 0.957527) +- (0, 0.118647)
      (Radboud-U, 0.460448) +- (0, 0.0322511)
      (Radboud-U $\rightarrow$\\ Karolinska-U, 0.843094) +- (0, 0.080393)
      (Radboud-L, 0.436546) +- (0, 0.0495801)
      (Radboud-L $\rightarrow$\\ Radboud-R, 0.677416) +- (0, 0.0264209)};
    \addplot [fill=caribbeangreen2, draw=caribbeangreen2, error bars/.cd, y dir=both, y explicit, error bar style={thick}, error bar style=black] coordinates {
      (PANDA, 0.459866) +- (0, 0.0294934)
      (Karolinska, 0.537132) +- (0, 0.0380013)
      (Radboud, 0.4484) +- (0, 0.0324074)
      (Radboud $\rightarrow$\\ Karolinska, 1.05841) +- (0, 0.137268)
      (Radboud-U, 0.478856) +- (0, 0.0659486)
      (Radboud-U $\rightarrow$\\ Karolinska-U, 1.2674) +- (0, 0.0421948)
      (Radboud-L, 0.482731) +- (0, 0.0405523)
      (Radboud-L $\rightarrow$\\ Radboud-R, 0.723062) +- (0, 0.0167067)};
    \addplot [fill=darkpastelpurple2, draw=darkpastelpurple2, error bars/.cd, y dir=both, y explicit, error bar style={thick}, error bar style=black] coordinates {
      (PANDA, 0.680077) +- (0, 0.0367422)
      (Karolinska, 0.742647) +- (0, 0.0282275)
      (Radboud, 0.6262) +- (0, 0.0420328)
      (Radboud $\rightarrow$\\ Karolinska, 1.3429) +- (0, 0.0864625)
      (Radboud-U, 0.685572) +- (0, 0.0546721)
      (Radboud-U $\rightarrow$\\ Karolinska-U, 1.43672) +- (0, 0.0216737)
      (Radboud-L, 0.681928) +- (0, 0.0541439)
      (Radboud-L $\rightarrow$\\ Radboud-R, 1.35618) +- (0, 0.0118524)};
  \end{axis}
  \end{tikzpicture}\vspace{-2.5mm}
    \begin{tikzpicture}
  \centering
  \begin{axis}[
        ybar, axis on top,
        title={},
        height=4.525cm, width=18.0cm, 
        bar width=0.35cm,
        ymajorgrids, tick align=inside,
        enlarge y limits={value=.1,upper},
        ymin=0, ymax=2.4,
        set layers,
        ytick={0.5, 1.0, 1.5, 2.0},
        axis x line*=bottom,
        axis y line*=left,
        y axis line style={opacity=0},
        tickwidth=0pt,
        enlarge x limits=true,
        extra y tick style={
            grid style={
                solid,
                line width = 2.5pt,
                /pgfplots/on layer=axis background,
            },
        },
        tick style={
            grid style={
                dashed,
                /pgfplots/on layer=axis background,
            },
        },
        legend style={
            at={(0.5,-0.30)},
            anchor=north,
            legend columns=-1,
            /tikz/every even column/.append style={column sep=0.325cm},
            font=\footnotesize
        },
        ylabel={MAE ($\downarrow$)},
        ylabel style={font=\footnotesize},
        symbolic x coords={
           PANDA,
           Karolinska,
           Radboud, 
           Radboud $\rightarrow$\\ Karolinska,
           Radboud-U, 
           Radboud-U $\rightarrow$\\ Karolinska-U,
           Radboud-L, 
           Radboud-L $\rightarrow$\\ Radboud-R},
       xtick=data,
       x tick label style={text width = 2.25cm, align=center, font=\footnotesize},
    ]
    \addplot [fill=skyblue2, draw=skyblue2, line width=0mm, error bars/.cd, y dir=both, y explicit, error bar style={thick}, error bar style=black] coordinates { 
      (PANDA, 0.393966) +- (0, 0.0235252)
      (Karolinska, 0.323346) +- (0, 0.0282688)
      (Radboud, 0.467) +- (0, 0.039792)
      (Radboud $\rightarrow$\\ Karolinska, 0.780401) +- (0, 0.0871312)
      (Radboud-U, 0.502488) +- (0, 0.0466045)
      (Radboud-U $\rightarrow$\\ Karolinska-U, 0.864608) +- (0, 0.0806058)
      (Radboud-L, 0.452209) +- (0, 0.0672544)
      (Radboud-L $\rightarrow$\\ Radboud-R, 0.691372) +- (0, 0.0433038)};
    \addplot [fill=caribbeangreen2, draw=caribbeangreen2, error bars/.cd, y dir=both, y explicit, error bar style={thick}, error bar style=black] coordinates {
      (PANDA, 0.543295) +- (0, 0.0367517)
      (Karolinska, 0.559191) +- (0, 0.0422058)
      (Radboud, 0.5892) +- (0, 0.0536522)
      (Radboud $\rightarrow$\\ Karolinska, 1.15655) +- (0, 0.103995)
      (Radboud-U, 0.606219) +- (0, 0.0389373)
      (Radboud-U $\rightarrow$\\ Karolinska-U, 1.34475) +- (0, 0.0302089)
      (Radboud-L, 0.573494) +- (0, 0.03514)
      (Radboud-L $\rightarrow$\\ Radboud-R, 0.845328) +- (0, 0.0183696)};
    \addplot [fill=darkpastelpurple2, draw=darkpastelpurple2, error bars/.cd, y dir=both, y explicit, error bar style={thick}, error bar style=black] coordinates {
      (PANDA, 0.773372) +- (0, 0.0314381)
      (Karolinska, 0.765257) +- (0, 0.0312073)
      (Radboud, 0.7968) +- (0, 0.0416)
      (Radboud $\rightarrow$\\ Karolinska, 1.96327) +- (0, 0.0948864)
      (Radboud-U, 0.824627) +- (0, 0.0553252)
      (Radboud-U $\rightarrow$\\ Karolinska-U, 1.89608) +- (0, 0.0477629)
      (Radboud-L, 0.76988) +- (0, 0.0697698)
      (Radboud-L $\rightarrow$\\ Radboud-R, 1.44445) +- (0, 0.0176621)};
    \legend{
            \textit{ABMIL},
            \textit{Mean Feature},
            \textit{kNN}
            }
  \end{axis}
  \end{tikzpicture}\vspace{-1.0mm}
    \caption{\textbf{Comparison of downstream ISUP grade classification models for additional feature extractors (MAE).} Performance of the three ISUP grade models \textit{ABMIL}, \textit{Mean Feature}, and \textit{kNN} when using different feature extractors: Phikon-v2 \textbf{(top row)}, UNI2-h, H-optimus-0, H0-mini, Prov-GigaPath, and Virchow \textbf{(bottom row)}. All results are reported as mean $\pm$ standard deviation over $10$ cross-validation folds. This figure complements Figure~\ref{fig:main_results_attmil_mean-pool_knn_mae} by presenting the same comparison for the remaining six feature extractors, and presents the same comparison as Figure~\ref{fig:main_results_attmil_mean-pool_knn_kappa_remaining6} but using MAE (lower is better) instead of quadratically weighted Cohen's kappa.}
    \label{fig:main_results_attmil_mean-pool_knn_mae_remaining6}
\end{figure*}

\end{document}